\documentclass[12pt,leqno,english]{extarticle}

\input{glyphtounicode}\pdfgentounicode=1
\usepackage[utf8]{inputenc}
\usepackage[english]{babel}
\usepackage[letterpaper,hmargin=1.25in,top=1.5in,textheight=570pt]{geometry}
\usepackage{amsmath,amssymb,amsthm}
\usepackage{booktabs,tabularx,needspace,placeins,xurl}
\usepackage{caption}
\usepackage{graphicx}
\usepackage{setspace}
\usepackage{etoolbox}
\usepackage{enumitem}
\usepackage[authoryear]{natbib}
\usepackage[colorlinks=true,citecolor=blue,linkcolor=black,urlcolor=black,filecolor=black]{hyperref}
\usepackage{url}
\hypersetup{pdftitle={Observable Self-Restraint in Repeated Cheap Talk},pdfauthor={Haohui Wang},pdfsubject={},pdfkeywords={}}

\patchcmd{\abstract}{\small}{}{}{}

\newtheorem{theorem}{Theorem}
\newtheorem{benchmark}{Benchmark}
\newtheorem{proposition}{Proposition}
\newtheorem{lemma}{Lemma}
\newtheorem{corollary}{Corollary}
\newtheorem{assumption}{Assumption}
\newtheorem{definition}{Definition}
\newtheorem{remark}{Remark}

\newcommand{\admissible}{\widetilde{\mathcal V}}

\newcommand{\enteringhistory}{\nu}
\newcommand{\initialhistory}{0^{N-1}}

\begin{document}
\begin{center}
{\LARGE Observable Self-Restraint in Repeated Cheap Talk\par}
\vskip 1em
{\large Haohui Wang\par}
\vskip 0.3em
{\normalsize Department of Economics, University of Washington\par}
{\normalsize \href{mailto:haohuiw@uw.edu}{haohuiw@uw.edu}\par}
\vskip 1em
{\large November 8, 2026\par}
\vskip 0.4em
\end{center}
\vskip 0.5em
\begin{abstract}
A long-lived sender seeks to persuade short-lived receivers to take his preferred action. Receivers observe his past advice but cannot verify its accuracy. Restrictions on recommendation histories alone can support a partially informative equilibrium. Under rolling quotas, which cap favorable recommendations within a fixed memory window, favorable advice is more costly when past favorable recommendations are more numerous or more recent. In an informative equilibrium, a violation can itself convey enough information about the current state for the receiver to follow the recommendation. After some sequences of states, the sender therefore breaks his own restriction, ending informative communication on the equilibrium path. As the sender becomes patient, suitable rolling quotas approach his normalized equilibrium payoff ceiling.
\end{abstract}

\clearpage

\FloatBarrier
\section{Introduction}\label{sec:introduction}

An informed adviser may benefit from making favorable recommendations more often than his audience would prefer, and such conflicts of interest limit informative cheap talk \citep{crawford1982strategic}. The assessments underlying recommendations by financial analysts, venture accelerators, and other professional advisers are difficult to verify, often even after the fact---a defining feature of credence goods \citep{darby1973free,dulleck2006doctors}. Yet an adviser can make his recommendations public, leaving a record that his audience can observe. Can he use this record to create incentives for self-restraint and thereby make his advice informative?

I study a long-lived sender who privately observes an independently drawn state each period and advises a sequence of myopic receivers, who observe the recommendation record but receive no feedback about states. The sender announces a nonbinding rule whose permitted histories receivers monitor; a rolling quota, for example, caps favorable recommendations, or endorsements, within every moving window of periods. Under such a quota, each endorsement uses capacity that could serve a later opportunity, so the sender weighs its current gain against its cost, the continuation value it displaces. The restriction thus links otherwise independent decisions and creates incentives for self-restraint. Following \citet{lin2024credible}, the rule is \emph{credible} if no undetectable deviation, such as recommending a mediocre opportunity and leaving less capacity for later ones, is profitable. I call a credible rule \emph{sustainable} if no observable violation is profitable either. A violation ends informative communication in the equilibria I study, but it can reveal a sufficiently valuable current opportunity and secure a favorable response, so breakdown can occur along the equilibrium path.

The first set of results concerns how the record shapes incentives for self-restraint. Under a rolling quota, advice becomes more selective when past endorsements are more numerous or more recent, since a more recent endorsement keeps capacity occupied longer (Figure~\ref{fig:multistate-linking}, top). Unlike a lifetime budget, whose state is the number of units left, a rolling quota has the record itself as its state. I order records by temporal precedence: one record precedes another if every recent window contains weakly fewer endorsements. Closure under this order ranks continuation values and, under an upper-tail obedience condition, reduces receiver obedience to a check at the empty record, where the cost of an endorsement is lowest. The costs that set the cutoffs are differences of continuation values. Ordering continuation values does not by itself order these costs; rolling quotas supply the additional structure. Under a rolling quota, a geometrically discounted count of past endorsements need not preserve this ranking, whatever its discount factor.

The second set of results shows how quota size and memory length move these cutoffs. The quota determines how much capacity the sender has, and the memory how long each endorsement occupies it. At the empty record, a larger quota therefore lowers the cost of an endorsement and extends favorable advice to more states, whereas longer memory does the reverse. Whatever the memory, this cost is bounded by a quantity that falls to zero as the quota grows, because additional capacity pushes the opportunity displaced by a current endorsement further into the future. At each discount factor, only finitely many quotas are therefore \emph{implementable}, meaning that receivers follow the sender's optimal advice within the restriction, and choosing the best rolling quota reduces to a finite search over quotas, each paired with its shortest implementing memory.

The third set of results identifies the trade-off between disciplining advice and sustaining communication. Just as obedience can be checked at the empty record, sustainability can be checked at the binding record with the most recent endorsements, where the sender has the least to lose from a violation. Too little capacity, or too long a wait for its renewal, leaves too little continuation value to deter a violation; too much capacity, or too rapid renewal, leaves too little continuation cost to keep endorsements selective enough for receivers to follow. The quotas that are both implementable and sustainable therefore form an interval at each memory, and the memories form an interval at each quota (Figure~\ref{fig:multistate-linking}, bottom).

Finally, I compare these restrictions with the sender's payoff ceiling, which in the binary-action setting equals the bound of \citet{kuvalekar2022goodwill}, hereafter KLR. KLR characterize how much an expert can earn without feedback; the results above concern how an observable restriction disciplines advice and when the sender keeps to it. When obedience requires a positive cost of self-restraint, I express the ceiling in closed form as the best credible, receiver-obedient partitional payoff minus the expected cost of self-restraint, a cost that persists however patient the sender becomes. Under the upper-tail condition, a rolling quota attains the ceiling when it first becomes implementable, because its empty-record cost then equals the minimum required for obedience; greater patience raises that cost and lowers its normalized payoff. The best rolling-quota payoff therefore follows a sawtooth that converges to the ceiling as the sender becomes patient (Figure~\ref{fig:klr-sawtooth}). Restrictions admitting only part of the next quota layer can narrow the shortfall at a fixed discount factor.

To illustrate, consider a startup accelerator that recommends ventures to outside investors. Each period, it privately assesses a new venture, good or bad with equal probability and independently across periods, and a different investor observes the recommendation and decides whether to fund. The table reports the expected present value of each party's payoff from the current venture, with the accelerator's payoff listed first.

\begin{center}
\begin{tabular}{@{}lcc@{}}
\toprule
Venture prospects & Fund & Do not fund\\
\midrule
Good & $(1,1)$ & $(0,0)$\\
Bad & $(1/4,-2)$ & $(0,0)$\\
\bottomrule
\end{tabular}
\end{center}

Returns arrive years later and are noisy, so investors see the record long before outcomes can reveal its reliability. Without information, the investor does not fund, and in a one-shot interaction the accelerator would recommend funding both types if the investor followed its advice. Suppose instead that the accelerator publishes its recommendation history and announces a quota of at most one endorsement in any two consecutive periods: it endorses a venture exactly when the venture is good and no endorsement was issued in the preceding period. Investors evaluate a violating endorsement for what it reveals about the current venture, and communication becomes uninformative from the following period on. When $\beta_S\ge(\sqrt{21}-1)/5$, endorsing a bad venture does not justify forgoing a future endorsement of a good one, so endorsements are informative and investors follow them. For $(\sqrt{21}-1)/5\le\beta_S\le\sqrt3-1$, however, the accelerator violates the quota when two good ventures arrive consecutively: investors infer that the venture is good and fund it, communication ends, and this occurs eventually with probability one (Appendix~\ref{app:accelerator-calculation}).

Section~\ref{sec:model} presents the model, Section~\ref{sec:quota} the record order and the implementation test, Section~\ref{sec:rolling-design} rolling-quota cutoffs and design, Section~\ref{sec:sustainability-analysis} sustainability, and Section~\ref{sec:optimization} the comparison with the payoff ceiling.

\paragraph{Related literature.}
The paper connects to linking mechanisms that discipline reports by allocating scarce opportunities across decisions \citep{jackson2007overcoming,frankel2014aligned,frankel2016discounted,ball2026quota}; receivers impose action quotas in \citet{bergemann2026managing}. In these papers, a designer or the receiver imposes the quota. In \citet{lin2024credible}, the sender commits to an observable distribution of messages, and their finite-sample formulation allocates fixed quotas across a batch of states; here the quota applies to rolling message windows and is self-enforced by the value of future communication. In a one-shot cheap-talk game, \citet{chakraborty2007comparative} show that comparisons across simultaneously observed issues support credible rankings, and with i.i.d.\ issues, per-issue payoffs under a complete ranking approach their full-information values as the number of issues grows. Here recommendations are linked across time rather than within a single message, so the timing of past advice shapes the selectivity of current advice (Section~\ref{subsec:history-cutoffs}).

Linking can also be sustained without commitment when the parties are long-lived. \citet{escobar2013efficiency} show that patient players whose private types follow independent, possibly persistent Markov chains can generically approximate any Pareto-efficient payoff above a stationary minmax value by linking their cheap-talk reports over time. In delegation, \citet{lipnowski2020repeated} show that a long-lived principal who never observes project quality optimally delegates through a dynamic capital budget, tolerating some bad projects as a reward for the agent's restraint, while the agent's autonomy inevitably shrinks over time. Here receivers are myopic and cannot reward restraint with later concessions; it is paid for with forgone future capacity, which renews as endorsements leave the record.

A direct precedent for disciplining advice through a record without feedback about treatment appropriateness is \citet{frankel2014experts}. In their model, customers reward less profitable treatments with future business. Here the sender announces his own quota: forgone capacity disciplines advice within the restriction, while the loss of future informative communication provides an incentive to preserve it. With heterogeneous experts, discounted quotas within blocks make every expert type indifferent between treatments during the truthful phase of each block, so the continuation cost of the more profitable treatment equals its premium; their discounted counts weight a play by its date from the start of a block, so an earlier play uses more of the quota. Indifference also drives the credible comparative statements of \citet{chakraborty2010persuasion}. Here sender gains vary across states, an earlier endorsement releases capacity sooner, and the continuation cost lies between the gains of adjacent states, selecting which states receive favorable advice; with equal gains, selective advice would require exact indifference and would bring the sender no gain over babbling.

\citet{kuvalekar2022goodwill} provide the closest comparison. They study a long-lived decision maker whose actions are private, which removes her dynamic incentives, so the expert's attainable payoffs are those with a myopic receiver; as here, the receiver receives no feedback about states. Their characterization does not require the upper-tail condition used here: the highest payoff attainable with uncapped money burning bounds the sender's equilibrium payoff, and their Corollary~1 shows that this bound is attained exactly at every discount factor above some threshold below one. Their general construction uses public randomization, and their project application holds a fixed state cutoff while the implementation scale declines. Rolling quotas instead renew capacity through the message record and keep implementation binary without public randomization, but their cutoffs vary with history, so obedience must hold at cells more selective than the initial obedient tail, which the upper-tail condition guarantees.

Elsewhere, the sender is disciplined by forces absent here. A long-lived receiver can condition future responses on the message record \citep{renault2013dynamic,margaria2018dynamic,meng2021value,jain2026dynamic}. Feedback about past accuracy can discipline a sender facing short-lived receivers \citep{best2024persuasion,pei2023repeated}, and in credence-goods markets, consumers' occasional rejections can reveal the problem's severity \citep{fong2022trust}. Here the recommendation record is the sole source of discipline.

\FloatBarrier
\section{Model and Equilibrium}
\label{sec:model}

\subsection{Environment}

I consider an infinite-horizon repeated cheap-talk game between a long-lived sender and a myopic receiver. In each period $t=1,2,\ldots$, a state $\theta_t$ is drawn independently from a full-support prior over $\Theta=\{\theta_1,\ldots,\theta_n\}$, with $\mu_i:=\Pr(\theta_t=\theta_i)$. The sender privately observes $\theta_t$ and sends a recommendation $m_t\in\{0,1\}$, after which the receiver chooses an action $a_t\in\{0,1\}$. States are unverifiable, and the receiver receives no feedback about their realizations, either directly or through realized payoffs. The sender can make his recommendation history public, allowing the receiver to condition her action on past recommendations and the current one. Each recommendation is labeled by the action it recommends, but its informational content can vary with the history: under the accelerator's announced one-in-two rule, recommendation~$0$ after $0$ reveals a bad venture, whereas $0$ after $1$ is forced and conveys no information.

The stage payoffs are $u_S(\theta_t,a_t)$ and $u_R(\theta_t,a_t)$. Write
$g_i:=u_S(\theta_i,1)-u_S(\theta_i,0)$ and $r_i:=u_R(\theta_i,1)-u_R(\theta_i,0)$, and order states so that $g_1<\cdots<g_n$, with $g_n>0$. Because states are independent, subtracting the sender's state-specific action-$0$ payoff preserves all incentive comparisons, so I normalize $u_S(\theta_i,0)=0$ and $u_S(\theta_i,1)=g_i$. The receiver strictly prefers action $0$ under the prior: $\sum_i\mu_ir_i<0$. The sender discounts future payoffs by $\beta_S\in(0,1)$, while the receiver maximizes her current expected payoff; equivalently, she is a sequence of short-lived receivers, so her action must be optimal given her beliefs at every history.

\begin{assumption}[Tie-breaking]\label{ass:tie-breaking}
Whenever a player is indifferent among optimal choices, the player selects the choice that gives the other player the highest current expected payoff, evaluated using the information available to the choosing player. Sender optimality includes current and discounted continuation payoffs; receiver optimality concerns the current payoff. Any remaining ties are resolved by a fixed ordering, with action or recommendation $0$ selected when both $0$ and $1$ remain tied.
\end{assumption}

The sender cannot commit to a signaling strategy, but he can announce a nonbinding, history-dependent strategy whose induced restriction on recommendation histories the receiver monitors. A violation is publicly detectable; for the announcement to be credible, the sender must have no profitable deviation that escapes detection.

\subsection{Receiver Obedience and the Relevant State Tails}\label{sec:binary-environment}

If an informative one-shot equilibrium exists, pooling states by the sender's preferred action is also an equilibrium and attains his statewise bound $\mathbb E[\max_a u_S(\theta,a)]$, so repetition cannot raise his payoff. I therefore maintain the absence of an informative one-shot equilibrium, except where the one-shot benchmark is explicitly identified. Assume that at least one nonempty upper tail of the sender-gain ordering induces action $1$. Define
\begin{equation}
k_R:=\min\left\{k:\sum_{i=k}^n\mu_i r_i\geq0\right\},
\qquad \underline c_R:=g_{k_R-1}.
\label{eq:finite-receiver-threshold}
\end{equation}
The prior preference implies $k_R\geq2$. The \emph{upper-tail obedience condition (UT)}, imposed where stated, requires every more selective upper tail to remain receiver-obedient:
\begin{equation*}
\sum_{i=k}^n\mu_i r_i\geq0
\qquad\text{for every }k_R<k\le n.
\tag{UT}\label{eq:multiple-upper-tail}
\end{equation*}
Single crossing of receiver gains from negative to positive in this ordering, for instance nondecreasing receiver gains with $r_n\geq0$, is sufficient for (UT).
Under (UT), a nonempty favorable cell selected at cost $c$, consisting of the states with $g_i>c$ and, by Assumption~\ref{ass:tie-breaking}, tied states with $r_i>0$, is receiver-obedient exactly when $c\geq\underline c_R$, and the absence of an informative one-shot equilibrium is equivalent to $\underline c_R>0$: the largest obedient upper tail excludes a state in which the sender strictly prefers action~$1$, so a positive continuation cost is needed to deter favorable recommendations there. Let $F(c):=\sum_{i=1}^{n}\mu_i(g_i-c)_+$, which is continuous and strictly decreasing for $c<g_n$. This function measures the expected gain net of a common continuation cost. When $\underline c_R>0$, define
\begin{equation}
\beta^\ast:=\frac{\underline c_R}{\underline c_R+F(\underline c_R)}.
\label{eq:patience-threshold-primitives}
\end{equation}
This threshold has the same form as the patience threshold in the project application of \citet[Proposition~1]{kuvalekar2022goodwill}, with $\underline c_R$ and $F(\underline c_R)$ corresponding to their cutoff and value. Section~\ref{subsec:implementable-quotas} establishes implementation with finite rolling quotas above it.

The two-state examples set $\Theta=\{L,H\}$ with $p=\mu(H)\in(0,1)$, sender gains $g_L=\ell\in[0,1)$ and $g_H=1$, and receiver gains $r_L<0<r_H$ with $(1-p)r_L+pr_H<0$; then (UT) holds and $\underline c_R=\ell$.

\subsection{History Restrictions and Monitoring}
\label{subsec:announced-arrangements}\label{subsec:public-monitoring}

Fix a positive integer $N$, the memory length. At the beginning of period $t$, the full public history is $\nu_t=(m_1,\ldots,m_{t-1})$, and its most recent $N$ recommendations form the rolling history $\nu_t^N=(m_{t-N},\ldots,m_{t-1})$, with pre-game entries specified below. Let $\mathcal H=\bigcup_{t\geq1}\{0,1\}^{t-1}$. The sender's announced signaling strategy is $\phi:\Theta\times\mathcal H\to\Delta(\{0,1\})$, and the receiver's response is $\sigma:\{0,1\}\times\mathcal H\to\{0,1\}$.

\begin{definition}[Induced history set]
\label{def:induced-history-set}
The history set induced by a signaling strategy $\phi$, denoted $\mathcal V^{N,\phi}\subseteq\{0,1\}^N$, is the set of length-$N$ rolling histories that occur with positive probability under $\phi$ and the state process.
\end{definition}

For the announced strategy $\phi$, write $\widetilde{\mathcal V}^{N}=\mathcal V^{N,\phi}$; the receiver monitors the announcement by checking whether each realized rolling history belongs to this set. For $m\in\{0,1\}$, let $T_m$ drop the oldest entry of a history and append $m$, $T_m(x_1,\ldots,x_k)=(x_2,\ldots,x_k,m)$, with $T_m$ the identity on the empty history, so that $\nu_{t+1}^N=T_{m_t}\nu_t^N$. A departure from the announcement is publicly detected in period $t$ if this rolling history falls outside $\widetilde{\mathcal V}^{N}$; I call such a departure a \emph{violation}. Departures that keep every realized rolling history in $\widetilde{\mathcal V}^{N}$ are undetectable. A public history is a \emph{pre-breakdown history} if no violation has been detected; the full public history records a past violation after the rolling window has moved beyond it.

I focus on stationary strategies with finite rolling-history dependence: the sender chooses the memory $N$ as part of the announcement, and at pre-breakdown histories the signaling rule, the receiver's response, and the admissibility of recommendations depend on the past only through $\nu_t^N$. With finite memory, permission to recommend returns as earlier recommendations leave the rolling window, so capacity renews; Section~\ref{sec:binary-memory} shows that every finite memory yields a strictly higher payoff than a lifetime cap with the same quota.

\begin{assumption}[Initial history]
\label{ass:initial-history}
Vacant pre-game positions are filled with recommendation $0$, so that $\nu_1^N=0^N\in\widetilde{\mathcal V}^{N}$.
\end{assumption}

This initialization represents a new relationship with no recommendation record. A detected violation triggers permanent babbling: the sender sends a recommendation that does not depend on the state, and the receiver takes action $0$ thereafter, which is sequentially rational and gives the sender zero. This continuation follows every violation, including one that the sender's actual policy prescribes on path (Section~\ref{subsec:equilibrium}). For the violation period itself, I consider two receiver responses. Under \emph{immediate reversion} (IR), the receiver already takes action $0$. Under \emph{belief updating} (BU), she best responds to the posterior that the violation induces, so an anticipated violation can be informative. The sender's normalized continuation payoff is
\[
\Pi(\theta_t,\nu_t;\phi,\sigma)
=(1-\beta_S)\mathbb E_t^\phi\left[
\sum_{i=0}^{\infty}\beta_S^i
u_S\bigl(\theta_{t+i},\sigma(m_{t+i},\nu_{t+i})\bigr)
\,\middle|\,\theta_t,\nu_t\right].
\]

\label{sec:stable}
Sustained informative communication requires credibility, which compares the prescribed rule with deviations that respect the restriction, and sustainability, which compares maintaining the restriction with violating it.

\subsection{Credibility}
\label{subsec:credibility}

As in the static setting of \citet{lin2024credible}, the sender can deviate from the announced strategy without being detected. Credibility requires the announced strategy to maximize the sender's continuation payoff among deviations that keep every rolling history within the induced set, starting from the public history at which incentives are evaluated and holding the receiver's response fixed. Let $\mathfrak S$ be the set of sender continuation strategies, including history-dependent and randomized strategies. For an announcement $\phi^C$ and a pre-breakdown public history $\nu_t$, define
\[
\Phi(\phi^C;\nu_t)
=\left\{\widehat\phi\in\mathfrak S:
\Pr_{\nu_t}^{\widehat\phi}\!\left(
\nu_s^N\in\mathcal V^{N,\phi^C}\ \text{for every }s>t
\right)=1\right\},
\]
where $\Pr_{\nu_t}^{\widehat\phi}$ is the law of the continuation process initialized at $\nu_t$, even when $\widehat\phi$ would not reach that history from the initial record. Undetectability thus constrains the deviation from the history under consideration, rather than only along its path from the start of the relationship.

\begin{definition}[Credibility]
\label{def:credibility}
A strategy profile $(\phi^C,\sigma)$ is \emph{credible} if, for every period $t$, every state $\theta_t$, and every pre-breakdown public history $\nu_t$, $\phi^C$ maximizes $\Pi(\theta_t,\nu_t;\widehat\phi,\sigma)$ over $\widehat\phi\in\Phi(\phi^C;\nu_t)$.
\end{definition}
Credibility evaluates perpetual compliance with the announcement; detectable violations are considered separately.

\subsection{Receiver Obedience and Equilibrium}
\label{subsec:receiver-obedience}\label{subsec:equilibrium}

\begin{definition}[Receiver obedience]
\label{def:receiver-obedience}
A strategy profile $(\phi^C,\sigma)$ is \emph{receiver-obedient} if, at every pre-breakdown public history $\nu_t$ and after every recommendation $m_t$ sent with positive probability under $\phi^C$ at that history, $\sigma(m_t,\nu_t)=m_t$ and
\[
\sum_{\theta\in\Theta}\mu_t^{\phi^C}(\theta\mid\nu_t,m_t)\bigl[u_R(\theta,m_t)-u_R(\theta,1-m_t)\bigr]\ge0,
\]
where the posterior $\mu_t^{\phi^C}(\cdot\mid\nu_t,m_t)$ is obtained from $\phi^C$ by Bayes' rule.
\end{definition}

\begin{definition}[Perfect Bayesian equilibrium]
\label{def:equilibrium}
Given an announced strategy $\phi^C$, a profile $(\phi,\sigma)$ and beliefs $\mu^\phi$ form a perfect Bayesian equilibrium if (i) at every state $\theta_t$ and public history $\nu_t$, $\phi$ maximizes the sender's continuation payoff over all continuation strategies in $\mathfrak S$, including detectable violations; (ii) at every public history and after every message, $\sigma$ is optimal under $\mu^\phi$; and (iii) $\mu^\phi$ follows from $\phi$ by Bayes' rule whenever the message has positive probability.
\end{definition}

When the actual strategy perpetually follows the announcement, $\phi=\phi^C$, sender rationality requires credibility and the absence of profitable detectable violations; by the one-period deviation principle, checking both at every history suffices. Pure signaling strategies are without loss under perpetual compliance: credibility makes the sender indifferent among the recommendations that a mixed rule supports, so selecting one at each state and history leaves every continuation value and credibility unchanged, and where the rule mixes in state $\theta_i$, selecting recommendation~$1$ exactly when $r_i>0$ weakly raises the receiver gain pooled into recommendation~$1$ and weakly lowers it for recommendation~$0$, preserving obedience. Under belief updating, the actual strategy may instead include a violation followed by babbling. I call a strategy that specifies both the admissible recommendations and the decision to violate a \emph{completed policy}. An \emph{implementation with anticipated breakdown} is a receiver-obedient completed policy, with obedience evaluated under the actual strategy, that satisfies sender rationality; the announcement still defines which histories count as violations.

\subsection{Sustainability}
\label{subsec:sustainability}\label{subsubsec:receiver-response}

A credible, receiver-obedient rule need not be sustainable. Credibility deters undetectable misreports, whereas an observable violation gives the sender the current payoff from the receiver's response, followed by babbling.

\begin{definition}[Sustainability]
\label{def:sustainability}
A credible, receiver-obedient profile $(\phi^C,\sigma,\mu)$ is \emph{sustainable} if, at every pre-breakdown public history $\nu_t$, every state $\theta_t$, and every message $m_t$ with $T_{m_t}\nu_t^N\notin\mathcal V^{N,\phi^C}$,
\begin{equation}
\Pi(\theta_t,\nu_t;\phi^C,\sigma)\;\ge\;
(1-\beta_S)\,u_S\bigl(\theta_t,\sigma(m_t,\nu_t)\bigr),
\label{eq:sustainability}
\end{equation}
with equality permitted only when Assumption~\ref{ass:tie-breaking} selects compliance.
\end{definition}

Because a violating message has zero probability under $\phi^C$, Bayes' rule does not pin down the receiver's response, and sustainability depends on it. Under immediate reversion, the receiver treats a violation as uninformative, so it yields zero, the payoff from the undetectable deviation of recommending $0$ from now on; when that recommendation is always admissible, credibility implies sustainability. Under belief updating, the receiver follows a violating recommendation if it is optimal given the states in which violating is worthwhile, as in the accelerator. The violation then acts as an exit option, so sustainability can require more patience than credibility. When compliance is optimal under both responses, they generate the same play and payoffs.

\FloatBarrier
\section{History Restrictions and Recommendation Incentives}
\label{sec:quota}

A history restriction links current recommendations to future feasible advice. This section isolates the general conditions that order those opportunities and reduce implementation to the history with no inherited favorable recommendations.

\subsection{Continuation Costs and Local Credibility}
\paragraph{Continuation values.}
\label{subsec:continuation-history-sets}

Fix a pure stationary strategy $\phi$ and let receivers obey it. Before the current state is observed, the sender's normalized continuation payoff from rolling history $\nu$ under perpetual compliance satisfies
\begin{equation}
\Pi^\phi(\nu)
=\sum_{i=1}^n\mu_i\left[
(1-\beta_S)g_i\phi(\theta_i,\nu)
+\beta_S\Pi^\phi\!\left(T_{\phi(\theta_i,\nu)}\nu\right)
\right].
\label{eq:one-period-continuation}
\end{equation}
Unnormalized values are denoted by $V$; multiplying by $1-\beta_S$ gives normalized payoffs. The rolling history evolves as a finite Markov chain whose transitions and payoffs the recommendations control, so the sender's optimization within a history restriction is a finite discounted Markov decision problem. Because states are independent, a pure stationary policy attains the optimum over all causal deviations that keep rolling histories within the restriction.

A set \(\mathcal V\subseteq\{0,1\}^N\) is an \emph{admissible history set} if it contains the initial rolling history $0^N$ and, from every rolling history in \(\mathcal V\), at least one recommendation keeps the next rolling history in \(\mathcal V\). Every induced history set is admissible. Let \(\Pi^*_{\mathcal V}(\nu^N):=\max_{\phi}\Pi^\phi(\nu^N)\) for \(\nu^N\in\mathcal V\), the maximum taken over pure stationary strategies that keep every rolling history in \(\mathcal V\); a strategy is \emph{optimal within} \(\mathcal V\) if it attains \(\Pi^*_{\mathcal V}\) at every rolling history in \(\mathcal V\). Such a strategy exists because there are finitely many rolling histories and payoffs are discounted.

\begin{lemma}[Credible strategies and admissible history sets]
\label{lem:credible-sets}
For a pure stationary strategy \(\phi\), let \(\sigma^\phi\) obey \(\phi\) at every pre-breakdown history, with either response to violations. Then \((\phi,\sigma^\phi)\) is credible if and only if \(\phi\) is optimal within \(\mathcal V^{N,\phi}\); and if \(\phi\) is optimal within some admissible history set \(\mathcal V\), then \(\mathcal V^{N,\phi}\subseteq\mathcal V\) and \((\phi,\sigma^\phi)\) is credible.
\end{lemma}

Appendix~\ref{app:credible-sets-proof} proves the lemma.

\paragraph{Entering histories and base sets.}\label{subsec:base-histories}\label{sec:base}\label{sec:tests}\label{sec:deviation}

Fix an admissible history set $\mathcal V$. In period $t$, admissibility depends only on the \emph{entering history} $\nu_t^{N-1}=(m_{t-N+1},\ldots,m_{t-1})$, whose coordinates run from oldest to newest. Write $\nu$ for a generic entering history. Recommendation $m$ is admissible exactly when $(\nu,m)\in\mathcal V$, the next entering history is $T_m\nu$, and Assumption~\ref{ass:initial-history} gives the initial entering history $0^{N-1}$. The \emph{base set} $\mathcal B(\mathcal V):=\{\nu\in\{0,1\}^{N-1}:(\nu,1)\in\mathcal V\}$ collects the entering histories at which recommendation $1$ is admissible. For the quota $(N,Q)$, write $\admissible^{N,Q}:=\{\nu\in\{0,1\}^N:\sum_{j=1}^N\nu_j\le Q\}$; its base set is $\{\nu:|\nu|\le Q-1\}$, where $|\nu|$ is the number of ones in $\nu$. I call $\admissible^{N,Q}$ a \emph{full quota}, as opposed to restrictions that admit only part of the next quota layer. More general base sets restrict the timing of favorable recommendations as well as their number.

\begin{definition}[Prior-action closure]\label{def:prior-action-closure}
The set $\mathcal V$ is \emph{prior-action closed} if $T_0\nu\in\mathcal V$ for every $\nu\in\mathcal V$.
\end{definition}
Recommendation $0$ then always remains admissible, and $1$ is admissible exactly when the entering history belongs to $\mathcal B(\mathcal V)$. Under prior-action closure, credibility implies sustainability under immediate reversion (Section~\ref{subsec:sustainability}), so the analysis below characterizes equilibria under that response, and sustainable equilibria under belief updating once \eqref{eq:sustainability} holds. Prior-action closure holds for restrictions induced by obedient rules: recommending $1$ in every state would leave the receiver with her prior, so an obedient rule recommends $0$ with positive probability at every reached history, and appending $0$ keeps the rolling history in the induced set. An entering history is \emph{active} if it is reachable from $0^{N-1}$ through admissible recommendations. Write $\mathcal H(\mathcal V)$ for these histories and $\mathcal H^\circ=\mathcal H(\mathcal V)\cap\mathcal B(\mathcal V)$ for those permitting $1$. For a full quota, the active entering histories are $\mathcal H_{N,Q}=\{\nu:|\nu|\le Q\}$.

\paragraph{The local credibility test.}
Fix a pure stationary strategy $\phi$, and let receivers obey it. For $\nu\in\mathcal V^{N,\phi}$, let $A^\phi(\nu)=\{m:T_m\nu\in\mathcal V^{N,\phi}\}$ be the undetectable recommendations and $\mathcal V^\phi_\circ=\{\nu:A^\phi(\nu)=\{0,1\}\}$. At $\nu\in\mathcal V^\phi_\circ$, the \emph{continuation cost} of recommendation $1$ is
\begin{equation}
c^\phi(\nu)=\frac{\beta_S}{1-\beta_S}
\bigl[\Pi^\phi(T_0\nu)-\Pi^\phi(T_1\nu)\bigr],
\label{eq:local-continuation-cost}
\end{equation}
so recommending $1$ rather than $0$ in state $\theta_i$ at $\nu$, and following $\phi$ thereafter, changes the sender's normalized payoff by $(1-\beta_S)[g_i-c^\phi(\nu)]$.

\begin{proposition}[Local credibility test]\label{prop:local}
The profile $(\phi,\sigma^\phi)$ is credible if and only if, at every $\nu\in\mathcal V^\phi_\circ$ and in every state $\theta_i$, $\phi(\theta_i,\nu)=1$ implies $g_i\ge c^\phi(\nu)$ and $\phi(\theta_i,\nu)=0$ implies $g_i\le c^\phi(\nu)$.
\end{proposition}
At a history with only one undetectable recommendation, there is no alternative to test. Since the sender's problem within the restriction is a finite discounted Markov decision problem, the local test also rules out deviations that change recommendations at several histories, depend on the entire past, or randomize (Appendix~\ref{app:identity}; Appendix~\ref{app:occupation-details} gives an exact occupation-measure accounting of deviations and a necessary nesting diagnostic). For a threshold rule recommending $1$ exactly in $\{\theta_k,\ldots,\theta_n\}$, the test is $g_{k-1}\le c^\phi(\nu)\le g_k$ at every history where both recommendations are undetectable.

\subsection{Temporal Ordering and Continuation Values}
\label{sec:temporal-restrictions}

An inherited favorable recommendation constrains future recommendations until it leaves the rolling window. The following order compares histories by the number and timing of their inherited recommendations.

\begin{definition}[Temporal precedence and lower closure]
\label{def:temporal-precedence}
For entering histories $\nu',\nu\in\{0,1\}^{N-1}$, write $\nu'\preceq_T \nu$ if $\sum_{i=j}^{N-1}\nu'_i\le\sum_{i=j}^{N-1}\nu_i$ for every $j=1,\ldots,N-1$. The base set is \emph{temporally lower closed} if $\nu\in\mathcal B(\mathcal V)$ and $\nu'\preceq_T\nu$ imply $\nu'\in\mathcal B(\mathcal V)$.
\end{definition}

Thus $\nu'$ can be obtained from $\nu$ by deleting ones or moving them to earlier vacant positions; for example, $0101\preceq_T0011$, and $0^{N-1}$ is the least element. Lower closure requires a favorable recommendation to remain admissible when inherited recommendations are removed or made older, and it applies to the base set on all entering histories, including those not reached from $0^{N-1}$. Full quotas satisfy lower closure because $\nu'\preceq_T \nu$ implies $|\nu'|\le|\nu|$.

\paragraph{Recency weights and the scope of the order.}
For $N\geq2$, temporal precedence also has a weighted-record characterization:
\begin{equation}
\nu'\preceq_T\nu
\quad\Longleftrightarrow\quad
\sum_{i=1}^{N-1}w_i\nu'_i\leq\sum_{i=1}^{N-1}w_i\nu_i
\quad\text{for every }0\leq w_1\leq\cdots\leq w_{N-1}.
\label{eq:temporal-weight-duality}
\end{equation}
Indeed, writing $S_j(\nu)=\sum_{i=j}^{N-1}\nu_i$ gives
$\sum_iw_i\nu_i=w_1S_1(\nu)+\sum_{j=2}^{N-1}(w_j-w_{j-1})S_j(\nu)$; step weights recover each suffix inequality. Thus the order compares records consistently across every nonnegative recency weighting, and in particular a temporally lower record has a weakly lower recency-discounted count $\sum_i\delta^{N-1-i}\nu_i$ for every $\delta\in(0,1]$. A geometric count of this kind also ranks temporally incomparable records, but Section~\ref{subsec:history-cutoffs} shows that, at every discount rate, such a count can misorder continuation costs.

Fix a prior-action-closed admissible set $\mathcal V$ with a nonempty, temporally lower-closed base set, and hold receiver responses fixed at obedience. Let $\widehat V^{\mathrm{IR}}(\nu)$ be the sender's optimal unnormalized perpetual-compliance value within the restriction, so that $(1-\beta_S)\widehat V^{\mathrm{IR}}(\nu)=\Pi^*_{\mathcal V}(\nu^N)$ for any rolling history $\nu^N\in\mathcal V$ whose last $N-1$ entries form $\nu$; the superscript refers to immediate reversion, under which credibility implies sustainability for these restrictions. At active histories permitting $1$, let $\widehat c^{\mathrm{IR}}(\nu)=\beta_S[\widehat V^{\mathrm{IR}}(T_0\nu)-\widehat V^{\mathrm{IR}}(T_1\nu)]$.

\begin{lemma}[Temporal ordering of continuation values]
\label{lem:temporal-value-order}
Suppose $\mathcal V$ is prior-action closed and its base set is temporally lower closed. For active entering histories $\nu'\preceq_T\nu$,
\[
\widehat V^{\mathrm{IR}}(\nu')\geq\widehat V^{\mathrm{IR}}(\nu).
\]
In particular, $0^{N-1}$ maximizes the continuation value among active histories, and $\widehat V^{\mathrm{IR}}(T_0\nu)\geq\widehat V^{\mathrm{IR}}(\nu)$.
\end{lemma}

The comparison follows by simulating a feasible policy from the higher history on the same future states at the lower history. Appending the same recommendations preserves temporal precedence, so lower closure preserves feasibility and reproduces every stage payoff. Ordering the marginal cost of a recommendation requires more structure (Section~\ref{subsec:history-cutoffs}). Appendix~\ref{app:base-ordering-proofs} gives the proof.

\subsection{The Lowest-History Implementation Test}
\label{subsec:optimal-policy-obedience}

\begin{lemma}[Continuation-cost bounds]
\label{lem:general-lowest-cost}
Under prior-action closure and a nonempty, temporally lower-closed base set, $0\le\widehat c^{\mathrm{IR}}(0^{N-1})\le\widehat c^{\mathrm{IR}}(\nu)\le\beta_Sg_n<g_n$ at every active history $\nu$ permitting~$1$; hence the maximal-gain state receives recommendation~$1$ whenever it is admissible.
\end{lemma}

The lower bound combines Lemma~\ref{lem:temporal-value-order} with the Bellman equations at the initial and current histories. The upper bound follows by omitting at most one future favorable recommendation after using capacity today. Appendix~\ref{app:base-ordering-proofs} proves both comparisons.

\begin{proposition}[Receiver obedience of the optimal policy]
\label{prop:optimal-policy-obedience}
Suppose $\mathcal V$ is prior-action closed with a nonempty, temporally lower-closed base set, and (UT) holds. The selected sender-optimal perpetual-compliance policy is receiver-obedient at every active history if and only if $\widehat c^{\mathrm{IR}}(0^{N-1})\ge\underline c_R$, and it is then informative at $0^{N-1}$.
\end{proposition}
By Lemma~\ref{lem:general-lowest-cost}, every favorable cell is then at least as selective as the obedient tail at $0^{N-1}$, so (UT) carries obedience to every active history (Appendix~\ref{app:base-ordering-proofs}). By Lemma~\ref{lem:credible-sets}, the selected optimal policy is already credible. I call a restriction \emph{implementable under immediate reversion} when this policy is receiver-obedient; prior-action closure then also rules out profitable violations, so the profile can be completed to an equilibrium.

In the two-state examples, the selected policy, when implementable, is the \emph{separating rule}, which recommends $1$ exactly in $H$ whenever $1$ is admissible. Section~\ref{sec:lowest} extends the credibility test to prescribed upper-tail rules whose favorable states share a common sender gain and gives the parallel to \citet[Proposition~3]{lin2024credible}.

\FloatBarrier
\subsection{Prescribed Upper-Tail Rules}\label{sec:lowest}
The credibility test extends beyond two states when every prescribed favorable recommendation earns the same sender gain. In this subsection, allow ties within the recommended upper tail: for some $2\leq k\leq n$, suppose $g_1<\cdots<g_{k-1}<g_k=\cdots=g_n=G>0$, and write $p:=\sum_{i=k}^n\mu_i\in(0,1)$. Excluded states can have positive gains, the largest being $g_{k-1}$. Fix a prior-action-closed admissible set $\mathcal V$, and consider the prescribed rule that recommends $1$ exactly when $i\geq k$ and $\nu\in\mathcal H^\circ$. Since $p\in(0,1)$, it generates every admissible finite recommendation path from $0^{N-1}$ with positive probability. Its normalized continuation value and continuation cost under obedience depend only on the entering history; write them as $\Pi^\phi(\nu)$ and $c^\phi(\nu)$, and the cost as $c^\phi_{N,Q}(\nu)$ for a full quota $(N,Q)$.

\begin{proposition}[Lowest-history test]\label{prop:lowest}
Under the sender-gain conditions above, suppose $\mathcal V$ is prior-action closed and $\mathcal B(\mathcal V)$ is nonempty and temporally lower closed. The prescribed upper-tail rule satisfies $0\le c^\phi(0^{N-1})\le c^\phi(\nu)\le\beta_SG<G$ for $\nu\in\mathcal H^\circ$, and it is credible if and only if $g_{k-1}\le c^\phi(0^{N-1})$.
\end{proposition}
The upper bound gives every state $i\geq k$ a strict incentive to recommend $1$ whenever it is admissible, and the lower bounds make $0^{N-1}$ the most demanding history for withholding $1$ from the excluded states. Appendix~\ref{app:binary-prescribed-policy} proves the bounds by coupling the dates of favorable recommendations along common future state sequences. Temporal lower closure is substantive (Appendix~\ref{app:no-lower-closure}). Under the standing prior preference for action $0$, receiver obedience is equivalent to
\begin{equation}
\sum_{i=k}^n\mu_i r_i\geq0.
\label{eq:tail-obedience}
\end{equation}
The favorable cell is then obedient, and so are its complement and the forced zeros. Under immediate reversion, a credible rule satisfying \eqref{eq:tail-obedience} has no profitable detectable violation and supports an equilibrium respecting Assumption~\ref{ass:tie-breaking} if $g_{k-1}<c^\phi(0^{N-1})$ or $r_{k-1}\leq0$; the latter holds when $k=k_R$.

With two states, $n=k=2$, the common-gain condition holds automatically; the rule then separates states whenever $1$ is available and is the separating rule above. For the accelerator's quota $(2,1)$, $c^\phi_{2,1}(0)=\beta_Sp/(1+\beta_Sp)$; with $p=1/2$ and $\ell=1/4$, its credibility test under immediate reversion holds exactly when $\beta_S\ge2/3$.

\begin{corollary}[Agreement at maximal sender gains]\label{cor:common-gain-existence}
Under the sender-gain conditions of Proposition~\ref{prop:lowest} and \eqref{eq:tail-obedience}, if $\beta_S>g_{k-1}/[g_{k-1}+p(G-g_{k-1})]$, then the quota $(N,1)$ supports the prescribed credible, receiver-obedient rule under immediate reversion for every sufficiently large finite $N$. The sender strictly benefits relative to babbling. If \eqref{eq:tail-obedience} holds strictly, the receiver's expected current payoff strictly exceeds babbling whenever $1$ is available.
\end{corollary}

When $k=k_R$, $g_{k-1}=\underline c_R$ and $p(G-g_{k-1})=F(\underline c_R)$, so the threshold equals $\beta^\ast$. With strictly ordered gains, taking $k=n$ shows that agreement on action~$1$ at the maximal-gain state, $r_n\geq0$, suffices for sufficiently patient senders. Corollary~\ref{cor:common-gain-existence} parallels \citet[Proposition~3(1)--(2)]{lin2024credible}, who assume both players' payoffs are supermodular. Here the ordering of states makes the sender's payoff supermodular, the receiver need only accept the fixed tail, and neither receiver supermodularity nor (UT) is required.

\section{Recommendation Cutoffs: Timing, Quota Size, and Memory}
\label{sec:rolling-design}\label{sec:threshold}\label{sec:binary-memory}

Rolling quotas link recommendations through both a capacity limit and its renewal date. This section first compares records within one quota and then compares quotas and memories at the empty record.

\subsection{Recommendation Cutoffs Across Histories}
\label{subsec:history-cutoffs}\label{subsec:cost-ordering}

Fix $(N,Q)$ and date the current period $0$. For an entering record $\nu=(\nu_1,\ldots,\nu_{N-1})$, from oldest to newest, define its \emph{capacity-release profile}
\begin{equation}
R_\nu(t):=Q-\sum_{i=t+1}^{N-1}\nu_i,
\qquad 0\le t\le N-1.
\label{eq:release-profile}
\end{equation}
New endorsements in periods $0$ through $t$ can number at most $R_\nu(t)$; thereafter, every window of $N$ periods can contain at most $Q$ endorsements. Each inherited endorsement thus displaces future advice until it expires. A record $\nu'$ leaves at least as much capacity as $\nu$ at every date, $R_{\nu'}\ge R_\nu$, exactly when $\nu'\preceq_T\nu$. At a slack history, define the selected state cutoff
\begin{equation}
k_{N,Q}(\nu):=\min\{i:g_i>\widehat c^{\mathrm{IR}}_{N,Q}(\nu)
\text{ or }[g_i=\widehat c^{\mathrm{IR}}_{N,Q}(\nu)\text{ and }r_i>0]\}.
\label{eq:state-cutoff-index}
\end{equation}
The favorable cell is $A_{N,Q}(\nu)=\{\theta_i:i\geq k_{N,Q}(\nu)\}$. The maximal-gain state belongs to this cell by Lemma~\ref{lem:general-lowest-cost}; at a binding history set $A_{N,Q}(\nu)=\varnothing$.

\begin{proposition}[Cost ordering under rolling quotas]
\label{prop:quota-cost-order}
Under a rolling quota $(N,Q)$, $\nu'\preceq_T\nu$ implies $\widehat c^{\mathrm{IR}}(\nu')\le\widehat c^{\mathrm{IR}}(\nu)$ for active histories $\nu',\nu$ whenever $1$ is admissible at $\nu$. Hence, by \eqref{eq:state-cutoff-index}, $k_{N,Q}(\nu')\le k_{N,Q}(\nu)$ and $A_{N,Q}(\nu)\subseteq A_{N,Q}(\nu')$.
\end{proposition}

In words, deleting an inherited endorsement or moving it earlier weakly lowers the opportunity cost of an endorsement, so favorable advice becomes weakly more likely when inherited capacity is released earlier.

\paragraph{Values and their differences.}
Lemma~\ref{lem:temporal-value-order} orders continuation values under temporal lower closure alone, but a continuation cost is a difference of values, and ordering values does not order differences: under a temporally lower-closed restriction that is not a rolling quota, the cost can fall along the temporal order (Appendix~\ref{app:exclusion-examples}), although the empty record still has the lowest cost. The rolling constraint supplies the missing structure. Represent a record by its cumulative counts of endorsements. Each rolling constraint bounds a difference of two such counts, so the feasible records are closed under coordinatewise minima and maxima, forming a lattice on which the optimal value is supermodular \citep{topkis1998supermodularity}, and comparing the four successors of $\nu'\preceq_T\nu$ orders the costs (Appendix~\ref{app:quota-cost-order}). Appendix~\ref{sec:order-check} examines intersections of rolling quotas.

\paragraph{Records rather than budgets.}
Under a lifetime budget the state is the number of remaining units, and the shadow price of a unit falls as more units remain (Section~\ref{subsec:implementable-quotas}); a discounted quota \citep{frankel2016discounted,frankel2014experts} also summarizes the past by one number, under which an earlier play uses more of the quota. Under a rolling quota the state is the record, which is only partially ordered, and an earlier endorsement releases capacity sooner; temporal precedence is an order along which the shadow price of capacity is monotone. A cost change alters the favorable cell only when it crosses a sender gain, with equality governed by the tie-breaking rule; the next example exhibits these crossings.

\begin{figure}[!t]
\centering
\includegraphics[width=0.75\textwidth]{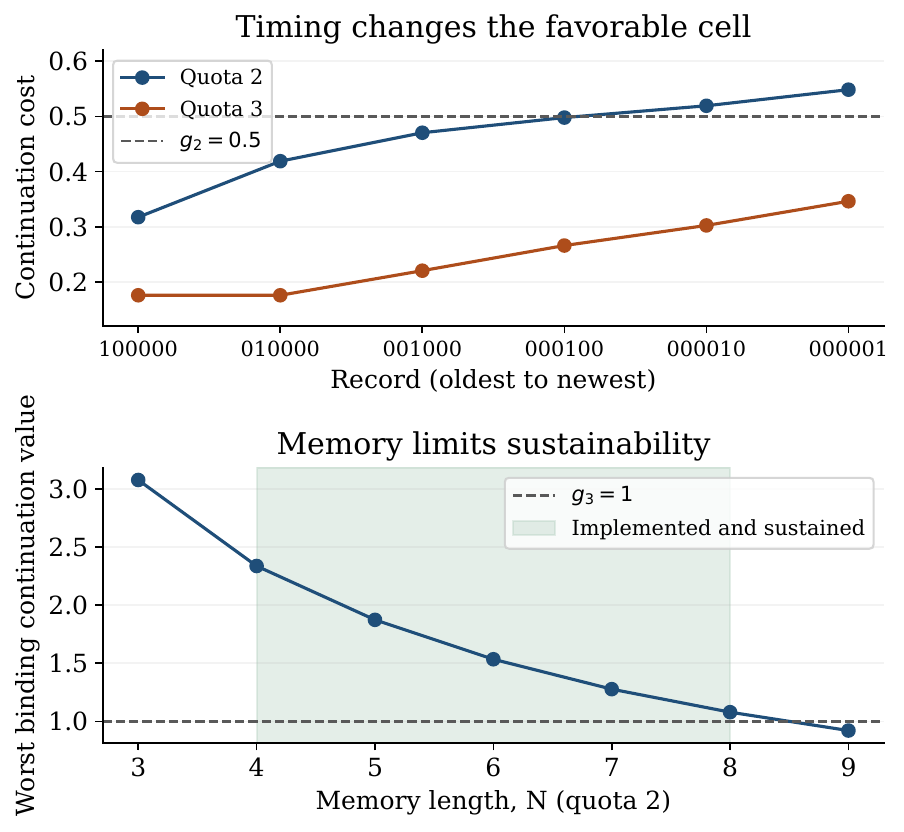}
\caption{The three-state example. Top: costs at records containing one endorsement; crossing $g_2=0.5$ removes $\theta_2$ from favorable advice. Bottom: the worst binding value at quota two; the shaded integer memories pass both implementation and sustainability.}
\label{fig:multistate-linking}
\end{figure}

\paragraph{A three-state example with changing recommendation cells.}
\label{par:multistate-example}
Let $\beta_S=0.9$ and
\[
\begin{aligned}
(\mu_1,\mu_2,\mu_3)&=(1/2,1/4,1/4),\\
(g_1,g_2,g_3)&=(0.15,0.50,1),\\
(r_1,r_2,r_3)&=(-1,0.20,1).
\end{aligned}
\]
The receiver's prior gain is $-0.20$, (UT) holds, and $k_R=2$, with $\underline c_R=0.15$. Both cells $\{\theta_2,\theta_3\}$ and $\{\theta_3\}$ are obedient; sending $1$ in every state is not. At memory $N=7$, the top panel of Figure~\ref{fig:multistate-linking} plots the optimal perpetual-compliance costs at records with one endorsement, and Table~\ref{tab:multistate-records} lists them with the favorable cells.

For quota two, moving the sole endorsement from $000100$ to $000010$ delays its expiration by one period. The cost rises from $0.4979$ to $0.5193$, crossing $g_2=0.50$, the favorable cell loses $\theta_2$, and the endorsement probability falls from $1/2$ to $1/4$. At the same recent record, increasing the quota to three restores advice in $\theta_2$. Both quotas are implementable at every active history and sustainable under belief updating; Section~\ref{subsec:memory-survival} reports the binding-history checks.

The temporal order leaves some records incomparable. The following pair shows that a geometrically discounted count need not preserve their cost ranking. At $(N,Q)=(7,3)$, the records $000110$ and $001001$ are incomparable. The first has the lower count $\sum_i\delta^{N-1-i}\nu_i$ at every $\delta>0$ other than $1$, whether recent or older endorsements receive more weight, since the difference is $(1-\delta)^2(1+\delta)$, yet its cost is $0.5150$ against $0.4957$: $\theta_2$ receives favorable advice after $001001$ but not after $000110$. The second record releases a unit one period sooner, although it holds its other unit one period longer.

Memory changes advice away from the empty record as well: at quota two, when the only past endorsement occurred in the preceding period, the cost is $0.4928$ at $N=5$ and $0.5256$ at $N=6$, so the recommended states shrink from $\{\theta_2,\theta_3\}$ to $\{\theta_3\}$, and both arrangements are implementable and sustainable. Appendix~\ref{app:multistate-computation} provides the numerical checks, and Appendix~\ref{app:continuous-example} a continuous-state example with the same pattern.

\subsection{Quota Size, Memory, and Sender Patience}
\label{subsec:quota-memory-comparisons}

A larger quota expands the sender's opportunities to recommend action $1$; longer memory delays the release of capacity. The selected optimal policy is credible by Lemma~\ref{lem:credible-sets}, so receiver obedience determines which quotas are implementable under immediate reversion; under belief updating, sustainable policies have the same values. At the empty record, the Bellman equation gives
\begin{equation}
(1-\beta_S)\widehat V^{\mathrm{IR}}_{N,Q}(0^{N-1})
=F\bigl(\widehat c^{\mathrm{IR}}_{N,Q}(0^{N-1})\bigr),
\label{eq:finite-lowest-identity}
\end{equation}
so a higher initial payoff corresponds to a lower continuation cost.

\begin{proposition}[Quota choice at a fixed memory]
\label{prop:binary-quota-monotonicity}
At a fixed memory $N$, $\widehat V^{\mathrm{IR}}_{N,Q}(0^{N-1})$ weakly increases and $\widehat c^{\mathrm{IR}}_{N,Q}(0^{N-1})$ weakly decreases with $Q$. Under (UT), the implementable positive quotas therefore form an initial segment $\{1,\ldots,Q^\star_{\mathrm{IR}}(N)\}$, where $Q^\star_{\mathrm{IR}}(N)$ is the largest $Q\le N$ with $\widehat c^{\mathrm{IR}}_{N,Q}(0^{N-1})\ge\underline c_R$ (zero if there is none), and its largest member maximizes the sender's payoff at memory $N$.
\end{proposition}

Every implementable quota weakly raises the receiver's expected payoff relative to babbling, since each favorable recommendation it sends has nonnegative expected receiver gain.

\begin{proposition}[Increasing memory at a fixed quota]
\label{prop:binary-memory}
Fix $Q\geq1$. On the integers $N\geq Q$, $\widehat V^{\mathrm{IR}}_{N,Q}(0^{N-1})$ weakly decreases and $\widehat c^{\mathrm{IR}}_{N,Q}(0^{N-1})$ weakly increases with $N$; under (UT), a quota implementable at some memory remains implementable at every longer memory.
\end{proposition}

Both propositions follow from policy inclusion and \eqref{eq:finite-lowest-identity} (Appendix~\ref{app:quota-memory-proofs}).

\begin{corollary}[Quota and memory effects on the initial recommendation cell]
\label{cor:quota-memory-cutoffs}
At the empty record, increasing $Q$ weakly lowers $k_{N,Q}$ and expands the favorable cell; increasing $N$ at fixed $Q$ weakly raises $k_{N,Q}$ and contracts that cell. These statements hold for the selected optimal policies before imposing obedience, and hence for any pair of implementable restrictions being compared.
\end{corollary}
\begin{proof}
Apply \eqref{eq:state-cutoff-index} to the cost comparisons in Propositions~\ref{prop:binary-quota-monotonicity} and~\ref{prop:binary-memory}.
\end{proof}

The initial cost rises with patience (Appendix~\ref{app:patience-comparisons}); the cell comparisons can be weak because the state space is discrete. Full support strengthens the comparisons whenever the quota changes feasible recommendation paths:

\begin{proposition}[Strict full-quota comparisons]
\label{prop:supp-strict}
For fixed $N$ and $0\le Q<N$,
\[
\widehat V^{\mathrm{IR}}_{N,Q+1}(0^{N-1})>
\widehat V^{\mathrm{IR}}_{N,Q}(0^{N-1}).
\]
For positive quotas the all-zero cost strictly decreases with $Q$.
For fixed $Q\ge1$ and $N\ge Q$,
\[
\widehat V^{\mathrm{IR}}_{N,Q}(0^{N-1})>
\widehat V^{\mathrm{IR}}_{N+1,Q}(0^N),
\]
and the all-zero cost strictly increases with $N$.
\end{proposition}
Appendix~\ref{app:strict-comparisons} proves the proposition. In the three-state example, the initial cutoff falls from $k=3$ at $(N,Q)=(7,1)$ to $k=2$ at $(7,2)$, and at quota one it rises from $k=2$ at $N=6$ to $k=3$ at $N=7$.

\subsection{Implementable Quotas and Optimal Design}
\label{subsec:implementable-quotas}

To bound the implementable quotas, consider the sender's optimal policy with at most $q$ favorable recommendations over the remaining lifetime, evaluated under perpetual compliance before imposing receiver obedience. Write $c(q)$ for its continuation cost when $q$ recommendations remain. Its normalized payoff is $F(c(q))$. The lifetime costs satisfy
\begin{equation}
F(c(1))=\frac{1-\beta_S}{\beta_S}c(1),\qquad
F(c(q))-F(c(q-1))=\frac{1-\beta_S}{\beta_S}c(q),
\quad q\geq2.
\label{eq:finite-lifetime-recursion}
\end{equation}
The lifetime benchmark $F(c(q))$ strictly increases to $F(0)$, while $c(q)$ strictly decreases to zero. For a fixed $Q$, increasing memory makes the normalized rolling-quota value converge to $F(c(Q))$ from above and its empty-record cost converge to $c(Q)$ from below (Lemma~\ref{lem:binary-memory-limit}). Call a positive quota \emph{eligible} if $c(Q)>\underline c_R$.

\begin{theorem}[A finite bound on implementable quotas]
\label{thm:binary-finite-quota}
Suppose (UT) holds and $\underline c_R>0$. A positive quota $Q$ is implementable at some finite memory if and only if $\underline c_R<c(Q)$. Hence $\overline Q^{\mathrm{IR}}:=\max(\{0\}\cup\{Q\ge1:\underline c_R<c(Q)\})$ is finite, and $Q^\star_{\mathrm{IR}}(N)$ weakly increases with $N$ to $\overline Q^{\mathrm{IR}}$.
\end{theorem}

Even without (UT), obedience at the empty record requires $\underline c_R\leq\widehat c^{\mathrm{IR}}_{N,Q}(0^{N-1})<c(Q)$, so $\overline Q^{\mathrm{IR}}$ bounds the implementable quotas; under (UT), the bound is attained at sufficiently long memory. Appendix~\ref{app:memory-derivations} gives the proofs and the binary closed form.

\begin{corollary}[Rolling-quota implementation above the patience threshold]\label{cor:existence}
Suppose (UT) holds and $\underline c_R>0$. If $\beta_S>\beta^\ast$, then for every sufficiently large finite $N$ the rolling quota $(N,1)$ supports an informative pure stationary equilibrium under immediate reversion.
\end{corollary}
\begin{proof}
The map $d\mapsto F(d)-(1-\beta_S)d/\beta_S$ is strictly decreasing and vanishes at $c(1)$ by \eqref{eq:finite-lifetime-recursion}, so $\beta_S>\beta^\ast$, which is equivalent to $F(\underline c_R)>(1-\beta_S)\underline c_R/\beta_S$, gives $c(1)>\underline c_R$. By Lemma~\ref{lem:binary-memory-limit} with $Q=1$, the quota-one cost at $0^{N-1}$ increases with $N$ to $c(1)$, so it exceeds $\underline c_R$ for all large $N$; Lemma~\ref{lem:credible-sets} and Proposition~\ref{prop:optimal-policy-obedience} then make the selected optimal policy credible, receiver-obedient, and informative. Under immediate reversion, a violation yields zero while compliance yields a positive continuation value.
\end{proof}

\paragraph{Choosing quota and memory.}\label{subsec:choosing-quota-memory}
\begin{corollary}[A finite search for the optimal full quota]
\label{prop:binary-finite-search}
Maintain the assumptions of Theorem~\ref{thm:binary-finite-quota}. At each eligible quota, the shortest implementing memory $N^{\mathrm{IR}}_Q:=\min\{N\ge Q:\widehat c^{\mathrm{IR}}_{N,Q}(0^{N-1})\ge\underline c_R\}$ maximizes the sender's payoff at quota $Q$; write $P^{\mathrm{IR}}_Q$ for the resulting value $\widehat V^{\mathrm{IR}}_{N,Q}(0^{N-1})$ at $N=N^{\mathrm{IR}}_Q$. An optimal full quota therefore solves $\max_{1\le Q\le\overline Q^{\mathrm{IR}}}P^{\mathrm{IR}}_Q$, a finite search; if $\overline Q^{\mathrm{IR}}=0$, no positive full quota is implementable.
\end{corollary}

A larger quota can require a discrete increase in its shortest implementing memory, so the largest eligible quota need not maximize payoff; Appendix~\ref{app:memory-derivations} bounds each memory search. Write $\Pi^*_{\mathrm{full}}(\beta_S):=(1-\beta_S)\max_{1\le Q\le\overline Q^{\mathrm{IR}}}P^{\mathrm{IR}}_Q$ for the best normalized payoff over implementable full quotas. In the two-state calibration with $p=1/2$, $\beta_S=0.9$, and $\ell=0.379$, quotas one to four are eligible, with shortest implementing memories $3$, $7$, $11$, and $20$; the pair $(11,3)$ gives the highest implementable payoff, $0.310282$, above the $0.307634$ of $(20,4)$ (Appendix~\ref{app:quota-numerics}).

\FloatBarrier
\section{Sustainability and Communication Breakdown}
\label{sec:sustainability-analysis}\label{sec:binary-sustainability}

Sustainability concerns the decision to preserve the restriction itself. Under belief updating, a favorable recommendation can secure the current gain even when it violates the announced quota, at the cost of all future informative communication.

\subsection{Sustainability at Binding Histories}
\label{subsec:binding-sustainability}

Under belief updating, a violation yields at most $g_n$ before babbling, so an implementable policy has no profitable exit if $\beta_S\widehat V^{\mathrm{IR}}_{N,Q}(T_0\nu)>g_n$ at every active history prohibiting~$1$. When a violating recommendation in $\theta_n$ is followed, as in the two-state examples, where it reveals $H$, the condition is also necessary, with equality allowed only when $r_n\le0$.

Write $\widehat W$ for the optimal value when the receiver follows a violating recommendation in its current period and babbles thereafter; Section~\ref{sec:anticipated-breakdown} compares this value with perpetual compliance. The following result uses the optimal policy in the two-action, $n$-state model. Let
\[
\nu^b_{N,Q}:=0^{N-1-Q}1^Q,\qquad 1\leq Q<N,
\]
be the binding entering history whose inherited favorable recommendations are most recent. Write $B_{N,Q}:=\widehat V^{\mathrm{IR}}_{N,Q}(\nu^b_{N,Q})$ for its continuation value.

\begin{proposition}[Sustainability of a full quota]
\label{prop:binary-sustainability}
For arbitrary finite-state primitives,
\begin{equation}
\min_{\nu\in\mathcal H_{N,Q}:\,|\nu|=Q}
\beta_S\widehat V^{\mathrm{IR}}_{N,Q}(T_0\nu)
=\widehat V^{\mathrm{IR}}_{N,Q}(\nu^b_{N,Q}).
\label{eq:binary-sustainability-minimum}
\end{equation}
Holding the prior, gains, and discount factor fixed,
\begin{equation}
B_{N+1,Q}\leq B_{N,Q},\qquad
B_{N,Q+1}\geq B_{N,Q}\quad (1\leq Q<N-1).
\label{eq:sustainability-quota-memory}
\end{equation}
Thus the minimum weakly decreases with memory and weakly increases with quota size. There is a unique $\beta^{\mathrm{sus}}_{N,Q}\in(0,1)$ satisfying
\[
\widehat V^{\mathrm{IR}}_{N,Q}
\bigl(\nu^b_{N,Q};\beta^{\mathrm{sus}}_{N,Q}\bigr)=g_n.
\]
Suppose the selected perpetual-compliance policy is implementable and a violating recommendation receives action $1$. No profitable exit exists at any active history if and only if $\beta_S\geq\beta^{\mathrm{sus}}_{N,Q}$. Selected compliance requires a strict inequality when $r_n>0$; at equality it is selected when $r_n\leq0$. Whenever no profitable exit exists, $\widehat W=\widehat V^{\mathrm{IR}}$.
\end{proposition}

The worst binding history has used the full quota as recently as possible. Every other binding history releases capacity at least as soon, so temporal value ordering reduces all exit comparisons to this record. Appendix~\ref{app:sustainability-proof} proves the proposition, including the reachability of the histories relevant for implementation.

\subsection{Quota Size, Memory, and the Survival of Communication}
\label{subsec:memory-survival}

Longer memory has opposing incentive effects. At the all-zero history it raises the continuation cost of a favorable recommendation and strengthens selection. At the most restrictive binding history it lowers the value of preserving the relationship and makes a current followed violation more attractive. Proposition~\ref{prop:binary-sustainability} also implies that the sustainability threshold weakly increases with memory: a longer record requires at least as much patience to keep the worst binding history's continuation value above $g_n$.

Increasing the quota has the reverse effects. It lowers the initial recommendation cost, but raises $B_{N,Q}$ and weakly lowers $\beta^{\mathrm{sus}}_{N,Q}$, even though the worst binding record itself contains an additional endorsement: from $\nu^b_{N,Q+1}$, every path feasible from $\nu^b_{N,Q}$ under quota $Q$ remains feasible under quota $Q+1$, because any window holds at most one additional inherited endorsement and has one additional unit of capacity.

\begin{corollary}[Intervals of quotas and memory lengths for sustained communication]
\label{cor:sustainable-design-intervals}
Suppose (UT) holds, $\underline c_R>0$, and $r_n>0$. Consider the receiver response that follows a first violating recommendation and babbles thereafter. At a fixed discount factor, define
\begin{align}
\mathcal I_N&:=\{Q:1\leq Q<N,\ \widehat c^{\mathrm{IR}}_{N,Q}(0^{N-1})\geq\underline c_R,\ B_{N,Q}>g_n\},\label{eq:sustained-quota-interval}\\
\mathcal J_Q&:=\{N>Q:\ \widehat c^{\mathrm{IR}}_{N,Q}(0^{N-1})\geq\underline c_R,\ B_{N,Q}>g_n\}.\label{eq:sustained-memory-interval}
\end{align}
These are exactly the quotas and memories for which the selected optimal perpetual-compliance policy is implementable and sustained under this response. Each is an interval of integers, possibly empty. At fixed $N$, sustainability imposes the lower quota bound and implementation the upper bound. At fixed $Q$, implementation imposes the lower memory bound and sustainability the upper bound. The latter bound is finite.
\end{corollary}

The result identifies the two constraints on linking decisions. Too little capacity, or too long a wait for its renewal, leaves too little continuation value to deter exit; too much capacity, or too rapid renewal, leaves too little opportunity cost to deter uninformative favorable advice. Appendix~\ref{app:sustainable-design-intervals} gives the proof.

\paragraph{Quota and memory comparisons in the three-state example.}
At $N=7$, Table~\ref{tab:multistate-sustainability} separates the two constraints. The sustained quota interval is $\mathcal I_7=\{2,3\}$. Quota one passes the within-restriction obedience test but fails the no-exit test. Quota four preserves sufficient continuation value but sends favorable recommendations in all three states at the empty record, violating receiver obedience. Larger quotas also fail implementation by Proposition~\ref{prop:binary-quota-monotonicity}.
\begin{table}[!htbp]
\centering
\caption{Quota choice at memory seven in the three-state example. The no-exit test is $B_{7,Q}>1$.}
\label{tab:multistate-sustainability}
\begin{tabular}{@{}cccccc@{}}
\toprule
$Q$ & Initial cost & Initial cutoff & $B_{7,Q}$ & Implementable & No exit\\
\midrule
1 & 0.5132 & 3 & 0.6468 & Yes & No\\
2 & 0.3172 & 2 & 1.2777 & Yes & Yes\\
3 & 0.1756 & 2 & 1.8844 & Yes & Yes\\
4 & 0.1077 & 1 & 2.4955 & No & Yes\\
\bottomrule
\end{tabular}
\end{table}

At quota two, the sustained memory interval is $\mathcal J_2=\{4,5,6,7,8\}$. Memory three fails implementation, with initial cost $0.1084<0.15$. Memory four passes both tests. At $N=8$, the worst binding value is $1.0799>1$; at $N=9$ it is $0.9221<1$, and the memory comparison rules out every longer memory. The bottom panel of Figure~\ref{fig:multistate-linking} plots this comparison.

The same opposition appears in the patience thresholds of the two-state calibration (Appendix~\ref{app:patience-threshold-numerics}).

\FloatBarrier
\subsection{Anticipated Breakdown and Current Advice}
\label{sec:anticipated-breakdown}

Anticipated exit changes current advice. For the same prior-action-closed restriction with a nonempty, temporally lower-closed base set, let $\widehat W$ be the optimal value when a violating endorsement is followed and communication ends thereafter, and let $\widehat c^W(\nu):=\beta_S[\widehat W(T_0\nu)-\widehat W(T_1\nu)]$ at slack histories. \begin{lemma}[Receiver obedience with anticipated breakdown]
\label{lem:binary-breakdown-cost}
Suppose (UT) holds. The selected optimal policy anticipating breakdown is receiver-obedient
at every active history if and only if
\[
\widehat c^W(0^{N-1})\ge\underline c_R.
\]
Moreover,
$\widehat c^W(0^{N-1})\le\widehat c^{\mathrm{IR}}(0^{N-1})$.
\end{lemma}
The comparison holds because perpetual compliance remains available, so $\widehat W\geq\widehat V^{\mathrm{IR}}$. At the empty record, the two Bellman identities give
\[
F(\widehat c^W(0^{N-1}))=(1-\beta_S)\widehat W(0^{N-1})
\geq(1-\beta_S)\widehat V^{\mathrm{IR}}(0^{N-1})
=F(\widehat c^{\mathrm{IR}}(0^{N-1})).
\]
The values are positive, so the relevant costs are below $g_n$; strict decrease of $F$ implies $\widehat c^W(0^{N-1})\leq\widehat c^{\mathrm{IR}}(0^{N-1})$. The prospect of exit can therefore weaken selection before breakdown. Appendix~\ref{app:breakdown-proof} proves the lemma, and Appendix~\ref{app:completed-quota-choice} studies quota choice when breakdown is anticipated.

In the accelerator example, perpetual compliance requires $\beta_S\geq2/3$, whereas the policy anticipating breakdown requires $\beta_S\geq(\sqrt{21}-1)/5$. Between the latter threshold and $\sqrt3-1$, including both endpoints, two consecutive good ventures induce an informative violation and breakdown occurs almost surely. Appendix~\ref{app:accelerator-calculation} verifies these claims.

\FloatBarrier
\section{Payoff Benchmarks and Selective Restrictions}
\label{sec:optimization}

This section evaluates the restrictions against the no-feedback payoff benchmark.

\subsection{The Payoff Ceiling and the KLR Benchmark}
\label{subsec:obedience-ceiling}

KLR show that without feedback the sender's equilibrium payoff never exceeds the highest payoff attainable with uncapped money burning. In the present finite-state, binary-action environment, define
\begin{equation}
\overline\Pi:=F(\underline c_R)
=\sum_{i=k_R}^{n}\mu_i(g_i-g_{k_R-1}).
\label{eq:ceiling-value}
\end{equation}
\begin{benchmark}[Binary-action payoff ceiling]\label{thm:obedience-ceiling}
Suppose some upper tail supports action~$1$. At every discount factor, the sender's normalized continuation payoff at any reached public history is at most $\overline\Pi$, almost surely, in every perfect Bayesian equilibrium, including with unrestricted messages, public randomization, and history dependence. The same ceiling bounds every credible, receiver-obedient pure stationary policy under perpetual compliance. It requires neither (UT) nor prior-action closure and, when $\underline c_R>0$, equals KLR's uncapped-money-burning bound.
\end{benchmark}
For a credible, receiver-obedient pure stationary policy, type $k_R-1$ weakly prefers recommendation~$0$ wherever both recommendations are used, so every favorable recommendation carries cost at least $\underline c_R$. Appendix~\ref{app:general-ceiling} proves the general bound and its money-burning interpretation.

\paragraph{Comparison with persuasion benchmarks.}
Suppose $\underline c_R>0$, and let $\Pi^{\mathrm{KG}}$ denote the sender's optimal normalized payoff under full commitment, as in \citet{kamenica2011bayesian}. With strictly ordered sender gains, a partitional signal is credible in the sense of \citet{lin2024credible} exactly when its favorable cell is an upper tail, and receiver obedience requires the tail to begin at or above $k_R$. The obedient tail $\{\theta_{k_R},\ldots,\theta_n\}$ therefore yields the highest payoff among credible, receiver-obedient partitional signals, $\Pi^{\mathrm{cp}}:=\sum_{i=k_R}^{n}\mu_i g_i$. By \eqref{eq:ceiling-value},
\begin{equation}
\overline\Pi
=\underbrace{\Pi^{\mathrm{cp}}}_{\text{credible-partition payoff}}
-\underbrace{\underline c_R\sum_{i=k_R}^{n}\mu_i}_{\text{expected cost of self-restraint}},
\label{eq:ceiling-decomposition}
\end{equation}
so full support gives $\overline\Pi<\Pi^{\mathrm{cp}}\leq\Pi^{\mathrm{KG}}$.

\begin{corollary}[Finite-memory recurrence at the patience threshold]
\label{cor:patience-threshold}
Suppose $\underline c_R>0$. If a finite-memory restriction supports an informative, credible, receiver-obedient pure stationary policy under perpetual compliance, then $\beta_S>\beta^\ast$.
\end{corollary}

The strict inequality reflects finite memory: after favorable advice, a sequence of zeros restores an informative history, so continuation value remains positive; KLR's project application instead attains its bound at the threshold with a declining implementation scale (Appendix~\ref{app:ceiling-patience}). With Corollary~\ref{cor:existence}, $\beta^\ast$ is exactly the patience boundary for informative finite-memory implementation under (UT).

\subsection{How Rolling Quotas Approach the Bound}
\label{subsec:quota-convergence}

The payoff identity \eqref{eq:finite-lowest-identity} measures how far a restriction falls short of the bound. An implementable full quota earns $F(\widehat c^{\mathrm{IR}}_{N,Q}(0^{N-1}))$ and $\overline\Pi=F(\underline c_R)$, so its shortfall
\[
\overline\Pi-(1-\beta_S)\widehat V^{\mathrm{IR}}_{N,Q}(0^{N-1})
=F(\underline c_R)-F\bigl(\widehat c^{\mathrm{IR}}_{N,Q}(0^{N-1})\bigr)\ \ge 0
\]
is the value of restraint beyond what obedience requires; in the two-state model it equals $p\,[\widehat c^{\mathrm{IR}}_{N,Q}(0^{N-1})-\ell]$. A full quota therefore attains the bound exactly when its cost at the empty record equals $\underline c_R$.

\begin{figure}[!tb]
\centering
\includegraphics[width=\textwidth]{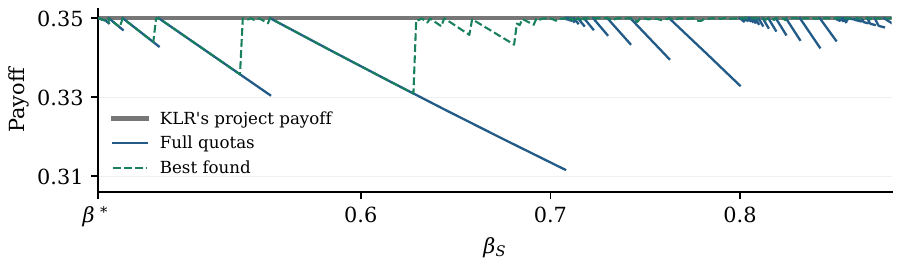}
\caption{Payoffs at $p=1/2$, $\ell=0.3$, and $r_L=-2$. The dashed curve includes searches through $N=128$; search details appear in Appendix~\ref{app:computational-materials}.}
\label{fig:klr-sawtooth}
\end{figure}

\begin{theorem}[Convergence to the payoff bound]
\label{thm:obedience-ceiling-convergence}
Suppose (UT) holds. If $\beta_S>\beta^\ast$, then
\[
\overline\Pi-\frac{1-\beta_S}{\beta_S}\,\underline c_R<\Pi^*_{\mathrm{full}}(\beta_S)\le\overline\Pi,
\]
and the lower bound is positive exactly when $\beta_S>\beta^\ast$. Hence $\Pi^*_{\mathrm{full}}(\beta_S)\to\overline\Pi$ as $\beta_S\uparrow1$, and so does the supremum of normalized initial payoffs over all credible, receiver-obedient pure stationary policies under finite-memory restrictions and perpetual compliance.
\end{theorem}

The lower bound comes from the largest eligible quota with its shortest implementing memory, a pair that need not maximize the payoff (Section~\ref{subsec:choosing-quota-memory}); Appendix~\ref{app:ceiling-convergence} gives the proof. (UT) enters through implementation, not the bound: away from the empty record, costs are at least as high, so favorable cells can be more selective than the obedient tail. The fixed cutoff in KLR's project application avoids this requirement. The convergence extends to quotas that remain sustainable under belief updating.
\begin{corollary}[Convergence under belief updating]\label{cor:bu-ceiling-convergence}
Maintain the assumptions of Theorem~\ref{thm:obedience-ceiling-convergence}. For every $\varepsilon>0$ there are a finite memory $N$ and $\beta_\varepsilon<1$ such that, for $\beta_S>\beta_\varepsilon$, the selected policy at quota $Q^\star_{\mathrm{IR}}(N)$ is sustainable under belief updating and earns a normalized initial payoff above $\overline\Pi-\varepsilon$.
\end{corollary}
Appendix~\ref{app:bu-convergence} gives the proof.

\begin{remark}[Patience and the shortfall of full quotas]\label{rem:sawtooth}
Under (UT) and $\underline c_R>0$, a full quota reaches the ceiling when it first becomes implementable: its empty-record cost then equals the minimum required for obedience. Greater patience raises that cost and lowers the same quota's payoff. The best full-quota payoff therefore declines on intervals containing no entry threshold and returns to the ceiling whenever a quota--memory pair becomes implementable, producing the sawtooth in Figure~\ref{fig:klr-sawtooth}.
\end{remark}
Appendix~\ref{app:sawtooth-proof} proves these comparisons. Discounted payoffs nonetheless rise with patience, because optimal admissible policies can be chosen to earn only nonnegative gains and implementable pairs remain implementable. KLR's project application instead fixes the favorable cutoff at the obedience boundary and lets the implementation scale decline, attaining its ceiling at and above its threshold \citep[Proposition~1]{kuvalekar2022goodwill}. In Figure~\ref{fig:klr-sawtooth}, $(2,1)$ attains $\overline\Pi=0.35$ at $\beta_S=6/7$; at $\beta_S=0.87$, the best full quota earns $0.34843$.

In the accelerator example, $\overline\Pi=3/8$. The one-in-two rule is a single tooth of this sawtooth: it attains $3/8$ at $\beta_S=2/3$, where it becomes implementable under immediate reversion, and its payoff $1/(2+\beta_S)$ then declines toward $1/3$, whereas suitable quotas approach $3/8$ under immediate reversion by Theorem~\ref{thm:obedience-ceiling-convergence}.

\paragraph{Selective restrictions.}
Much of the shortfall of full quotas reflects the coarseness of an integer quota and memory. Restrictions that remain temporally lower closed but admit only part of the next quota layer adjust the empty-record cost more finely. In the calibration of Figure~\ref{fig:klr-sawtooth}, at $\beta_S=0.87$, such a restriction with memory twelve has empty-record cost $0.30021$ and earns $0.34990$, and it is sustainable under belief updating, against $0.34843$ for the best full quota at any memory and a bound of $0.35$; it closes about 93 percent of the gap. At $p=1/2$, $\beta_S=0.95$, and $\ell=0.379$, a sustainable restriction with memory sixteen earns $0.3104999978$, against $0.298946$ for the best full quota with memory at most sixteen and a bound of $0.3105$ (Appendix~\ref{sec:quota-extension}). The shortfall of full quotas therefore need not represent a limit of discipline through recommendation records. Whether deterministic finite-memory restrictions under immediate reversion can approximate the bound arbitrarily closely at each fixed $\beta_S>\beta^\ast$ remains open.

\FloatBarrier
\section{Conclusion}\label{sec:conclusion}

This paper shows that an observable restriction on the recommendation record can make unverifiable advice informative by linking decisions over time. Under a rolling quota, the cost of an endorsement depends on when past endorsements release capacity, so the timing of past advice shapes current advice. Quota size and memory create a trade-off: stronger restraint disciplines recommendations but reduces the value of preserving communication. Designing the restriction therefore requires balancing the incentive to keep advice selective enough for receivers to follow against the incentive to keep giving it.

\appendix

\FloatBarrier
\section{History Restrictions: Proofs and Supporting Results}\label{app:single-history}
The replication package reproduces every numerical result in the paper; Appendix~\ref{app:computational-materials} lists its scripts.

\subsection{Proof of Lemma~\ref{lem:credible-sets}}
\label{app:credible-sets-proof}

At each pre-breakdown history \(\nu_t\), credibility compares \(\phi\) with deviations in \(\Phi(\phi;\nu_t)\), which keep every subsequent rolling history in \(\mathcal V^{N,\phi}\). Along them, \(\sigma^\phi\) prescribes obedience, so each deviation's payoff is given by \eqref{eq:one-period-continuation}, and the response to violating messages plays no role. Since it suffices to consider pure stationary deviations (Appendix~\ref{app:identity}; \citealp{Blackwell1965}), \((\phi,\sigma^\phi)\) has no profitable undetectable deviation exactly when \(\phi\) attains \(\Pi^*_{\mathcal V^{N,\phi}}\) at every rolling history in \(\mathcal V^{N,\phi}\), which is the first claim.

For the second, let \(\phi\) be optimal within \(\mathcal V\). Since \(\phi\) keeps every rolling history in \(\mathcal V\) and \(\mathcal V\) contains the initial rolling history, \(\mathcal V^{N,\phi}\subseteq\mathcal V\). Fix \(\nu^N\in\mathcal V^{N,\phi}\) and a pure stationary strategy \(\phi'\) that keeps every rolling history in \(\mathcal V^{N,\phi}\). Starting from \(\nu^N\), \(\phi'\) never leaves \(\mathcal V^{N,\phi}\), so redefining it outside \(\mathcal V^{N,\phi}\) to keep rolling histories in \(\mathcal V\), which admissibility permits, leaves \(\Pi^{\phi'}(\nu^N)\) unchanged and makes \(\phi'\) feasible for \(\mathcal V\). Hence \(\Pi^{\phi'}(\nu^N)\le\Pi^*_{\mathcal V}(\nu^N)=\Pi^\phi(\nu^N)\), so \(\phi\) is optimal within \(\mathcal V^{N,\phi}\), and \((\phi,\sigma^\phi)\) is credible by the first claim.

\subsection{Optimal Admissible Policies and the Local Test}
\label{app:optimal-implementation}\label{app:identity}

Use the optimal value and cost of Section~\ref{sec:temporal-restrictions}. The Bellman equation is
\begin{equation}
\widehat V^{\mathrm{IR}}(\nu)=\beta_S\widehat V^{\mathrm{IR}}(T_0\nu)+
\begin{cases}
F(\widehat c^{\mathrm{IR}}(\nu)),&\nu\in\mathcal H^\circ,\\
0,&\nu\notin\mathcal H^\circ.
\end{cases}
\label{eq:base-policy-bellman}
\end{equation}
Its unique bounded solution, together with Assumption~\ref{ass:tie-breaking}, selects the favorable cell
\begin{equation}
C_1^*(\nu)
=\{\theta_i:g_i>\widehat c^{\mathrm{IR}}(\nu)\}
\cup\{\theta_i:g_i=\widehat c^{\mathrm{IR}}(\nu),\ r_i>0\}.
\label{eq:selected-favorable-cell}
\end{equation}
This policy is optimal against every admissible causal strategy, including strategies that use randomization or the sender's private history. Minimality of $k_R$ gives $r_{k_R-1}<0$, so \eqref{eq:selected-favorable-cell} excludes that state at cost $\underline c_R=g_{k_R-1}$, leaving the obedient tail from $k_R$. At a smaller cost, the selected nonempty cell is an earlier, disobedient tail; at every larger cost below $g_n$, it is a more selective tail and remains obedient under (UT).

\begin{proof}[Proof of Proposition~\ref{prop:local}]
By Lemma~\ref{lem:credible-sets}, $(\phi,\sigma^\phi)$ is credible if and only if $\phi$ is optimal within $\mathcal V^{N,\phi}$ from every induced history. Holding the obedient response fixed, the sender's problem within $\mathcal V^{N,\phi}$ is a finite discounted Markov decision problem whose state is the rolling history before the current draw, so randomized or history-dependent strategies do no better than stationary deterministic ones \citep{Blackwell1965}. A stationary policy is optimal from every state if and only if, at every $\nu\in\mathcal V^{N,\phi}$ and every $\theta_i$, its recommendation maximizes $(1-\beta_S)u_S(\theta_i,m)+\beta_S\Pi^\phi(T_m\nu)$ over the undetectable recommendations $m\in A^\phi(\nu)$. The condition is vacuous where only one recommendation is undetectable. Where both are, the payoff difference between recommending $1$ and $0$ is $(1-\beta_S)[g_i-c^\phi(\nu)]$ by \eqref{eq:local-continuation-cost}, which gives the stated inequalities.
\end{proof}

\subsection{Temporal Value Ordering and the Lowest-History Test}
\label{app:base-ordering-proofs}\label{app:lowest}

\begin{proof}[Proof of Lemma~\ref{lem:temporal-value-order}]
\emph{Value ordering.} Fix an admissible causal policy starting at $\nu$ and a lower history $\widetilde\nu\preceq_T\nu$. From $\widetilde\nu$, maintain a virtual record starting at $\nu$ and apply the policy to that record and the same future states. Appending the same message preserves every suffix-count inequality, zero continuation is always feasible, and temporal lower closure makes every virtual recommendation $1$ feasible on the actual path, so the state-action payoffs are identical. Taking the supremum over policies gives $\widehat V^{\mathrm{IR}}(\widetilde\nu)\geq\widehat V^{\mathrm{IR}}(\nu)$. In particular, $\widehat V^{\mathrm{IR}}(0^{N-1})\geq\widehat V^{\mathrm{IR}}(\nu)$ and $\widehat V^{\mathrm{IR}}(T_0\nu)\geq\widehat V^{\mathrm{IR}}(\nu)$, and $\widehat c^{\mathrm{IR}}(\nu)\ge0$ because $T_0\nu\preceq_TT_1\nu$.
\end{proof}

\begin{proof}[Proof of Lemma~\ref{lem:general-lowest-cost}]
\emph{Upper bound.} Date the current admissible recommendation at zero. Starting from $T_1\nu$, simulate an optimal continuation from $T_0\nu$, sending zeros until its first future recommendation $1$. If that recommendation occurs at date $\tau\ge1$, omit it and thereafter use the virtual messages. Immediately after the omission, the actual record differs only by having the date-zero recommendation in place of the newer date-$\tau$ one, or by having one fewer recommendation if the date-zero message has expired, so it temporally precedes the virtual record, and lower closure keeps all subsequent messages feasible. The two continuations differ in at most one realized payoff, whose date-zero discounted value is at most $\beta_Sg_n$, so $\beta_S\widehat V^{\mathrm{IR}}(T_1\nu)\geq\beta_S\widehat V^{\mathrm{IR}}(T_0\nu)-\beta_Sg_n$, that is, $\widehat c^{\mathrm{IR}}(\nu)\le\beta_Sg_n$. For $N=1$, the successor histories coincide and the cost is zero.

\emph{Lower bound.} Since $T_0(0^{N-1})=0^{N-1}$, subtracting the Bellman equation at $\nu$ from the one at $0^{N-1}$ yields
\[
F(\widehat c^{\mathrm{IR}}(0^{N-1}))-F(\widehat c^{\mathrm{IR}}(\nu))
=(1-\beta_S)[\widehat V^{\mathrm{IR}}(0^{N-1})-\widehat V^{\mathrm{IR}}(\nu)]
+\beta_S[\widehat V^{\mathrm{IR}}(T_0\nu)-\widehat V^{\mathrm{IR}}(\nu)]\geq0.
\]
Since $\widehat c^{\mathrm{IR}}(0^{N-1})<g_n$ and full support makes $F$ strictly decreasing below $g_n$, the comparison follows.
\end{proof}

\begin{proof}[Proof of Proposition~\ref{prop:optimal-policy-obedience}]
Sender optimality follows from the Bellman equation. A cost below $\underline c_R$ at $0^{N-1}$ gives a nonempty favorable cell that violates receiver obedience. If the inequality holds, Lemma~\ref{lem:general-lowest-cost} makes every favorable cell at least as selective as the obedient upper tail at $0^{N-1}$, and (UT) establishes obedience to $1$. The complement of the selected cell \eqref{eq:selected-favorable-cell} has receiver gain $\sum_i\mu_ir_i-\sum_{\theta_i\in C_1^*(\nu)}\mu_ir_i<0$, and forced zeros are evaluated under the prior. The cell at $0^{N-1}$ is nonempty and proper, so communication there is informative. Discounted Bellman optimality covers all admissible continuation deviations (Appendix~\ref{app:identity}).
\end{proof}

\subsection{Pairwise Cost Ordering under Rolling Quotas}
\label{app:quota-cost-order}

\begin{proof}[Proof of Proposition~\ref{prop:quota-cost-order}]
For $N=1$, both successors coincide and the cost is zero. Let $L=N-1\geq1$, and let $s_j(\nu)$ count the ones in the $j$ most recent positions of $\nu$, with $s_0(\nu)=0$. Represent a history by $x=(x_0,x_0+s_1(\nu),\ldots,x_0+s_L(\nu))$ with $x_0\in\mathbb Z$. An entering history is active exactly when it contains at most $Q$ ones, and any such history can be generated from $0^{N-1}$, so its lifted representation belongs to
\[
\mathcal D=\left\{x\in\mathbb Z^{L+1}:
x_j-x_{j-1}\in\{0,1\}\ (1\leq j\leq L),\ x_L-x_0\leq Q\right\}.
\]
Write $J(x)=\widehat V^{\mathrm{IR}}(\nu)$; adding a common integer to all coordinates leaves $J$ unchanged. Sending $a\in\{0,1\}$ chooses $y=x_0-a$ and gives the next vector $(y,x_0,\ldots,x_{L-1})$, with the new window constraint $x_{N-1}-y\leq Q$. Every difference constraint $x_v-x_u\leq q$ is preserved by coordinatewise minima and maxima: if it holds for $x$ and $x'$, then $x_v\wedge x'_v\leq(x_u\wedge x'_u)+q$ and $x_v\vee x'_v\leq(x_u\vee x'_u)+q$. This also applies to the two inequalities defining each binary increment, so $\mathcal D$ is closed under both operations.

Take optimal policies from $x,x'\in\mathcal D$, coupled with the same future states. Construct policies from $x\wedge x'$ and $x\vee x'$ by choosing $y\wedge y'$ and $y\vee y'$ whenever the virtual policies choose $y,y'$. The inequalities $x_0-1\leq y\leq x_0$ ensure that each constructed recommendation belongs to $\{0,1\}$, and the new window constraint has the same difference form, so both constructed choices are feasible. After updating, the constructed vectors remain the minimum and maximum of the virtual vectors, and induction continues the construction forever; this is the step that uses the rolling-quota structure. The two constructed recommendations have the same sum as the virtual ones, since $(x_0\wedge x'_0-y\wedge y')+(x_0\vee x'_0-y\vee y')=(x_0-y)+(x'_0-y')$, so with a common current state their combined current gain is identical. Each constructed policy can simulate both virtual records using only observed current and past states, so it is causal. Discounting, taking expectations, and optimizing from the constructed initial histories gives
\begin{equation}
J(x)+J(x')\leq J(x\wedge x')+J(x\vee x').
\label{eq:quota-lifted-value-inequality}
\end{equation}

For a history $\nu$, represent its successor after $a$ by
\[
X^a(\nu)=(-a,0,s_1(\nu),\ldots,s_{L-1}(\nu)).
\]
If $\nu'\preceq_T \nu$ are active and $\nu$ permits $1$, the bound $|\nu|\leq Q-1$ also permits $1$ at $\nu'$. All four successors belong to $\mathcal D$, and $X^0(\nu')\wedge X^1(\nu)=X^1(\nu')$ and $X^0(\nu')\vee X^1(\nu)=X^0(\nu)$. Applying \eqref{eq:quota-lifted-value-inequality} yields $\widehat V^{\mathrm{IR}}(T_0\nu')+\widehat V^{\mathrm{IR}}(T_1\nu)\leq\widehat V^{\mathrm{IR}}(T_1\nu')+\widehat V^{\mathrm{IR}}(T_0\nu)$, and rearranging and multiplying by $\beta_S$ proves the cost ordering. The argument uses common i.i.d.\ states, bounded measurable gains, and optimal causal policies, so it applies to finite and continuous state spaces.
\end{proof}

\paragraph{Failure beyond rolling quotas.}\label{app:exclusion-examples}\label{app:cost-order-counterexample}
Temporal lower closure alone does not order costs. For $N=4$, $\mathcal B=\{000,100,010,001,110\}$, $\beta_S=0.99$, $p=0.9$, $g_H=1$, and $g_L=0.1$, the optimal separating policy has $c(010)\approx0.89535>0.49228\approx c(001)$ although $010\preceq_T001$. Receiver gains $r_H=1$ and $r_L=-10$ make the policy obedient.

\FloatBarrier
\section{Rolling-Quota Design: Proofs and Additional Comparisons}
\label{app:optimization}
\subsection{Quota and Memory Comparisons}\label{app:quota-memory-proofs}

\begin{proof}[Proof of Proposition~\ref{prop:binary-quota-monotonicity}]
Every recommendation path satisfying $(N,Q)$ also satisfies $(N,Q+1)$, so the optimal value weakly increases with $Q$. For $Q\ge1$, recommendation $1$ is admissible at $0^{N-1}$, and \eqref{eq:finite-lowest-identity} orders the costs, because $F$ is strictly decreasing below $g_n$ and $\widehat c^{\mathrm{IR}}(0^{N-1})<g_n$ by Lemma~\ref{lem:general-lowest-cost}. Under (UT), Proposition~\ref{prop:optimal-policy-obedience} makes the implementable positive quotas those with $\widehat c^{\mathrm{IR}}_{N,Q}(0^{N-1})\ge\underline c_R$, an initial segment whose largest member has the largest value.
\end{proof}

\begin{proof}[Proof of Proposition~\ref{prop:binary-memory}]
Every path satisfying $(N+1,Q)$ also satisfies $(N,Q)$, since each window of $N$ periods lies in one of $N+1$ periods, and both restrictions start from the all-zero record. The optimal value therefore weakly decreases with $N$, \eqref{eq:finite-lowest-identity} orders the costs, and Proposition~\ref{prop:optimal-policy-obedience} preserves implementation.
\end{proof}

\subsection{Memory Limits and Finite Quota Search}\label{app:memory-derivations}

\begin{lemma}[The fixed-quota limit]\label{lem:binary-memory-limit}
Let $c(q)$ be defined by \eqref{eq:finite-lifetime-recursion}. Then $F(c(q))$ strictly increases to $F(0)$, and $c(q)$ strictly decreases to zero. For every $Q\geq1$ and finite $N\geq Q$,
\begin{equation}
0<(1-\beta_S)\widehat V^{\mathrm{IR}}_{N,Q}(0^{N-1})-F(c(Q))
\leq F(0)\beta_S^N.
\label{eq:finite-memory-bound}
\end{equation}
Thus, as $N$ grows, the normalized rolling value weakly decreases to $F(c(Q))$, and its all-zero cost weakly increases to $c(Q)$ from strictly below.
\end{lemma}
\begin{proof}
Let $x_q$ be the optimal normalized value with $q$ lifetime units and $x_0=0$. The Bellman equation is $x_q=\sum_i\mu_i\max\{\beta_Sx_q,(1-\beta_S)g_i+\beta_Sx_{q-1}\}$. Subtracting $\beta_Sx_q$ gives $x_q=F(c(q))$, where $c(q)=\beta_S(x_q-x_{q-1})/(1-\beta_S)$, yielding \eqref{eq:finite-lifetime-recursion}.

Deleting the first recommendation of a $q$-unit policy bounds the extra unit's unnormalized value by $g_n$, so $c(q)\leq\beta_Sg_n<g_n$. Every maximal-gain state therefore uses a remaining unit, and any optimizer exhausts its quota almost surely. An extra unit permits another maximal-gain recommendation afterward, making $x_q$ strictly increasing and $c(q)$ strictly decreasing. Following the unrestricted positive-gain policy until $q$ recommendations gives $x_q\uparrow F(0)$ by dominated convergence, hence $c(q)\downarrow0$.

The lifetime-$Q$ optimizer is rolling-quota feasible. After exhaustion, waiting $N$ periods and making another maximal-gain recommendation strictly improves its payoff. Conversely, truncating the rolling optimizer after dates $0,\ldots,N-1$ gives a lifetime-$Q$ policy, and the discarded normalized payoff is at most $F(0)\beta_S^N$. These comparisons prove \eqref{eq:finite-memory-bound}. The empty-record identity \eqref{eq:finite-lowest-identity} and strict decrease of $F$ give the cost limit from strictly below; Proposition~\ref{prop:binary-memory} supplies memory monotonicity.
\end{proof}

\paragraph{Binary closed form.} Let $\rho:=p\beta_S/[1-(1-p)\beta_S]\in(0,1)$. The policy accepting the first $Q$ high states has marginal cost $\rho^q$ at each remaining quota $q\le Q$. If $\ell<\rho^Q$, these costs lie in $(\ell,1)$, so the policy is optimal. Conversely, $c(Q)>\ell$ and decreasing $c(q)$ imply that every remaining quota accepts only high states, giving $c(Q)=\rho^Q$. Thus, for $0<\ell<1$,
\[
\overline Q^{\mathrm{IR}}=\left\lceil\frac{\log\ell}{\log\rho}\right\rceil-1.
\]

\begin{proof}[Proof of Theorem~\ref{thm:binary-finite-quota}]
By Lemma~\ref{lem:binary-memory-limit}, the empty-record cost increases with memory to $c(Q)$ from strictly below. Under (UT), Proposition~\ref{prop:optimal-policy-obedience} therefore gives a finite implementing memory exactly when $\underline c_R<c(Q)$. Since $c(Q)\downarrow0$ and $\underline c_R>0$, only finitely many quotas qualify. Memory monotonicity then makes the largest implementable quota increase to the largest eligible quota.
\end{proof}

\begin{proof}[Proof of Corollary~\ref{prop:binary-finite-search}]
Memory monotonicity makes the shortest implementing memory optimal at each eligible quota. By Lemma~\ref{lem:binary-memory-limit}, the empty-record test holds whenever $F(0)\beta_S^N\leq F(\underline c_R)-F(c(Q))$, whose right-hand side is positive. This supplies a finite stopping bound for each of the finitely many eligible quotas.
\end{proof}

\FloatBarrier
\section{Sustainability: Proofs}
\label{sec:supp-breakdown}
\subsection{Sustainability at Binding Histories}\label{app:sustainability-proof}
\begin{proof}
At a binding history $\nu$, the only admissible recommendation is zero, so the value at $\nu$ is $\beta_S\widehat V^{\mathrm{IR}}_{N,Q}(T_0\nu)$. Moving an inherited recommendation earlier weakly enlarges the set of feasible future recommendation paths: any policy feasible from the more restrictive history can be simulated on the same future states from the less restrictive one, with the same rewards path by path, so taking suprema orders the optimal values. Every binding history temporally precedes $\nu^b_{N,Q}$, which proves \eqref{eq:binary-sustainability-minimum}.

For the memory comparison, represent the initial $Q$ recommendations by dates $-Q,\ldots,-1$, which describe $\nu^b_{N,Q}$ for every $N>Q$. Every future path satisfying a longer-window quota also satisfies the shorter-window quota with these inherited recommendations, so policy inclusion makes the displayed minimum weakly decrease with $N$. For the quota comparison at fixed $N$, $\nu^b_{N,Q+1}$ adds one inherited recommendation at date $-(Q+1)$. Couple any feasible continuation under quota $Q$ with the same future states and recommendations under quota $Q+1$: until that recommendation expires, each window contains at most one additional one while the permitted count is one higher, so the coupled path is feasible and earns the same payoffs. Taking suprema gives $B_{N,Q+1}\geq B_{N,Q}$. 

Write $v(\beta)=\widehat V^{\mathrm{IR}}_{N,Q}(\nu^b_{N,Q};\beta)$. Optimal policies can be chosen to take only nonnegative gains, because sending zero releases at least as much capacity as sending one. Finite-state discounted dynamic programming makes $v$ continuous on $(0,1)$. For $\beta_2>\beta_1$, evaluate a $\beta_1$-optimal policy at $\beta_2$. Its first $N-Q\geq1$ rewards are zero, and it earns a strictly positive reward at a later finite date with positive probability. Its discounted payoff therefore strictly increases, proving $v(\beta_2)>v(\beta_1)$. The unrestricted payoff bounds the value by
\[
0\leq v(\beta)\leq\frac{F(0)\beta^{N-Q}}{1-\beta},
\]
so $v(\beta)\to0$ as $\beta\downarrow0$. For a lower bound, send zero through date $N-1$, then recommend one only in maximal-gain states at dates $N,2N,3N,\ldots$. This policy is feasible for every $Q\geq1$ and yields
\[
v(\beta)\geq\frac{\mu_n g_n\beta^N}{1-\beta^N}\longrightarrow\infty
\quad\text{as }\beta\uparrow1.
\]
Continuity and strict increase give the unique root $\beta^{\mathrm{sus}}_{N,Q}$.

At a binding history, a followed violation in state $i$ yields $g_i$ and zero continuation, whereas compliance yields $\beta_S\widehat V^{\mathrm{IR}}(T_0\nu)$, so the maximal-gain state binds the no-exit condition. If the perpetual-compliance policy is implementable, receiver obedience gives a nonempty zero cell at every slack history and the maximal-gain state recommends one whenever feasible, so every full-quota history is reached with positive probability along a finite compliance path; the minimum condition therefore characterizes the absence of profitable exits from the initial history and from every active history. At equality only the maximal-gain state is indifferent, and Assumption~\ref{ass:tie-breaking} gives the stated boundary convention. Finally, if no exit is profitable, $\widehat V^{\mathrm{IR}}$ solves the completed Bellman equation, and contraction gives $\widehat W=\widehat V^{\mathrm{IR}}$.
\end{proof}
\subsection{The Intervals of Sustainable Designs}\label{app:sustainable-design-intervals}
\begin{proof}[Proof of Corollary~\ref{cor:sustainable-design-intervals}]
By Proposition~\ref{prop:optimal-policy-obedience}, implementation is equivalent to the initial cost being at least $\underline c_R$. By Proposition~\ref{prop:binary-sustainability} and $r_n>0$, selected perpetual compliance under a followed first violation is equivalent to $B_{N,Q}>g_n$. At fixed $N$, the first inequality selects an initial segment of quotas, whereas the second selects a terminal segment by \eqref{eq:sustainability-quota-memory}. Their intersection is an interval. At fixed $Q$, implementation selects a terminal segment of memories, whereas the no-exit test selects an initial segment. Their intersection is again an interval. Finally, at $\nu^b_{N,Q}$ the first $N-Q$ current and future recommendations must be zero, giving $B_{N,Q}\leq F(0)\beta_S^{N-Q}/(1-\beta_S)\to0$ as $N\to\infty$ for fixed $Q$ and $\beta_S<1$. Since $g_n>0$, the sustained memories have a finite upper bound. The same comparisons imply that $\beta^{\mathrm{sus}}_{N,Q}$ weakly decreases with $Q$ and weakly increases with $N$.
\end{proof}

\FloatBarrier
\subsection{The Accelerator Example}
\label{app:accelerator-calculation}

The accelerator example has $p=1/2$, $\ell=1/4$, the quota $(2,1)$, and receiver payoffs $u_R(H,1)=1$, $u_R(L,1)=-2$, and $u_R(\theta,0)=0$. The entering history is the preceding recommendation, $\nu\in\{0,1\}$, starting at $\nu=0$; after a violation, the sender's continuation value is zero.

\paragraph{Perpetual compliance.} The separating rule's values solve $V_0=\tfrac12[1+\beta_SV_1]+\tfrac12\beta_SV_0$ and $V_1=\beta_SV_0$, so

\begin{equation}
V_0=\frac{1}{2-\beta_S-\beta_S^2},\quad c_0=\beta_S(V_0-V_1)=\frac{\beta_S}{2+\beta_S}.
\label{eq:accelerator-compliance-values}
\end{equation}
Reserving the endorsement in $L$ at $\nu=0$ is optimal if and only if $c_0\ge1/4$, that is, $\beta_S\ge2/3$, and endorsing in $H$ is optimal because $c_0<1$. At $\nu=1$, waiting gives $\beta_SV_0$ and a followed violation at most $1$, so no violation is profitable if and only if $\beta_SV_0\ge1$, that is, $\beta_S\ge\sqrt3-1$; at equality, Assumption~\ref{ass:tie-breaking} selects the violation, which benefits the receiver. The rule is therefore credible and sustainable for $\beta_S>\sqrt3-1$. At $\nu=0$, the recommendations reveal the state; at $\nu=1$, recommendation $0$ is uninformative and the prior expected gain from funding is $-1/2$, so investors decline.

\paragraph{Anticipated breakdown.} If the sender instead endorses in $H$ at $\nu=1$ and then babbles, the values solve $V_0=\tfrac12[1+\beta_SV_1]+\tfrac12\beta_SV_0$ and $V_1=\tfrac12+\tfrac12\beta_SV_0$, so
\begin{equation}
V_0=\frac{2+\beta_S}{4-2\beta_S-\beta_S^2},
\qquad
c_0=\frac{\beta_S^2}{4-2\beta_S-\beta_S^2}.
\label{eq:accelerator-breakdown-values}
\end{equation}
Reserving in $L$ at $\nu=0$ is optimal if and only if $c_0\ge1/4$, that is, $\beta_S\ge(\sqrt{21}-1)/5$; violating in $H$ at $\nu=1$ is strictly optimal if and only if $\beta_SV_0<1$, that is, $\beta_S<\sqrt3-1$; and waiting in $L$ at $\nu=1$ is strictly optimal because $\beta_SV_0=c_0+\beta_SV_1>1/4$. At the endpoints, Assumption~\ref{ass:tie-breaking} selects $0$ in $L$ at the slack history and the violation in $H$ at the binding history. Recommendation $1$ is then sent only in $H$, including when it violates the quota, and $0$ reveals $L$, so investors' responses are optimal, and the completed policy is implementable but unsustainable. Among $k$ disjoint pairs of periods, the probability that none contains two good ventures is $(3/4)^k\to0$, so breakdown occurs almost surely.

\FloatBarrier
\section{Proofs for the Payoff Benchmarks}
\subsection{Derivation of the KLR Benchmark in the Binary-Action Model}\label{app:general-ceiling}\label{app:ceiling-proof}
\begin{proof}[Proof of Benchmark~\ref{thm:obedience-ceiling}]
Evaluate values after any public random draw and before the current state. Because there is no feedback and states are independent, the sender's optimal continuation value after a public message is common across his current types. Let $M<\infty$ be the supremum of on-path unnormalized continuation values (an essential supremum under nonatomic randomization). At a reached history, message $m$ induces action~$1$ with probability $\alpha_m$ and continuation value $V_m$; with $d_m:=\beta_S(M-V_m)$, nonnegative for every on-path $m$, type $i$'s payoff from $m$ is $\beta_SM+\alpha_mg_i-d_m$.

Let $j:=k_R-1$ and $b_i:=\Pr(\alpha_m>0\mid\theta_i)$. By incentive compatibility and strictly ordered gains, $b_i>0$ implies $b_k=1$ for every $k>i$: a positive-action message optimal for type $i$ strictly beats every zero-action message for every higher type. If $b_j=1$, let $i_0\le j$ be the lowest type with $b_{i_0}>0$; with $R_k:=\sum_{i\ge k}\mu_ir_i$, the receiver gain pooled over messages with $\alpha_m>0$ is $b_{i_0}R_{i_0}+(1-b_{i_0})R_{i_0+1}<0$ by minimality of $k_R$ (if $i_0=j$, then $b_{i_0}=1$ and the second term vanishes), whereas receiver optimality makes each such message's gain nonnegative. Hence $b_j<1$: type $j$ optimally sends some message with $\alpha_m=0$, worth $-d_m\le0$, so $d_m\ge\alpha_mg_j$ for every on-path $m$. Every type's surplus therefore satisfies $\alpha_mg_i-d_m\le(g_i-g_j)_+$, and averaging over states gives a value of at most $\beta_SM+F(\underline c_R)=\beta_SM+\overline\Pi$ at every reached history. Taking the supremum, $(1-\beta_S)M\le\overline\Pi$.

For a credible, receiver-obedient pure stationary policy under perpetual compliance, let $V$ be its unnormalized value and $\nu$ a reached entering history. If only recommendation~$0$ is used at $\nu$, then $V(\nu)=\beta_SV(T_0\nu)$. Otherwise recommendation~$1$ is used, so $T_1\nu$ is admissible and the continuation cost $c(\nu):=\beta_S[V(T_0\nu)-V(T_1\nu)]$ is defined. Receiver obedience then requires recommendation~$0$ to be used as well, since the whole state space is an upper tail beginning below $k_R\geq2$. Both recommendations are undetectable, and Proposition~\ref{prop:local} puts every state with $g_i>c(\nu)$ in the favorable cell. If $\theta_{k_R-1}$ recommended~$1$, the cell would be an upper tail beginning below $k_R$, which minimality of $k_R$ makes disobedient. Hence $\theta_{k_R-1}$ recommends~$0$, Proposition~\ref{prop:local} gives $c(\nu)\geq g_{k_R-1}=\underline c_R$, and
\[
V(\nu)=\beta_SV(T_0\nu)+\sum_{i:\,\phi(\theta_i,\nu)=1}\mu_i\bigl(g_i-c(\nu)\bigr)\leq\beta_SV(T_0\nu)+F(\underline c_R).
\]
In both cases $T_0\nu$ is reached and $V(\nu)\leq\beta_SV(T_0\nu)+F(\underline c_R)$, because $F\geq0$. Since gains are bounded, the supremum $\overline V$ of $V$ over reached entering histories is finite, and $\overline V\leq\beta_S\overline V+F(\underline c_R)$ gives $(1-\beta_S)\overline V\leq\overline\Pi$.
\end{proof}

\paragraph{Identification with uncapped money burning.}
Suppose $\underline c_R>0$. In the corresponding static money-burning game, prescribe recommendation~$1$ and burning $\underline c_R$ for types $i\geq k_R$, and recommendation~$0$ with no burning for types $i<k_R$. The receiver follows both recommendations: the favorable tail has nonnegative expected gain, and its complement has strictly negative expected gain. Types above $k_R-1$ strictly prefer favorable advice, earning $g_i-\underline c_R>0$, while lower types prefer recommendation~$0$. Type $k_R-1$ is indifferent and selects~$0$ because minimality of $k_R$ implies $r_{k_R-1}<0$. After any unused message--burning pair, assign probability one to type $k_R-1$ and choose action~$0$; such a deviation yields at most zero. The equilibrium payoff is therefore
\[
\sum_{i=k_R}^{n}\mu_i(g_i-\underline c_R)=F(\underline c_R).
\]
The preceding incentive-compatibility argument, with $d_m$ equal to the nonnegative amount burned, bounds every static money-burning equilibrium by the same expression. Thus $F(\underline c_R)$ equals KLR's uncapped-money-burning value without requiring (UT). If $\underline c_R<0$, the upper-bound inequality still holds, but $F(\underline c_R)>F(0)=\sum_i\mu_i(g_i)_+$ exceeds the sender's statewise payoff bound and cannot be the attainable money-burning value.

\subsection{The Finite-Memory Boundary at the Patience Threshold}\label{app:ceiling-patience}
\begin{proof}
Let $\nu$ be a reached history at which recommendation~$1$ is used. By the proof of Benchmark~\ref{thm:obedience-ceiling}, $c(\nu)\ge\underline c_R>0$, and by Proposition~\ref{prop:local} every state recommending $1$ has $g_i\ge c(\nu)>0$, so every favorable recommendation earns a positive gain and $V(0^{N-1})>0$. Obedience gives recommendation~$0$ positive probability at every reached history, so $N-1$ zeros lead from $T_1\nu$ to $0^{N-1}$ and $V(T_1\nu)>0$. The bound of Benchmark~\ref{thm:obedience-ceiling} at $T_0\nu$, which is reached because $0$ is used at $\nu$, then gives $\underline c_R\le c(\nu)<\beta_SV(T_0\nu)\le\beta_S\overline\Pi/(1-\beta_S)$, that is, $\beta_S>\beta^\ast$.
\end{proof}
\subsection{Rolling-Quota Convergence}\label{app:ceiling-convergence}
\begin{proof}
Let $Q:=\overline Q^{\mathrm{IR}}$. The proof of Corollary~\ref{cor:existence} gives $c(1)>\underline c_R$, so $Q\ge1$, and $c(Q+1)\le\underline c_R$ by the definition of $Q$. By \eqref{eq:finite-lifetime-recursion} at $q=Q+1$ and monotonicity of $F$,
\[
F(c(Q))=F(c(Q+1))-\frac{1-\beta_S}{\beta_S}\,c(Q+1)\ \ge\ \overline\Pi-\frac{1-\beta_S}{\beta_S}\,\underline c_R .
\]
Lemma~\ref{lem:binary-memory-limit} gives $(1-\beta_S)P^{\mathrm{IR}}_Q>F(c(Q))$, proving the lower bound, and Benchmark~\ref{thm:obedience-ceiling} gives the upper bound, which also bounds the broader class.
\end{proof}

\subsection{Sender Patience}\label{app:patience-comparisons}
\begin{lemma}[Sender patience]\label{lem:binary-patience}
Fix the finite-state primitives, $N$, and $1\leq Q\leq N$. For $0<\beta_1<\beta_2<1$,
\begin{equation}
(1-\beta_2)\widehat V^{\mathrm{IR}}_{N,Q}(0^{N-1};\beta_2)
\leq(1-\beta_1)\widehat V^{\mathrm{IR}}_{N,Q}(0^{N-1};\beta_1).
\label{eq:binary-patience-comparison}
\end{equation}
The inequality is strict when $Q<N$. At $Q=N$, both sides equal $F(0)$. The all-zero cost is therefore strictly increasing in patience for $Q<N$; under (UT), the maximal IR-implementable quota is weakly increasing.
\end{lemma}
\begin{proof}
Write $V_j(\nu)=\widehat V^{\mathrm{IR}}_{N,Q}(\nu;\beta_j)$, and let $\phi_2$ be an optimizer at $\beta_2$. The Bellman inequality for $V_1$ gives
\[
\mathbb E^{\phi_2}[u_t\mid \nu_t]\leq V_1(\nu_t)-\beta_1\mathbb E^{\phi_2}[V_1(\nu_{t+1})\mid \nu_t].
\]
Multiplying by $\beta_2^t$, taking expectations from $0^{N-1}$, and telescoping yields
\begin{equation}
V_2(0^{N-1})\leq V_1(0^{N-1})+(\beta_2-\beta_1)
\mathbb E_{0^{N-1}}^{\phi_2}\left[\sum_{t\geq0}\beta_2^t V_1(\nu_{t+1})\right]
\leq\frac{1-\beta_1}{1-\beta_2}V_1(0^{N-1}).
\label{eq:binary-discount-telescoping}
\end{equation}
The terminal term vanishes by boundedness. The second inequality uses $V_1(\nu)\leq V_1(0^{N-1})$ at all histories, since removing inherited recommendations preserves every feasible future path. If $Q<N$, an initial run of $Q$ maximal-gain states has positive probability, receives $Q$ recommendations under $\phi_2$, and reaches a binding history $\nu_Q$. At that history,
\[
V_1(\nu_Q)=\beta_1V_1(T_0\nu_Q)\leq\beta_1V_1(0^{N-1})<V_1(0^{N-1}).
\]
The second inequality in \eqref{eq:binary-discount-telescoping} is therefore strict. At $Q=N$, the sender chooses every positive-gain state without a capacity cost. The identity with $F$ orders costs and, under (UT), implementation.
\end{proof}

\subsection{Patience and the Shortfall of Full Quotas}\label{app:sawtooth-proof}
For each $1\le Q<N$ with a nonempty implementability set, define $\beta_{N,Q}:=\inf\{\beta\in(0,1):\widehat c^{\mathrm{IR}}_{N,Q}(0^{N-1};\beta)\ge\underline c_R\}$.
\begin{proof}[Proof of Remark~\ref{rem:sawtooth}]
Fix $1\le Q<N$ and write $\widehat c(\beta):=\widehat c^{\mathrm{IR}}_{N,Q}(0^{N-1};\beta)$. The optimal value of the finite discounted problem is continuous in the discount factor, so $\widehat c$ is continuous; Lemma~\ref{lem:binary-patience} makes it strictly increasing, and Lemma~\ref{lem:general-lowest-cost} gives $\widehat c(\beta)\le\beta g_n\to0$ as $\beta\to0$. Hence, when nonempty, $\{\beta:\widehat c(\beta)\ge\underline c_R\}=[\beta_{N,Q},1)$ with $\widehat c(\beta_{N,Q})=\underline c_R$, and Proposition~\ref{prop:optimal-policy-obedience} identifies this set with implementability. By \eqref{eq:finite-lowest-identity}, the normalized initial payoff is $F(\widehat c(\beta))$, which equals $F(\underline c_R)=\overline\Pi$ at $\beta_{N,Q}$ and is strictly smaller at larger $\beta$, because $F$ is strictly decreasing below $g_n$.

Let $\beta_1<\beta_2$ lie in $[a,b)$, whose interior contains no $\beta_{N,Q}$, and let $(N,Q)$ attain $\Pi^*_{\mathrm{full}}(\beta_2)$ (Corollary~\ref{prop:binary-finite-search}). Then $\beta_{N,Q}\le\beta_2$ and $\beta_{N,Q}\notin(a,b)$, so $\beta_{N,Q}\le a\le\beta_1$. The pair is therefore implementable at $\beta_1$, and $\Pi^*_{\mathrm{full}}(\beta_1)\ge F(\widehat c(\beta_1))>F(\widehat c(\beta_2))=\Pi^*_{\mathrm{full}}(\beta_2)$.
\end{proof}

\FloatBarrier
\section{Proofs for Prescribed Upper-Tail Rules}

\label{app:binary-prescribed-policy}\label{app:common-gain-existence}

The values and costs of a prescribed rule are evaluated before imposing credibility; write $V^\phi:=\Pi^\phi/(1-\beta_S)$, so that $c^\phi(\nu)=\beta_S[V^\phi(T_0\nu)-V^\phi(T_1\nu)]$.

\begin{proof}[Proof of Proposition~\ref{prop:lowest}]
\emph{Value ordering.} Fix active histories $\nu'\preceq_T\nu$ and a common future state sequence. Match every inherited one in $\nu'$ to a distinct, weakly later inherited one in $\nu$. Each successive future favorable recommendation occurs weakly earlier under the prescribed rule from $\nu'$. Otherwise, take the first index $j$ at which the $j$th favorable recommendation from $\nu'$ occurs later. Immediately before the $j$th recommendation from $\nu$, both paths have issued $j-1$ future ones; matching these by index makes the first entering history temporally lower. The common state belongs to the prescribed upper tail, and lower closure makes $1$ admissible on the first path, so the rule recommends $1$ there as well, a contradiction. Each favorable recommendation yields $G>0$, so $V^\phi(\nu')\geq V^\phi(\nu)$; in particular, $V^\phi(0^{N-1})\geq V^\phi(\nu)$ and $V^\phi(T_0\nu)\geq V^\phi(\nu)$.

\emph{Upper and lower cost bounds.} Date the current message at $0$. For $\nu\in\mathcal H^\circ$, follow the prescribed rule from $T_0\nu$ and $T_1\nu$ along a common future state sequence, and let $s_j$ and $t_j$ be the dates of their $j$th future favorable recommendations, with a missing date set to infinity. Since $T_0\nu\preceq_T T_1\nu$, the preceding argument gives $s_j\leq t_j$. The dates also satisfy $t_j\leq s_{j+1}$: match the extra current one on the second path to the first future one on the first path, and inductively $t_i$ to $s_{i+1}$; if $t_j>s_{j+1}$ were the first failure, the second entering history would temporally precede the first immediately before $s_{j+1}$, and lower closure would permit $1$ on the second path, contradicting the delay. Let $K^0_t$ and $K^1_t$ count future favorable recommendations through date $t$ on the two paths. Then $0\leq K^0_t-K^1_t\leq1$, and summation by parts gives
\[
c^\phi(\nu)=G(1-\beta_S)\sum_{t=1}^{\infty}\beta_S^t\,\mathbb E[K^0_t-K^1_t]\in[0,\beta_SG].
\]
For $N=1$, both successor histories are empty and the cost is zero.

\emph{The minimum cost.} The nonempty lower-closed base set contains $0^{N-1}$. At every active history permitting $1$, the policy Bellman equation is $V^\phi(\nu)=\beta_S V^\phi(T_0\nu)+p[G-c^\phi(\nu)]$. Since $T_0(0^{N-1})=0^{N-1}$, subtracting the equations gives
\[
p[c^\phi(\nu)-c^\phi(0^{N-1})]
=(1-\beta_S)[V^\phi(0^{N-1})-V^\phi(\nu)]+\beta_S[V^\phi(T_0\nu)-V^\phi(\nu)]\geq0.
\]

\emph{Credibility.} Every prescribed favorable state has gain $G>c^\phi(\nu)$ and strictly prefers $1$. Every excluded state has gain at most $g_{k-1}$, so $g_{k-1}\leq c^\phi(0^{N-1})$ makes $0$ optimal at every active history permitting $1$, and Proposition~\ref{prop:local} rules out all profitable undetectable deviations. If the inequality fails, state $k-1$ profitably recommends $1$ at $0^{N-1}$.
\end{proof}

For quota $(N,1)$ with $N\ge2$, write $z=0^{N-1}$ and $A_N:=\sum_{t=1}^{N-1}\beta_S^t$. A rule recommending $1$ at $z$ exactly when a state with common gain $G$ occurs, with probability $p$, satisfies
\[
(1-\beta_S)V^\phi(z)=\frac{pG}{1+pA_N},\qquad
c^\phi_{N,1}(z)=\frac{pGA_N}{1+pA_N}.
\]
Indeed, its Bellman equation is $V^\phi(z)=p(G+\beta_S^NV^\phi(z))+(1-p)\beta_SV^\phi(z)$, and $1-(1-p)\beta_S-p\beta_S^N=(1-\beta_S)(1+pA_N)$; the cost is $\beta_S(1-\beta_S^{N-1})V^\phi(z)=(1-\beta_S)A_NV^\phi(z)$.

\begin{proof}[Proof of Corollary~\ref{cor:common-gain-existence}]
By the preceding formula, $c^\phi_{N,1}(0^{N-1})=pGA_N/(1+pA_N)$, which increases strictly to $\beta_SpG/(1-\beta_S+\beta_Sp)$ as $N\to\infty$. The limit exceeds $g_{k-1}$ exactly when $\beta_S>g_{k-1}/[g_{k-1}+p(G-g_{k-1})]$, so every sufficiently large finite $N$ gives $g_{k-1}<c^\phi_{N,1}(0^{N-1})<G$. Proposition~\ref{prop:lowest} gives strict sender incentives wherever $1$ is admissible, and \eqref{eq:tail-obedience} gives receiver obedience. At binding histories, compliance has a positive continuation value, whereas an immediate-reversion violation yields zero, so the profile is an equilibrium. The sender's normalized gain $pG/(1+pA_N)$ is positive; at the history permitting $1$, the receiver's expected current gain over babbling is $\sum_{i=k}^n\mu_i r_i$, and at binding histories she takes the prior action.
\end{proof}

\subsection{Failure without Temporal Lower Closure}\label{app:no-lower-closure}
Let $N=4$, $p=1/2$, and $\beta_S=0.9$, and exclude one word from the full quota $Q=3$: $\mathcal W=\admissible^{4,3}\setminus\{1011\}$. This set is coordinatewise downward closed and prior-action closed, but its base set is not temporally lower closed: $101\preceq_T011$, but $101$ is excluded while $011$ is admitted. Exact evaluation of the separating rule gives the value $155300/36163\approx4.294445$ and cost $5103/36163\approx0.141111$ at $000$; across the base set, the costs range from $729/8729\approx0.083515$, at $001$, to $140121/253141<1$. At $\ell=0.1$, the test at $000$ passes, but the low state profitably recommends $1$ at $001$. Coordinatewise downward closure and prior-action closure therefore do not suffice for the lowest-history reduction. At $\ell=0.07$, by contrast, every local incentive inequality holds, and the separating rule's payoff exceeds $3.844832$, the best credible full-quota payoff at this memory, attained by $Q=2$; the full quotas $Q=3$ and $Q=4$ fail credibility because their lowest-history costs are $729/13949\approx0.052262$ and zero. Excluding a history can thus improve the best credible payoff at a fixed memory.

\section{Strict Comparisons and Sender Patience}\label{sec:strict}

Section~\ref{app:strict-comparisons} proves Proposition~\ref{prop:supp-strict}. Lemma~\ref{lem:binary-patience} in Appendix~\ref{app:patience-comparisons} establishes the patience comparison. Greater patience raises the weight on the opportunities displaced by a recommendation. At a fixed restricted quota, the normalized initial payoff falls because the initial empty record is the most favorable history; this comparison concerns normalized values and allows the policy to adjust with patience.

\subsection{The Lower Patience Boundary}
Let $\beta_{N,Q}$ denote the entry threshold defined in Appendix~\ref{app:sawtooth-proof}. Under (UT) and $\underline c_R>0$, the best full-quota payoff approaches $\overline\Pi$ as $\beta_S\downarrow\beta^\ast$, although no full quota is implementable at $\beta^\ast$.
\begin{proof}
For the limit, write $c(1;\beta)$ for the lifetime cost in \eqref{eq:finite-lifetime-recursion}; it is continuous in $\beta$, and $c(1;\beta^\ast)=\underline c_R$ by \eqref{eq:patience-threshold-primitives}, so no quota is eligible at $\beta^\ast$ (Theorem~\ref{thm:binary-finite-quota}). For $\beta>\beta^\ast$, $c(1;\beta)>\underline c_R$, and Lemma~\ref{lem:binary-memory-limit} with $Q=1$ makes the quota-one cost at $0^{N-1}$ increase with $N$ to $c(1;\beta)$. Hence $\beta_{N,1}$ is nonincreasing in $N$ and converges to $\beta^\ast$, with $\beta_{N,1}>\beta^\ast$ by Lemma~\ref{lem:binary-memory-limit}. For $\beta\in[\beta_{N+1,1},\beta_{N,1})$, the pair $(N+1,1)$ is implementable, and its cost is at most $\widehat c^{\mathrm{IR}}_{N+1,1}(0^{N};\beta_{N,1})<c(1;\beta_{N,1})$, so $\Pi^*_{\mathrm{full}}(\beta)\ge F(c(1;\beta_{N,1}))\to F(\underline c_R)=\overline\Pi$ as $N\to\infty$. Together with the upper bound of Benchmark~\ref{thm:obedience-ceiling}, this proves the limit.

\end{proof}

\subsection{Strict Quota and Memory Comparisons}\label{app:strict-comparisons}
\begin{proof}[Quota comparison]
Fix arbitrary gains $g_1<\cdots<g_n$, $g_n>0$, a full-support prior, and $0<\beta_S<1$. Simulate an optimal $Q$-quota policy from the all-zero history under quota $Q+1$. During dates $0,\ldots,N-1$, insert a single recommendation $1$ at the first maximal-gain state where the virtual policy sends $0$, if such a date occurs. Otherwise retain all virtual messages forever. Every window has at most one extra recommendation, so the policy is feasible. On the event that all first $N$ states have maximal gain, a $Q<N$ policy must send at least one zero and the insertion occurs. Writing $p_{\max}=\mu_n>0$ gives
\[
\widehat V^{\mathrm{IR}}_{N,Q+1}(0^{N-1})-\widehat V^{\mathrm{IR}}_{N,Q}(0^{N-1})
\geq\beta_S^{N-1}g_np_{\max}^{N}>0.
\]
The initial value at $Q=0$ is zero. For positive quotas, the identity $(1-\beta_S)\widehat V(0^{N-1})=F(\widehat c(0^{N-1}))$ and strict decrease of $F$ imply strict cost decrease. Under (UT), the obedience characterization for the optimal policy then gives an initial segment of implementable quotas and optimality of its largest member.

\end{proof}

\begin{proof}[Memory comparison]
Every message path satisfying the $(N+1,Q)$ restriction satisfies $(N,Q)$, so its optimal value is weakly lower. To prove strictness, simulate the longer-window optimizer with its virtual record under the shorter window. With probability $p_{\max}^{Q}$, the first $Q$ states have maximal gain. Lemma~\ref{lem:general-lowest-cost} implies that all receive recommendation $1$. Under the longer window, capacity then remains exhausted through date $N$, whereas the shorter window releases the first unit at date $N$.

On this event, if the date-$N$ state also has maximal gain, insert a recommendation $1$ at that date. Send the virtual zeros up to the first subsequent virtual recommendation $1$, omit that recommendation, and thereafter copy the virtual policy. Before the omission, zero continuation is feasible. After the omission, the shorter-window actual record temporally precedes the virtual record because the inserted message is older than the omitted message. All subsequent messages are feasible. The gain at the insertion date is at least $(1-\beta_S)g_n$; if no later recommendation occurs, it is $g_n$. Thus
\[
\widehat V^{\mathrm{IR}}_{N,Q}(0^{N-1})-\widehat V^{\mathrm{IR}}_{N+1,Q}(0^N)
\geq p_{\max}^{Q+1}\beta_S^N(1-\beta_S)g_n>0.
\]
The all-zero identity strictly orders the costs. Under (UT), the obedience characterization for the optimal policy proves preservation of implementation. 
\end{proof}

\paragraph{Binary policy identities and comparisons.}
For the prescribed high-state rule, with the notation of Section~\ref{app:notation}, the all-zero policy equation gives
\begin{equation}
\begin{split}
c^{C,\mathrm{IR}}_{N,Q}(0^{N-1})
&=1-\frac{1-\beta_S}{p}V^{C,\mathrm{IR}}_{N,Q}(0^{N-1}),\\
d^{C,\mathrm{IR}}_{N,Q}(0^{N-1})
&=\ell-1+\frac{1-\beta_S}{p}V^{C,\mathrm{IR}}_{N,Q}(0^{N-1}).
\end{split}
\label{eq:binary-lowest-identity}
\end{equation}
The prescribed policy value is independent of $\ell$. Hence its maximal credible quota weakly decreases with $\ell$. Whenever credible, the rule attains the optimal value $\widehat V^{\mathrm{IR}}$.
For the prescribed binary policy, the same strict payoff comparison follows by the acceptance-count induction: a larger quota weakly advances every accepted high state, and $N$ initial high states yield an additional acceptance. The all-zero identity \eqref{eq:binary-lowest-identity} orders its costs even for noncredible candidates. At the unrestricted quota,
\begin{equation}
V^{C,\mathrm{IR}}_{N,N}(0^{N-1})=\frac{p}{1-\beta_S},\qquad c^{C,\mathrm{IR}}_{N,N}(0^{N-1})=0.
\label{eq:binary-unrestricted}
\end{equation}

The acceptance-time comparison also makes the prescribed value strictly decrease with memory, including for noncredible candidates. For its patience comparison, replace the first Bellman inequality in the general proof by the Bellman equality for the fixed policy and use the binary value ordering. The same strictness argument applies, with unrestricted normalized payoff $p$.

\section{Rolling Restrictions and Record Orders}\label{sec:order-check}
This section supports the discussion following Proposition~\ref{prop:quota-cost-order}.

\paragraph{Intersections of rolling quotas.} An intersection of rolling quotas with windows of at most $N$ periods caps, for each $k\leq N$, the number of favorable recommendations in every $k$ consecutive periods by some $q_k$, taken nondecreasing in $k$ without loss. On active histories, whose inherited windows already satisfy their caps, a favorable recommendation is admissible exactly when $s_{k-1}(\nu)\leq q_k-1$ for all $k$, where $s_j(\nu)$ counts the favorable recommendations among the last $j$ entries of $\nu$. Extended by the same formula to every entering history, this base set is temporally lower closed, and recommendation $0$ is always admissible. In the lifted representation used in the proof of Proposition~\ref{prop:quota-cost-order} (Appendix~\ref{app:quota-cost-order}), each cap is a difference constraint $x_v-x_u\leq q_k$ on cumulative counts, both on inherited records and when a recommendation is appended. The set of lifted records is therefore closed under coordinatewise minima and maxima. If $\nu'\preceq_T\nu$ are active and $1$ is admissible at $\nu$, the suffix-sum comparison makes $1$ admissible at $\nu'$, and the four successors are again active. The proof then applies verbatim, so continuation costs are ordered along $\preceq_T$ at active records under every such intersection.

\paragraph{Enumeration.} For memories $N=4,\ldots,7$, Table~\ref{tab:order-check} reports an exhaustive enumeration; for $N\leq3$, every such restriction is an intersection of rolling quotas. A restriction is given by a nonempty, temporally lower-closed base set $\mathcal B\subseteq\{0,1\}^{N-1}$, with recommendation $0$ always admissible; restrictions that permit the same recommendation paths from the empty record are identified. A restriction is classified as an intersection of rolling quotas if, on the records it reaches, its base set coincides with a set $\{\nu:s_j(\nu)\leq b_j\text{ for all }j\}$, with $s_j$ as above. For each restriction, the optimal perpetual-compliance policy is computed by value iteration for a fixed list of primitives: two or three states with random gains and priors, discount factors between $0.4$ and $0.99$, and the primitives of the examples in the main text. Every pair $\nu'\preceq_T\nu$ of active records with $1$ admissible at $\nu$ is checked for $\widehat c^{\mathrm{IR}}(\nu')>\widehat c^{\mathrm{IR}}(\nu)$. For restrictions without a reversal, a two-state grid over $p$, $\ell$, and $\beta_S\leq0.99$ is added.

\begin{table}[!htbp]
\centering
\caption{Cost reversals among temporally lower-closed restrictions.}
\label{tab:order-check}
\begin{tabular}{@{}ccccccc@{}}
\toprule
&&\multicolumn{2}{c}{Rolling intersections}&\multicolumn{3}{c}{Other restrictions}\\
\cmidrule(lr){3-4}\cmidrule(l){5-7}
$N$ & Total & Number & Reversed & Number & Reversed & Smallest\\
\midrule
4 & 8 & 7 & 0 & 1 & 1 & 0.4031\\
5 & 19 & 13 & 0 & 6 & 6 & 0.0532\\
6 & 60 & 22 & 0 & 38 & 38 & 0.0064\\
7 & 324 & 40 & 0 & 284 & 282 & 0.0017\\
\bottomrule
\end{tabular}
\end{table}

``Reversed'' counts restrictions with a cost reversal. Counts are specific to each memory: a restriction with memory $N$ reappears, in an equivalent form, among those with longer memories, so the rows should not be added. ``Smallest'' is the smallest, across those restrictions, of the largest cost difference found; each exceeds the numerical error of the value iteration by many orders of magnitude. The two exceptions at $N=7$ each differ from the smallest intersection of rolling quotas containing their reachable base set at a single record, $101101$ and $110011$, respectively, and neither produced a reversal in the random draws or on the grid. Among temporally lower-closed restrictions with $N\leq7$, cost ordering thus nearly characterizes intersections of rolling quotas; the enumeration does not settle whether the two exceptions order costs for all primitives. The script \path{replicate_order_check.py} reproduces the table and writes \path{order_check_results.json}.

\section{Exact Deviation Accounting and a Nesting Diagnostic}

\subsection{Exact deviation accounting}
\label{app:occupation-details}
\label{app:continuation-derivations}
\label{app:recurring-derivation}
Fix the receiver's response and a stationary reference policy $\phi$, with bounded value $V^\phi$. Add an absorbing state $\dagger$ with zero value and rewards after breakdown. Let $g(\theta,\nu,m)$ be the current sender payoff and $\widetilde T_m\nu$ the next history, equal to $\dagger$ after a violation. Define the reference policy's state-conditional value and the advantage of message $m$ by
\[
\begin{aligned}
V^\phi(\theta,\nu)&=g(\theta,\nu,\phi(\theta,\nu))+
\beta_S V^\phi(\widetilde T_{\phi(\theta,\nu)}\nu),\\
A^\phi(\theta,\nu;m)&=g(\theta,\nu,m)+\beta_S V^\phi(\widetilde T_m\nu)-V^\phi(\theta,\nu).
\end{aligned}
\]
For a mixed reference, the first expression is averaged over its prescribed messages. Independence of states gives $\mathbb E[V^\phi(\theta,\nu)\mid \nu]=V^\phi(\nu)$.

\begin{lemma}[Exact occupation identity]
\label{lem:binary-occupation}
For any causal alternative $\widehat\phi$ and initial history $x$,
\begin{equation}
V^{\widehat\phi}(x)-V^\phi(x)
=\mathbb E_x^{\widehat\phi}\left[\sum_{t\geq0}\beta_S^t
A^\phi(\theta_t,\nu_t;m_t)\right].
\label{eq:binary-main-occupation}
\end{equation}
The summands are zero after breakdown.
\end{lemma}
\begin{proof}
Conditional on information before the date-$t$ state draw, the expected subtracted term is $V^\phi(\nu_t)$. Summing to any $K\geq1$ and telescoping therefore gives
\[
\mathbb E_x^{\widehat\phi}\left[\sum_{t=0}^{K-1}\beta_S^t A^\phi(\theta_t,\nu_t;m_t)\right]
=\mathbb E_x^{\widehat\phi}\left[\sum_{t=0}^{K-1}\beta_S^t g(\theta_t,\nu_t,m_t)
+\beta_S^K V^\phi(\nu_K)\right]-V^\phi(x).
\]
Bounded rewards and $\beta_S<1$ make the terminal term vanish. The argument allows randomization and dependence on the sender's private history.
\end{proof}

\paragraph{The local credibility test.}
The sender's compliance problem is a finite discounted Markov decision problem \citep{Blackwell1965}. The occupation identity of Lemma~\ref{lem:binary-occupation} gives the comparison with alternative strategies explicitly.
For the setting of Proposition~\ref{prop:local}, restrict attention to undetectable recommendations and multiply the occupation identity by $1-\beta_S$. For a pure stationary alternative $\phi'$ selecting an element of $A^\phi(\nu)$ at every induced history and state, define
\[
\Lambda_{\phi'}(\nu\mid x)=\sum_{s=0}^{\infty}\beta_S^s
\Pr_{\phi'}(\nu_s=\nu\mid\nu_0=x).
\]
Then, for every $x\in\mathcal V^{N,\phi}$,
\begin{equation}
\begin{aligned}
\Pi^{\phi'}(x)-\Pi^\phi(x)
={}&(1-\beta_S)\sum_{\nu\in\mathcal V^\phi_\circ}
\Lambda_{\phi'}(\nu\mid x)\\
&\quad\times\sum_{i=1}^n\mu_i
[\phi'(\theta_i,\nu)-\phi(\theta_i,\nu)]
[g_i-c^\phi(\nu)].
\end{aligned}
\label{eq:local-performance-difference}
\end{equation}
Indeed, the normalized one-step advantage is
\[
(1-\beta_S)[m-\phi(\theta_i,\nu)][g_i-c^\phi(\nu)]
\]
when both recommendations are undetectable, and zero when only one is available. Substitution in Lemma~\ref{lem:binary-occupation} and grouping expected discounted visits by rolling history gives \eqref{eq:local-performance-difference}. The visitation weights are computed under the deviation and include the histories reached because recommendations change.

Switching only tied recommendations makes every changed term in \eqref{eq:local-performance-difference} zero. It therefore preserves every sender continuation value under the fixed obedient receiver response. Receiver obedience is a separate condition on the resulting recommendation cells.

\paragraph{Finite-state recurring deviations.}
Fix a pure reference policy $\phi$ sending $0$ at $(\theta_i,0^{N-1})$, where $1$ is admissible. Let $\phi^{D_i}$ change only those visits to message $1$, and let $c^\phi(0^{N-1})=\beta_S[V^\phi(T_0(0^{N-1}))-V^\phi(T_1(0^{N-1}))]$. Its local advantage is $g_i-c^\phi(0^{N-1})$. Write $\Lambda_i(0^{N-1}\mid x)$ for the expected discounted number of visits to the state-history pair $(\theta_i,0^{N-1})$ under the alternative, starting before the state draw at $x$.
\begin{corollary}[Exact evaluation of recurring deviations]\label{prop:binary-recurring}
The alternative's gain conditional on the initial state-history pair $(\theta_i,0^{N-1})$ is
\begin{equation}
\Delta_i=(1+K_i)[g_i-c^\phi(0^{N-1})],\qquad
K_i=\beta_S\Lambda_i(0^{N-1}\mid T_1(0^{N-1})),\qquad
0\leq K_i\leq\frac{\beta_S^N}{1-\beta_S^N}.
\label{eq:finite-recurring}
\end{equation}
It has exactly the sign of the one-period advantage.
\end{corollary}
\begin{proof}
The occupation identity gives $V^{D_i}(x)-V^\phi(x)=\Lambda_i(0^{N-1}\mid x)[g_i-c^\phi(0^{N-1})]$. Conditional on the initial deviation, add its current advantage to the discounted continuation difference at $T_1(0^{N-1})$. Every tested visit sends $1$, preventing a return to the all-zero entering history for at least $N$ periods. Hence $K_i\leq\sum_{j\geq1}\beta_S^{jN}$. The proof applies to any finite state space and does not require an optimal reference policy. For a fixed policy, the occupation multiplier depends on its transition law; it need not stay fixed when the policy is reoptimized after changing payoffs.
\end{proof}

For the binary high-state reference, take $\theta_i=L$ and $g_i=\ell$ in \eqref{eq:finite-recurring}. For a completed reference, count returns only before breakdown.

\subsection{A necessary nesting diagnostic}
\label{subsec:nested-cells}

For a proposed pure admissible policy $\phi$, use $C_1^\phi(\nu)$ for its favorable recommendation cell, now indexed by entering history. Cost ordering imposes a necessary nesting condition on these sets. Write $C_1^*(\nu)$ for the corresponding set under the selected optimal policy $\phi^*$. At histories permitting both messages, sender optimality implies
\begin{equation}
\{\theta_i:g_i>\widehat c^{\mathrm{IR}}(\nu)\}
\subseteq C_1^*(\nu)
\subseteq\{\theta_i:g_i\ge\widehat c^{\mathrm{IR}}(\nu)\}.
\label{eq:optimal-cell-bounds}
\end{equation}
At a history prohibiting $1$, set $C_1^*(\nu)=\varnothing$.

\begin{corollary}[Necessary nesting condition]
\label{cor:nested-favorable-cells}
Under a rolling quota $(N,Q)$, for active histories $\nu',\nu$,
\[
\nu'\preceq_T \nu
\quad\Longrightarrow\quad C_1^*(\nu)\subseteq C_1^*(\nu').
\]
Under a general prior-action-closed restriction with nonempty temporally lower-closed base set,
\[
C_1^*(\nu)\subseteq C_1^*(0^{N-1})
\]
at every active history.
\end{corollary}

The cost comparisons in Proposition~\ref{prop:quota-cost-order} and Lemma~\ref{lem:general-lowest-cost} yield the nested cells. The rolling-quota bound, or temporal lower closure in the second case, preserves availability of $1$ at the lower history. 

To apply the test to a proposed policy $\phi$ whose induced set is the rolling quota under consideration, compare $C_1^\phi(\nu)$ and $C_1^\phi(\nu')$ at every active pair $\nu'\preceq_T \nu$. If a state receives $1$ at $\nu$ and $0$ at $\nu'$, the policy violates a necessary condition for the selected optimum. Under a general temporally lower-closed restriction, the necessary comparison is $C_1^\phi(\nu)\subseteq C_1^\phi(0^{N-1})$. Passing these nesting comparisons leaves the local incentive inequalities in Proposition~\ref{prop:local} to be checked.

\paragraph{Nesting is not sufficient.} Take quota $(3,2)$, $\mu=(1/4,1/2,1/4)$, $g=(0.4,0.9,1)$, $r=(-3,1/2,1)$, and $\beta_S=0.9$. The rule recommends $1$ in $\{\theta_2,\theta_3\}$ at $00$ and only in $\theta_3$ at $01$ and $10$, with $0$ forced at $11$. It induces the full quota, is receiver-obedient, and passes every nesting comparison. Its costs include
\[
c^\phi(00)=\frac{72171}{177740}\approx0.406,
\qquad c^\phi(01)=\frac{5931}{17774}\approx0.334.
\]
The local inequalities hold at $00$, but $c^\phi(01)<g_2$, so state $\theta_2$ profitably recommends $1$ at $01$.

Ordering states by sender gain expresses each favorable cell as an upper tail. Under rolling quotas, the threshold weakly rises along temporal order on active histories. For a general temporally lower-closed restriction, every threshold is at least as selective as the threshold at $0^{N-1}$.

\paragraph{Nested recommendation cells.} If the two costs differ, every state assigned $1$ at the higher cost strictly prefers $1$ at the lower cost. If the costs coincide, the common receiver-favoring convention assigns every tied state identically. For rolling quotas, $|\nu'|\leq|\nu|$ preserves availability of $1$ between comparable active histories. For a general temporally lower-closed base set, lower closure makes $1$ available at $0^{N-1}$. These observations prove Corollary~\ref{cor:nested-favorable-cells}.

\section{Anticipated Breakdown: Completed Policies}

\subsection{The Completed-Policy Bellman Equation}

\label{app:breakdown-proof}
Under belief updating, the sender may plan to violate. Fix a prior-action-closed restriction with a nonempty, temporally lower-closed base set, and let the receiver follow admissible recommendations, follow a violating recommendation $1$ in its period, and take action $0$ thereafter. Let $\widehat W(\nu)$ be the sender's optimal unnormalized value against these responses, and at slack histories let $\widehat c^W(\nu)=\beta_S[\widehat W(T_0\nu)-\widehat W(T_1\nu)]$. A followed violation yields the current gain and zero continuation. The selected policy sends $1$ above the relevant cutoff, and at equality exactly when $r_i>0$; receiver obedience is then checked under this actual policy.

Let $\mathcal H^\circ$ denote the active entering histories permitting $1$, let $F(c)=\sum_i\mu_i(g_i-c)_+$, and define $e(\nu)=\beta_S\widehat W(T_0\nu)$ at binding histories. The optimal unnormalized value allowing terminal violations satisfies
\begin{equation}
\widehat W(\nu)=\beta_S\widehat W(T_0\nu)+
\begin{cases}
F(\widehat c^W(\nu)),&\nu\in\mathcal H^\circ,\\
F(e(\nu)),&\nu\notin\mathcal H^\circ.
\end{cases}
\label{eq:finite-completed-bellman}
\end{equation}
The sender recommends $1$ when the current gain exceeds the relevant cutoff, and also at equality when $r_i>0$; otherwise the sender selects $0$. Under the conditions of Lemma~\ref{lem:binary-breakdown-cost}, the cost bounds are
\begin{equation}
\begin{aligned}
0\leq\widehat c^W(0^{N-1})&\leq\widehat c^W(\nu)&& (\nu\in\mathcal H^\circ),\\
\widehat c^W(0^{N-1})&\leq e(\nu)&& (\nu\notin\mathcal H^\circ).
\end{aligned}
\label{eq:binary-breakdown-cost-bound}
\end{equation}

\begin{proof}[Proof of Lemma~\ref{lem:binary-breakdown-cost}]
The completed Bellman operator is a contraction: at a binding history the violating choice earns $g_i$ and moves to a zero-value absorbing state. Its unique fixed point is the optimal value $\widehat W$ over causal policies. The sender can always send zeros, so values are nonnegative.

First, completed values preserve temporal ordering. Simulate a policy starting from a higher history at a lower history with the same states and a virtual record. Before a virtual violation, all messages are feasible on the lower path. At a virtual violation, the lower path can receive the same action-$1$ payoff. If the message is also a violation there, both paths terminate. If it is admissible, send zeros forever afterward. The lower path can therefore reproduce the higher path's payoffs, implying
\[
\widehat W(\widetilde \nu)\geq\widehat W(\nu)\quad(\widetilde \nu\preceq_T \nu).
\]
Thus $\widehat W(0^{N-1})\geq\widehat W(\nu)$, $\widehat W(T_0\nu)\geq\widehat W(\nu)$, and slack-history costs are nonnegative. The subtraction used in Lemma~\ref{lem:general-lowest-cost} now gives
\[
F(\widehat c^W(0^{N-1}))-F(\widehat c^W(\nu))
=(1-\beta_S)[\widehat W(0^{N-1})-\widehat W(\nu)]
+\beta_S[\widehat W(T_0\nu)-\widehat W(\nu)]\geq0.
\]
Since the base set is nonempty, $0^{N-1}$ permits recommendation $1$ and has positive value. Hence $\widehat c^W(0^{N-1})<g_n$, and strict decrease of $F$ below $g_n$ proves the slack-history comparison.

The optimal single-unit lifetime policy has normalized benchmark payoff $F(c(1))$ and satisfies $F(c(1))=(1-\beta_S)c(1)/\beta_S$ by \eqref{eq:finite-lifetime-recursion}. From every history, wait until this stopping policy recommends $1$, take its payoff admissibly or by violation, and send zeros thereafter if still active. Therefore $(1-\beta_S)\widehat W(\nu)\geq F(c(1))$ for every history. At $0^{N-1}$,
\[
F(\widehat c^W(0^{N-1}))=(1-\beta_S)\widehat W(0^{N-1})
\geq F(c(1)).
\]
Since $c(1)<g_n$, strict decrease of $F$ gives $\widehat c^W(0^{N-1})\leq c(1)$. At every binding history,
\[
e(\nu)=\beta_S\widehat W(T_0\nu)
\geq\frac{\beta_S}{1-\beta_S}F(c(1))
=c(1)\geq\widehat c^W(0^{N-1}).
\]
Finally, $\widehat W\geq\widehat V^{\mathrm{IR}}$, because perpetual compliance is available. Applying the two all-zero Bellman identities gives $\widehat c^W(0^{N-1})\leq\widehat c^{\mathrm{IR}}(0^{N-1})$.

If the all-zero cost is below $\underline c_R$, its nonempty favorable tail is not obedient. If it is at least $\underline c_R$, every nonempty favorable cell, including a violating cell, is a more selective upper tail and is obedient by (UT). Its complement is obedient to $0$ by the strict prior preference. Optimality of $\widehat W$ proves all sender incentive constraints, including optimal stopping. This proves the obedience characterization for the selected optimal policy.

For sustainability, the perpetual-compliance optimizer already maximizes over admissible messages. At a binding history, every violation is unprofitable exactly when $g_n\leq\beta_S\widehat V^{\mathrm{IR}}(T_0\nu)$. If these inequalities hold everywhere, $\widehat V^{\mathrm{IR}}$ also solves the completed Bellman equation; uniqueness gives $\widehat W=\widehat V^{\mathrm{IR}}$. If any inequality fails, the corresponding maximal-gain violation is profitable. Under Assumption~\ref{ass:tie-breaking}, equality also selects violation whenever the tied maximal-gain state has positive receiver gain. Thus equality of completed and perpetual-compliance values does not itself ensure selected perpetual compliance.
\end{proof}

\subsection{Binary Sustainability Calculations}
\paragraph{Binary specialization.}
For the prescribed two-state rule, with $V^C$ its completed value (Section~\ref{app:binary-completed-values}), define
\begin{equation}
s^C_{N,Q}:=\min_{|\nu|=Q}
\{\beta_S V^C_{N,Q}(T_0\nu)-1\}.
\label{eq:binary-sustainability-margin}
\end{equation}
Conditional on BU implementation, selected sustainability requires $s^C_{N,Q}>0$, since $r_H>0$. The prescribed perpetual-compliance rule satisfies
\[
\min_{|\nu|=Q}\beta_S V^{C,\mathrm{IR}}_{N,Q}(T_0\nu)
=V^{C,\mathrm{IR}}_{N,Q}(\nu^b_{N,Q}).
\]
To verify the prescribed-policy comparison even when it is not sender-optimal, couple future state sequences and index accepted high states chronologically. Each acceptance is the first high state after the previous acceptance and at least $N$ periods after the recommendation $Q$ places earlier, including inherited recommendations. Induction orders all acceptance dates when inherited messages are moved earlier or memory is shortened. The same fixed policy has a value continuous and strictly increasing from zero to infinity as patience rises. Hence the binary threshold used in the figures is the unique solution of
\begin{equation}
V^{C,\mathrm{IR}}_{N,Q}
\bigl(\nu^b_{N,Q};\beta^{\mathrm{sus,BU}}_{N,Q}(p)\bigr)=1.
\label{eq:binary-sustainability-root}
\end{equation}
At an implementable policy this prescribed value equals the optimum. The optimal-policy threshold $\beta^{\mathrm{sus}}_{N,Q}$ of Proposition~\ref{prop:binary-sustainability} and the prescribed-policy threshold need only be identified where the separating policy is optimal.

The forced initial wait also gives the necessary condition for selected binary sustainability
\begin{equation}
1<V^{C,\mathrm{IR}}_{N,Q}(\nu^b_{N,Q})
\leq\frac{p\beta_S^{N-Q}}{1-\beta_S},
\qquad p\beta_S^{N-Q}>1-\beta_S.
\label{eq:binary-sustainability-necessary}
\end{equation}

\subsection{Prescribed binary completions}\label{app:binary-completed-values}
Let $\phi^{\mathrm{BU}}$ retain the high-state recommendation at slack histories and choose optimally between continuation and violation at binding histories, using Assumption~\ref{ass:tie-breaking}. Write $V^C$ for its unnormalized value and $g^D(H,\nu)=1$ and $g^D(L,\nu)=\ell$ for the current payoff from a violating recommendation. Since babbling yields zero under the normalization in Section~\ref{sec:binary-environment}, the value solves
\begin{equation}
V^C(\nu)=
\begin{cases}
p[1+\beta_S V^C(T_1\nu)]+(1-p)\beta_S V^C(T_0\nu),&|\nu|<Q,\\[2pt]
\displaystyle\sum_{\theta\in\{L,H\}}\mu(\theta)
\max\{\beta_S V^C(T_0\nu),g^D(\theta,\nu)\},&|\nu|=Q.
\end{cases}
\label{eq:binary-breakdown-bellman}
\end{equation}
The operator is a contraction. At slack histories, define
\begin{equation}
c^C_{N,Q}(\nu)=\beta_S[V^C_{N,Q}(T_0\nu)-V^C_{N,Q}(T_1\nu)],
\qquad d^C_{N,Q}(\nu)=\ell-c^C_{N,Q}(\nu).
\label{eq:binary-main-cost}
\end{equation}
The prescribed slack choices are optimal exactly when $\ell\leq c^C(\nu)\leq1$. Together with optimal binding-history choices, these inequalities make $V^C$ the unrestricted optimum for the specified response, namely $\widehat W$.

Under the binary implementation conditions proved in Section~\ref{app:breakdown-proof}, low-state violations are strictly unprofitable, so an on-path violation reveals $H$. The high-state sender violates when $\beta_S V^C(T_0\nu)\leq1$. Thus selected perpetual compliance requires $\beta_S V^C(T_0\nu)>1$ at every binding history. Appendix~\ref{app:sustainability-proof} reduces this test to the most restrictive binding history and establishes reachability under compliance.

For a full quota, the prescribed completed high-state rule has $V^C\geq V^{C,\mathrm{IR}}$ and
\[
c^C(0^{N-1})=1-\frac{1-\beta_S}{p}V^C(0^{N-1})\leq c^{C,\mathrm{IR}}(0^{N-1}).
\]
If $\ell\leq c^C(0^{N-1})$, Lemma~\ref{lem:binary-memory-limit} implies $\ell<\rho^Q\leq\rho$ for $Q<N$, where $\rho=p\beta_S/[1-(1-p)\beta_S]$. From every history, waiting for the next high state and then recommending $1$, admissibly or by violation, gives $\beta_SV^C(\nu)\geq\rho$. Thus low-state breakdown is never optimal. Acceptance times of the prescribed high-state rule are weakly earlier from a temporally lower history. Coupling up to the higher path's optimal high-state stopping date, the lower path can also collect the terminal high-state payoff; if its recommendation is admissible, any subsequent continuation is nonnegative. This proves the same value ordering for $V^C$. Subtracting the slack-history Bellman equations gives $c^C(\nu)\geq c^C(0^{N-1})$.

For the upper bound, couple the high-state paths from $T_0\nu$ and $T_1\nu$. Their acceptance times interlace as in Section~\ref{app:binary-prescribed-policy}, so their cumulative acceptance counts differ by at most one. At an optimal stopping date for the first path, the second path can obtain the same terminal high-state payoff. Summation by parts bounds the discounted difference by $\beta_S$, giving $c^C(\nu)\leq\beta_S<1$. The prescribed completed rule is therefore sender-optimal whenever $\ell\leq c^C(0^{N-1})$; the converse follows from the low-state deviation at $0^{N-1}$. For $Q=N$, the cost is zero and the same conclusion holds with $\ell=0$. This establishes the binary test used in the numerical comparisons.

\subsection{Quota and Memory Choice with Exits}\label{app:completed-quota-choice}
\begin{corollary}[Quota choice with anticipated breakdown]
\label{cor:finite-completed-quota}
At a fixed memory $N$, $\widehat W_{N,Q}(0^{N-1})$ is nondecreasing and $\widehat c^W_{N,Q}(0^{N-1})$ is nonincreasing in positive quotas $Q$. Define
\[
Q_W^\star(N)=\max\left(\{0\}\cup
\left\{1\leq Q\leq N:\widehat c^W_{N,Q}(0^{N-1})\geq\underline c_R\right\}\right).
\]
Under (UT), the implementable positive quotas form the initial segment ending at $Q_W^\star(N)$, and its largest member maximizes the completed value. Moreover, $Q_W^\star(N)\leq Q^\star_{\mathrm{IR}}(N)$. At a fixed quota, the completed value weakly decreases with memory, so the shortest implementing memory maximizes value whenever it exists. Every implementable positive quota satisfies
\begin{equation}
F(c(Q+1))\leq F(\underline c_R).
\label{eq:finite-completed-quota-bound}
\end{equation}
For $\underline c_R>0$, only finitely many quotas satisfy this necessary condition.
\end{corollary}

\Needspace{5\baselineskip}
\begin{proof}
Under a larger quota, simulate the smaller-quota optimizer on the same future states until its terminal recommendation. All preceding recommendations remain admissible. Obtain the same action-$1$ payoff at the terminal recommendation; if it has become admissible, send zero forever afterward. This feasible policy reproduces the smaller-quota payoffs, including on paths with no exit. Thus the optimal completed value weakly increases with $Q$. The identity
\[
(1-\beta_S)\widehat W_{N,Q}(0^{N-1})
=F\bigl(\widehat c^W_{N,Q}(0^{N-1})\bigr)
\]
orders the costs. Under (UT), Lemma~\ref{lem:binary-breakdown-cost} gives the obedience test, the initial segment of implementable quotas, and the comparison with immediate reversion.

The same simulation reproduces a longer-window completed policy under a shorter window: admissible recommendations remain feasible until its terminal violation, whose payoff is followed by zero continuation if it becomes admissible. Hence the shorter-window optimum is weakly larger and its lowest-history cost weakly smaller. Among memories supporting implementation, the shortest therefore maximizes value.

Finally, simulate an optimal lifetime-$(Q+1)$ policy from the all-zero history. Its first $Q$ favorable recommendations are admissible; its last favorable recommendation receives action $1$, whether admissible or a violation, and is followed by zeros. This yields $\widehat W_{N,Q}(0^{N-1})\geq F(c(Q+1))/(1-\beta_S)$. The completed-policy ceiling in Corollary~\ref{cor:binary-payoff-ceiling} proves \eqref{eq:finite-completed-quota-bound}. Since $F(c(q))\uparrow F(0)$ and $F(\underline c_R)<F(0)$ when $\underline c_R>0$, the necessary quota range is finite. These simulation arguments permit the optimal state partition to vary with the restriction.
\end{proof}

\section{Payoff Benchmarks: Examples and Extensions}

\subsection{Exact Attainment and Benchmark Comparisons}\label{app:benchmark-details}
\paragraph{Implications for the accelerator.}
In the accelerator example, $p=1/2$, $g_H=1$, and $\underline c_R=\ell=1/4$, so $\Pi^{\mathrm{cp}}=1/2$ and, by \eqref{eq:ceiling-decomposition}, $\overline\Pi=1/2-(1/2)(1/4)=3/8$. By Benchmark~\ref{thm:obedience-ceiling}, no equilibrium gives the accelerator a normalized payoff above three-quarters of the credible-partition payoff, whatever its patience: the restrictions that deter endorsements of bad ventures also leave some good ventures unfunded. Under immediate reversion, Corollaries~\ref{cor:patience-threshold} and~\ref{cor:existence} give the patience threshold $\beta^\ast=2/5$ and finite rolling-quota implementation above it: for $\beta_S\in(2/5,2/3)$, the one-in-two rule fails credibility, but quota one with a sufficiently long window supports informative endorsements.

\paragraph{Exact attainment and the cost of self-restraint.}
Remark~\ref{rem:sawtooth} gives the general entry-threshold and patience comparison; the examples below illustrate it.
By \eqref{eq:finite-lowest-identity} and strict decrease of $F$, an implementable full quota attains $\overline\Pi$ at the all-zero history exactly when $\widehat c^{\mathrm{IR}}_{N,Q}(0^{N-1})=\underline c_R$. In the two-state model with $p=1/2$ and $\ell=0.3$, for instance, the quota $(2,1)$ gives normalized payoff $p/(1+p\beta_S)$ and all-zero cost $p\beta_S/(1+p\beta_S)$, which equals $\ell$ at $\beta_S=6/7$; there the payoff equals $p(1-\ell)=\overline\Pi=0.35$, although four quotas are eligible, and the rule is sustainable under belief updating because its binding-history value is $2.1>1$. Because Benchmark~\ref{thm:obedience-ceiling} bounds equilibrium payoffs even when messages, mixing, public randomization, and history dependence are unrestricted, Theorem~\ref{thm:obedience-ceiling-convergence} makes full quotas asymptotically optimal: these instruments cannot improve on them by an amount bounded away from zero as $\beta_S\uparrow1$.

At every discount factor, the Bellman equation at the all-zero history splits the payoff of an optimal full quota into the expected stage gain from the states it recommends $1$ at the all-zero history, $C_1^*(0^{N-1})$ in \eqref{eq:selected-favorable-cell}, and an on-path continuation loss of $\widehat c^{\mathrm{IR}}(0^{N-1})$ per favorable recommendation. As $\beta_S\uparrow1$, \eqref{eq:finite-lowest-identity} and Theorem~\ref{thm:obedience-ceiling-convergence} give $\widehat c^{\mathrm{IR}}(0^{N-1})\to\underline c_R$, and for sufficiently high $\beta_S$ the cell is $\{\theta_{k_R},\ldots,\theta_n\}$, with Assumption~\ref{ass:tie-breaking} excluding state $k_R-1$ at the boundary cost. The decomposition therefore approaches \eqref{eq:ceiling-decomposition}: the optimal quota induces the credible partition and pays the minimum cost of self-restraint. Unlike the on-path punishments in \citet{best2024persuasion}, which appropriately designed incomplete records make arbitrarily small, this cost stays bounded away from zero: without accuracy feedback, it must deter a favorable recommendation in the highest excluded state.

In the project application of \citet{kuvalekar2022goodwill}, the sender's payoff set is $\{0\}$ for $\delta<\bar\delta$ and $[0,\bar v]$ for $\delta\ge\bar\delta$, where $\bar v=\mathbb E[(\theta-\theta^*)_+]$, $\bar\delta=\theta^*/(\theta^*+\bar v)$, and $\mathbb E[\theta\mid\theta\ge\theta^*]=\kappa$, the project cost in their application; here $\underline c_R$, $\overline\Pi$, and $\beta^*$ play the roles of $\theta^*$, $\bar v$, and $\bar\delta$. Rolling quotas approach $\overline\Pi$ with a shortfall of at most $(1-\beta_S)\underline c_R/\beta_S$, and particular full quotas attain it exactly at some discount factors below one, such as $(2,1)$ at $\beta_S=6/7$ above.

\paragraph{Implications for quota design.}
In the accelerator example, suitable quotas bring the normalized payoff arbitrarily close to $\overline\Pi=3/8$ as the accelerator becomes patient, under either response to violations. The one-in-two rule does not. Its all-zero continuation cost $\beta_S/(2+\beta_S)$ equals $\underline c_R=1/4$ at $\beta_S=2/3$, where the rule attains the ceiling exactly under immediate reversion, but at higher patience the excess cost is incentive slack that the fixed rule cannot use. Under perpetual compliance, which is sustainable under belief updating for $\beta_S>\sqrt3-1$, its normalized payoff $1/(2+\beta_S)$ converges to $1/3$, two-thirds of the credible-partition payoff.

\subsection{The initial-history equality condition}
\begin{corollary}[The initial-history ceiling with anticipated breakdown]
\label{cor:binary-payoff-ceiling}
Let $W$ be the optimal completed value under belief updating for a prior-action-closed restriction permitting both recommendations at $0^{N-1}$. If the completed policy is receiver-obedient, then
\begin{equation}
W(0^{N-1})\leq\overline V:=\frac{\overline\Pi}{1-\beta_S}
=\frac{F(\underline c_R)}{1-\beta_S}.
\label{eq:finite-payoff-ceiling}
\end{equation}
Equality holds exactly when the completed cost at the empty record, $c^W(0^{N-1})$, equals $\underline c_R$. For a credible, receiver-obedient policy under perpetual compliance, the bound of Benchmark~\ref{thm:obedience-ceiling} likewise holds with equality at the empty record exactly when the cost there, $c(0^{N-1})$, equals $\underline c_R$.
\end{corollary}
\begin{proof}
Both recommendations are admissible at $0^{N-1}$, and recommending $0$ leaves the entering history unchanged. The completed-policy Bellman equation therefore gives
\[
(1-\beta_S)W(0^{N-1})=F(c^W(0^{N-1})).
\]
Receiver obedience requires $c^W(0^{N-1})\geq\underline c_R$. Since $c^W(0^{N-1})<g_n$ and $F$ is strictly decreasing below $g_n$, the identity gives the ceiling and its equality condition. Under perpetual compliance, the identity $(1-\beta_S)V(0^{N-1})=F(c(0^{N-1}))$ gives the stated equality condition for Benchmark~\ref{thm:obedience-ceiling}.
\end{proof}

\subsection{Convergence with Sustainable Quotas under Belief Updating}
\label{app:bu-convergence}
This section proves Corollary~\ref{cor:bu-ceiling-convergence}.

\begin{lemma}[The marginal value of quota capacity]
\label{lem:quota-deletion-bound}
For $0\leq Q<N$,
\begin{equation}
0\leq(1-\beta_S)
\bigl[\widehat V^{\mathrm{IR}}_{N,Q+1}(0^{N-1})
-\widehat V^{\mathrm{IR}}_{N,Q}(0^{N-1})\bigr]
\leq g_n\frac{(1-\beta_S)\beta_S^Q}{1-\beta_S^N}.
\label{eq:quota-deletion-bound}
\end{equation}
\end{lemma}
\begin{proof}
Simulate an optimal quota-$(Q+1)$ policy on the realized state sequence, retain a virtual record of its recommendations, and delete every $(Q+1)$-st favorable recommendation from the actual record, replacing it with~$0$. Any $N$ consecutive periods contain at most $Q+1$ virtual favorable recommendations; if they contain $Q+1$, these are consecutive entries in the virtual sequence of favorable recommendations, one of which is deleted. The actual record therefore satisfies quota $Q$, and the simulation is an admissible sender strategy in the smaller-quota problem. Numbering dates from zero, the first deletion occurs no earlier than date $Q$, and consecutive deletions are at least $N$ periods apart: otherwise an $N$-period window would contain their endpoints and the $Q$ virtual favorable recommendations between them, violating quota $Q+1$. Each deletion reduces the stage payoff by at most $g_n$, so the total discounted loss is at most $g_n\beta_S^Q/(1-\beta_S^N)$. Nesting of the feasible sets gives the lower bound.
\end{proof}

\begin{proof}[Proof of Corollary~\ref{cor:bu-ceiling-convergence}]
Fix $\varepsilon>0$ and choose an integer $N>1+\underline c_R/\overline\Pi$ with $g_n/N<\varepsilon$. Write $z=0^{N-1}$, $A_N:=\sum_{t=1}^{N-1}\beta_S^t$, and $V:=\widehat V^{\mathrm{IR}}_{N,1}$. Under quota one, an endorsement forces $N-1$ zeros, so $V(T_1z)=\beta_S^{N-1}V(z)$. Thus $c=(1-\beta_S)A_NV(z)$, and the empty-record Bellman equation gives $c=A_NF(c)$. Since $A_N\to N-1$ as $\beta_S\uparrow1$ and $(N-1)\overline\Pi>\underline c_R$, it follows that $A_NF(\underline c_R)>\underline c_R$ for all sufficiently high $\beta_S$, so the root exceeds $\underline c_R$. Hence $Q:=Q^\star_{\mathrm{IR}}(N)\geq1$, and $Q<N$ because the unrestricted quota has zero continuation cost. Quota $Q+1$ fails the all-zero obedience test, so \eqref{eq:finite-lowest-identity} gives $(1-\beta_S)\widehat V^{\mathrm{IR}}_{N,Q+1}(0^{N-1})>\overline\Pi$, and Lemma~\ref{lem:quota-deletion-bound} gives
\[
0\leq\overline\Pi-(1-\beta_S)\widehat V^{\mathrm{IR}}_{N,Q}(0^{N-1})
<g_n\frac{(1-\beta_S)\beta_S^Q}{1-\beta_S^N}
\leq g_n\frac{1-\beta_S}{1-\beta_S^N}\longrightarrow\frac{g_n}{N}<\varepsilon.
\]
For sustainability, from any entering history the sender can send $0$ for $N-1$ periods and then recommend $1$ only in state $\theta_n$ at one scheduled date every $N$ periods. This uses at most one unit per window and is feasible for every $Q\ge1$, so at every binding history $\beta_S\widehat V^{\mathrm{IR}}_{N,Q}(T_0\nu)\geq\mu_ng_n\beta_S^N/(1-\beta_S^N)$, which diverges as $\beta_S\uparrow1$ and therefore exceeds $g_n$ for all sufficiently high $\beta_S$. A common threshold for implementation, the payoff bound, and sustainability proves the corollary.
\end{proof}

\section{Selective History Restrictions}

\label{sec:quota-extension}\label{sec:binary-subsets}\label{subsec:extension-numerics}

Full quotas expand in discrete layers. Admitting selected histories from the next layer can use incentive slack without permitting every path allowed by the next full quota. The general value comparison applies to any finite state space.

\begin{proposition}[Value comparison for nested restrictions]
\label{prop:binary-exclusion-value}
In the two-action, $n$-state model, let $\mathcal W\subseteq\mathcal W'$ have the same memory, all-zero initialization, and feasible zero continuation. Holding receiver responses fixed at obedience,
\[
\widehat V^{\mathrm{IR}}_{\mathcal W}(0^{N-1})
\leq\widehat V^{\mathrm{IR}}_{\mathcal W'}(0^{N-1}).
\]
If both restrictions have nonempty, temporally lower-closed base sets, then
\[
\widehat c^{\mathrm{IR}}_{\mathcal W}(0^{N-1})
\geq\widehat c^{\mathrm{IR}}_{\mathcal W'}(0^{N-1}).
\]
Under (UT), implementation of the selected larger-set policy implies implementation of the selected smaller-set policy.
\end{proposition}

The first comparison follows from policy inclusion; the all-zero Bellman identity orders the costs. Under (UT), the lowest-history test gives the implementation implication. Section~\ref{app:restriction-derivations} provides the proof.

For a systematic two-state construction, start from an implementable $(N,Q)$ quota and select a temporally lower-closed collection of entering histories with $Q$ ones. Permit an additional favorable recommendation at those histories, together with the appended-zero continuations needed to preserve prior-action closure. Admitting every such class recovers $(N,Q+1)$. Along a nested sequence of these expansions, the separating rule's initial value strictly increases and its continuation cost strictly decreases. Credible members form an initial segment, so every payoff-maximizing member of the construction class is maximal under inclusion (Proposition~\ref{prop:finite-base-expansions}).

At $p=1/2$, $\beta_S=0.95$, and $\ell=0.379$, the memory-sixteen construction in Section~\ref{app:flexible-construction} yields a normalized payoff of approximately $0.3104999978$, compared with $0.298946$ for the best full quota with memory at most sixteen. The continuation-cost bounds verify credibility, and its positive sustainability margin verifies preservation of communication under belief updating. The comparison holds the memory bound fixed. It establishes an attainable improvement within the general restriction class.

\begin{table}[htbp]
\centering
\begin{tabular}{@{}lr@{}}
\toprule
Restriction & Normalized initial payoff\\
\midrule
Full quota $(16,4)$ & 0.277082\\
Best full quota with memory at most 16: $(14,4)$ & 0.298946\\
Selective restriction at memory 16 & 0.3104999978\\
Ceiling & 0.3105000000\\
\bottomrule
\end{tabular}
\caption{Payoffs at $p=1/2$, $\beta_S=0.95$, and $\ell=0.379$. The
comparison with $(14,4)$ holds the upper bound on memory fixed.}
\label{tab:selective-comparison}
\end{table}

Write $V^{C,\mathrm{IR}}$ for the unnormalized value of the prescribed two-state separating rule under perpetual compliance. When the rule is credible, its all-zero policy equation gives
\[
\overline\Pi-(1-\beta_S)V^{C,\mathrm{IR}}(0^{N-1})
=p\bigl[c^{C,\mathrm{IR}}(0^{N-1})-\ell\bigr].
\]
Thus unused incentive slack translates directly into a payoff gap. When memory is unrestricted, Theorem~\ref{thm:obedience-ceiling-convergence} bounds the advantage of any credible, receiver-obedient finite-memory restriction under perpetual compliance:
\begin{equation}
\Pi_{\mathcal W}-\Pi^*_{\mathrm{full}}(\beta_S)
<\frac{1-\beta_S}{\beta_S}\,\underline c_R,
\qquad \beta_S>\beta^*.
\label{eq:selective-improvement-bound}
\end{equation}
The construction addresses design at a fixed memory bound; the approximation bound addresses the patient limit with memory chosen freely.

\subsection{Nested Expansions}\label{app:restriction-derivations}
\begin{proof}[Proof of Proposition~\ref{prop:binary-exclusion-value}]
Every smaller-set policy is feasible in the larger set, so taking optimal values proves the first inequality for any finite state space. At the all-zero history, both Bellman equations give $(1-\beta_S)\widehat V=F(\widehat c)$. Strict decrease of $F$ below $g_n$ orders the costs, and Proposition~\ref{prop:optimal-policy-obedience} gives the implementation implication.
\end{proof}
The construction and prescribed-policy comparisons specialize to two states. The rule recommends one in $H$ whenever admissible and zero otherwise, with gains $g_H=1$ and $g_L=\ell$.

\paragraph{The prescribed rule under a full quota.}
For any restricted set $\mathcal W\subseteq\admissible^{N,Q}$ with zero continuation, its prescribed high-state payoff is no larger than the full-quota high-state payoff. To see this, fix a state sequence, let $s_t=\mathbf 1\{\theta_t=H\}$, and let $J_{\mathcal W}(t)$ and $J_Q(t)$ count recommendations one through date $t$, with zero counts at nonpositive dates. Every restricted path satisfies
\[
J_{\mathcal W}(t)\leq
\min\{J_{\mathcal W}(t-1)+s_t,\ Q+J_{\mathcal W}(t-N)\}.
\]
The full-quota rule satisfies this recursion with equality. Induction orders the cumulative counts, and summation by parts orders their discounted payoffs. Exclusions can still improve the best credible payoff by restoring incentives: Section~\ref{app:no-lower-closure} gives the example $\admissible^{4,3}\setminus\{1011\}$ at $p=1/2$, $\beta_S=0.9$, and $\ell=0.07$.

\paragraph{Zero-continuation classes.}\label{app:base-layer-construction}
For $1\leq Q<N$, choose a set $\mathcal I$ of entering histories with $Q$ ones, lower closed under temporal precedence within that layer. For each selected history $\nu$, include the successive appended-zero windows of $(\nu,1)$ that preserve its number of ones:
\begin{equation}
\mathcal H_\nu:=\left\{
(\nu_{r+1},\ldots,\nu_{N-1},1,0^r):
0\leq r\leq N-1,\ \sum_{i=1}^{r}\nu_i=0
\right\}.
\label{eq:extension-zero-class}
\end{equation}
The class ends immediately before an inherited one leaves the window. These classes partition the next quota layer:
\begin{equation}
\{\nu^N\in\{0,1\}^N:|\nu^N|=Q+1\}
=\bigsqcup_{\substack{\nu\in\{0,1\}^{N-1}\\|\nu|=Q}}\mathcal H_\nu.
\label{eq:extension-class-partition}
\end{equation}
A window with $r$ trailing zeros has a unique representation $(u,1,0^r)$ and generating history $(0^r,u)$. Thus
\[
\mathcal W_{\mathcal I}:=\admissible^{N,Q}
\cup\bigcup_{\nu\in\mathcal I}\mathcal H_\nu
\]
preserves prior-action closure. Deleting an inherited one returns to the original quota; moving it earlier stays in the selected lower set. Its base set is therefore temporally lower closed. Admitting every class recovers $(N,Q+1)$.

\begin{proposition}[Binary selective-history expansions]
\label{prop:finite-base-expansions}
Along a nested sequence of the sets $\mathcal W_{\mathcal I}$, every proper expansion strictly raises the prescribed separating rule's perpetual-compliance value at $0^{N-1}$ and strictly lowers its continuation cost there. The credible members form an initial segment. Among the credible members of this finite construction class, every payoff-maximizing member is maximal under inclusion.
\end{proposition}
\begin{proof}
First, under a prior-action-closed restriction with a temporally lower-closed base set, the separating rule maximizes discounted high-state rewards among policies constrained to recommend zero in $L$. Fix a finite state sequence and another such feasible policy. At the first disagreement with the separating rule, the separating rule accepts a feasible high state that the other policy rejects. Insert that recommendation and omit the other policy's next favorable recommendation, if one occurs. Until the omission, appended zeros are feasible. After the omission, the modified record temporally precedes the original record because the inserted recommendation is older than the omitted one. Temporal lower closure preserves all subsequent recommendations. The exchange advances a unit reward, or adds one when there is no later recommendation. Iterating proves finite-horizon optimality; bounded discounted tails give the infinite-horizon result. Policy inclusion therefore orders the separating-rule values across nested restrictions, without imposing credibility in advance.

For strictness, take $\nu$ admitted as a base history by the larger selection but not by the smaller one. From the initial history, realize $H$ at its one positions and $L$ at its zero positions, followed by another $H$. This event has positive probability. Every proper prefix contains at most $Q$ ones and is feasible under the original full quota, so both rules produce the record $\nu$. The larger restriction permits an additional favorable recommendation there. Insert it into the smaller-set rule and omit that rule's first subsequent favorable recommendation. The same exchange preserves feasibility and produces a strictly positive discounted gain, at least $1-\beta_S$ measured at the insertion date. The larger restriction's optimal high-state rule therefore has strictly greater expected value.

For every set in the construction, its all-zero policy equation gives
\[
(1-\beta_S)V^{C,\mathrm{IR}}(0^{N-1})
=p\bigl[1-c^{C,\mathrm{IR}}(0^{N-1})\bigr].
\]
Values and costs move in opposite directions. Proposition~\ref{prop:lowest} makes credibility equivalent to $\ell\leq c^{C,\mathrm{IR}}(0^{N-1})$, proving the initial-segment claim. Receiver obedience follows from separation where both recommendations are available and the prior preference where zero is forced. A credible proper expansion strictly raises payoff, so a maximizing credible selection must be maximal under inclusion.
\end{proof}

\paragraph{Weighted searches.}
The numerical construction ranks the candidate windows $(\nu,1)$ with $|\nu|=Q$ by either
\begin{equation}
W_{\beta_S}(\nu^N)=\sum_{i=1}^{N}\beta_S^{N-i}\nu_i,
\qquad W_{\mathrm{bin}}(\nu^N)=\sum_{i=1}^{N}2^{i-1}\nu_i,
\label{eq:extension-weights}
\end{equation}
breaking ties arbitrarily. Temporal precedence strictly lowers either weight when a one is moved earlier. If the ordered candidates are $(\nu^{(1)},1),\ldots,(\nu^{(K)},1)$, where $K=\binom{N-1}{Q}$, the sets
\begin{equation}
\admissible^N_k=\admissible^{N,Q}
\cup\bigcup_{j=1}^{k}\mathcal H_{\nu^{(j)}},
\qquad k=0,\ldots,K,
\label{eq:extension-nested-sets}
\end{equation}
form a nested chain of lower selections. Incomparable lower selections can be compared by their induced payoffs. Each candidate has a separate sustainability test. A search can retain the best candidate and also allow shorter-memory rules. General conditions for arbitrary approximation of the ceiling by deterministic, finite-memory restrictions at a fixed discount factor remain open; Theorem~\ref{thm:obedience-ceiling-convergence} establishes convergence as the sender becomes patient.

\subsection{A verified construction at memory sixteen}
\label{app:flexible-construction}
Set $p=1/2$, $\beta_S=0.95$, and $\ell=0.379$. The normalized ceiling is $\overline\Pi=0.3105$, and its unnormalized counterpart is $\overline V=6.21$.

For a concrete example, take $N=16$, for which $Q^\star_{\mathrm{IR}}(16)=4$.
Write the positions of the four ones in an entering history $\enteringhistory$ as
$i(\enteringhistory)=(i_1,i_2,i_3,i_4)$, numbering the fifteen positions from oldest
to newest. Define
\[
\mathcal J=\left\{
\begin{array}{lll}
(1,2,8,15),&(1,3,5,15),&(1,5,6,14),\\
(2,11,12,13),&(3,4,5,14),&(4,10,12,13),\\
(8,9,12,13),&(8,10,11,13),&(9,10,11,12)
\end{array}
\right\}.
\]
Select every $\enteringhistory$ with $|\enteringhistory|=4$ for which $i(\enteringhistory)$ is coordinatewise no
larger than some element of $\mathcal J$, and call this set $\mathcal I$.
The resulting admissible set is
\[
\admissible^{16}
=\admissible^{16,4}\cup\bigcup_{\enteringhistory\in\mathcal I}\mathcal H_{\enteringhistory}.
\]
It adds $725$ base histories and their zero-continuation classes,
containing $1{,}935$ distinct windows with five ones. By construction,
the base histories are lower closed. Numerical evaluation gives
\[
V^C_{16,4}(\initialhistory)\approx5.541630233,
\qquad
V^C(\initialhistory)\approx6.209999956.
\]
The first value uses memory exactly $16$. With memory at most $16$, the best full quota is $(14,4)$, whose value is approximately $5.978921$; the comparable gain from the construction is approximately $0.231079$ before normalization.
The computed continuation costs range from approximately $0.3790000044$
to $0.8694217953$ at histories permitting message $1$. Both incentive bounds, $\ell\le c^C(\enteringhistory)\le1$, are
satisfied, including numerical error. The Bellman residual gives a
value-error bound below $6.5\times10^{-12}$. Allowing for this error,
the gap to $\overline V$ is less than $4.41\times10^{-8}$.
The sustainability margin over histories at which message $1$ is prohibited is approximately $3.015730>0$. I compute this minimum over the active histories of the
constructed admissible set, including its added quota layer. The
$N=16$ extension is therefore sustainable as well as credible, and its
near-ceiling payoff is attainable under both IR and BU.

The construction therefore raises the normalized payoff by approximately
$0.011554$ relative to the best full quota using memory at most sixteen.
It gives an attainable payoff, without establishing optimality over all
restrictions at that memory. Whether deterministic finite-memory restrictions under immediate reversion can attain or approximate the ceiling arbitrarily closely at each fixed $\beta_S>\beta^\ast$ remains open; with message-dependent public lotteries, \citet[Corollary~1]{kuvalekar2022goodwill} give exact attainment above some discount factor below one.

\subsection{Searched Selective Restrictions}\label{sec:searched-selective}
\paragraph{Construction from the bottom of the temporal order.}
Fix $N$ and order the entering histories as
$\nu^{(1)},\ldots,\nu^{(2^{N-1})}$ so that temporal precedence implies
an earlier position. Every prefix
$\mathcal B_k:=\{\nu^{(1)},\ldots,\nu^{(k)}\}$ is temporally lower closed.
Permit recommendation~$1$ precisely at histories in $\mathcal B_k$ and
permit recommendation~$0$ at every history. Evaluate the separating rule,
which recommends $1$ in $H$ whenever admissible. In the two-state model,
credibility is equivalent to
\[
c_{\mathcal B_k}(0^{N-1})\geq\ell,
\qquad
(1-\beta_S)V_{\mathcal B_k}(0^{N-1})
=p\bigl[1-c_{\mathcal B_k}(0^{N-1})\bigr].
\]
Along the chain, expanding the base set weakly raises the prescribed
payoff and lowers the empty-record cost. Select the largest credible prefix.
This is equivalent to removing histories from the top of the same order
until credibility holds.

The numerical search uses two weights, with histories recorded oldest first:
\[
w_{\mathrm{bin}}(\nu)=\sum_{i=1}^{N-1}2^{i-1}\nu_i,
\qquad
w_{\beta_S}(\nu)=\sum_{i=1}^{N-1}\beta_S^{N-1-i}\nu_i.
\]
Each weight defines a chain by ordering all histories directly, and another
by ordering first by the number of ones and then by weight within each layer.
The latter chains give selective expansions of the next full-quota layer.
The reported comparison retains the best credible prefix across these four
chains and $2\leq N\leq12$, together with the unrestricted-memory full-quota
optimum for $\beta_S\leq0.88$ and certified full-quota lower bounds above
$0.88$. The chains do not exhaust all
temporally lower-closed restrictions.

\paragraph{Numerical comparison.}
With $p=1/2$, $g_H=1$, $g_L=0.3$, and $\beta_S=0.87$, the best constructed
rule uses $N=12$. It starts from quota $(12,3)$ and adds the $76$
lowest-discount-weight base histories containing three ones, together with
their zero continuations. Its normalized payoff is $0.3498956904$, compared
with $0.3484320557$ for the optimal full quota and a ceiling of $0.35$.
Its active-history costs lie between $0.3002086192$ and $0.7384874815$,
and its minimum binding continuation value is $1.1785402915>1$.
The construction is therefore credible and sustainable under belief updating.

The implementation code also evaluates every reachable history, verifies that active-history costs lie between $\ell$ and $1$, checks the Bellman residual, and compares the minimum binding continuation value with $g_H=1$. Thus the reported sustainability claim is checked directly, rather than inferred solely from the empty-record test. The search is over four chains, not over all temporally lower-closed restrictions.

\begin{figure}[p]
\centering
\includegraphics[width=\textwidth]{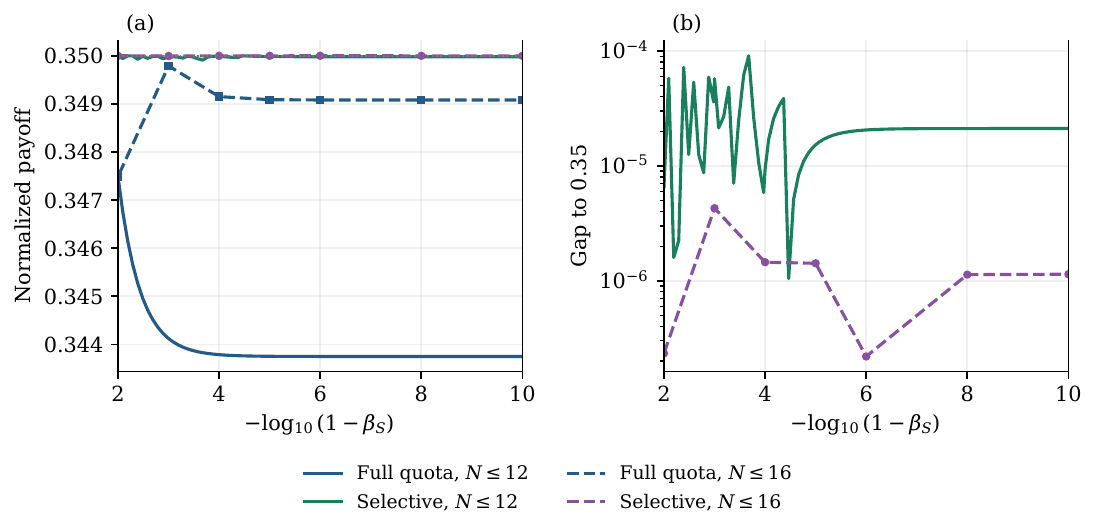}
\caption{Near-one payoffs with matched memory caps, at $p=1/2$ and $\ell=0.3$. Both classes face the same cap, twelve or sixteen. Selective restrictions are the best found along the four searched chains. The right panel reports their gap to $0.35$; increasing the cap reduces the numerical gap without establishing exact attainment.}
\label{fig:selective-matched-memory}
\end{figure}
\begin{figure}[p]
\centering
\includegraphics[width=.9\textwidth]{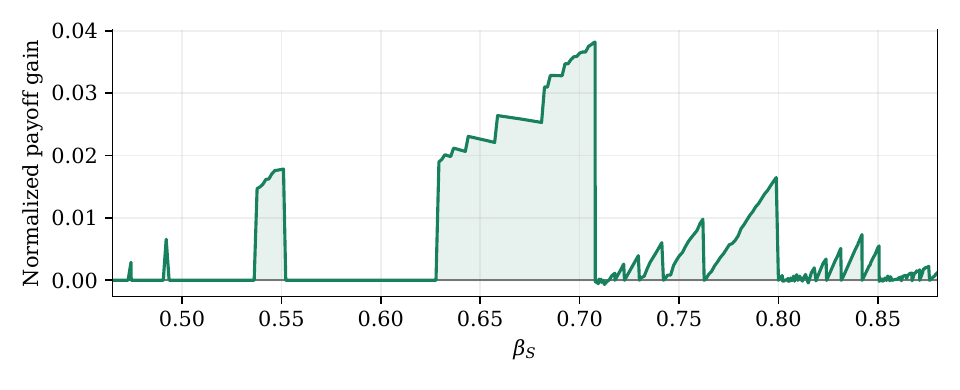}
\caption{Payoff gain from the searched selective restrictions with memory at most twelve over the unrestricted-memory full-quota optimum, for $0.465\le\beta_S\le0.88$, at $p=1/2$ and $\ell=0.3$. The selective curve is a lower bound for its class.}
\label{fig:selective-gain}
\end{figure}

\FloatBarrier
\section{Numerical Illustrations and Policy Evaluation}
\label{sec:supp-methods}\label{sec:supp-quota}
\subsection{Three-State Cutoffs and Sustainability}\label{app:multistate-computation}
The three-state example uses the primitives specified in Section~\ref{subsec:history-cutoffs}. For every $(N,Q)$, enumerate all entering records with at most $Q$ ones. Policy iteration solves the optimal perpetual-compliance problem, allowing the favorable cell to differ at every record. Given a policy with endorsement probability $p_\nu$ and expected current gain $b_\nu$, solve
\[
V(\nu)=b_\nu+\beta_S[(1-p_\nu)V(T_0\nu)+p_\nu V(T_1\nu)].
\]
At binding histories $p_\nu=b_\nu=0$. Improve the policy using the gain comparison with $\beta_S[V(T_0\nu)-V(T_1\nu)]$ and the maintained tie-breaking rule, and repeat until the policy is unchanged. No displayed policy comparison is tied.

The code \path{replicate_multistate.py} evaluates the Bellman residual $\varepsilon=\|TV-V\|_\infty$ after convergence. Contraction bounds the value error by $\varepsilon/(1-\beta_S)$ and the cost error by $2\beta_S\varepsilon/(1-\beta_S)$. It checks that every state-cost comparison is separated from zero by more than this error, verifies both receiver obedience inequalities at every active history, checks every temporally ordered pair, and verifies that the binding minimum equals $B_{N,Q}$. It also checks the quota and memory inequalities on the reported grid and the capacity-release representation by exhaustive enumeration for $N\leq6$. Numerical output is in \path{multistate_results.json}.

Table~\ref{tab:multistate-memory} reports both the empty record and the nonempty record $0^{N-2}1$ at quota two. The latter has exactly one endorsement in the preceding period. Thus it holds the age and number of inherited endorsements fixed as memory changes. The state cutoff at that record changes between $N=5$ and $N=6$; the initial cutoff remains two throughout the implementable rows. Values in this table are perpetual-compliance values even where the no-exit test fails.
\begin{table}[!htbp]
\centering
\caption{Memory at quota two in the three-state example. The starred row fails receiver obedience; the dagger marks failure of the no-exit test.}
\label{tab:multistate-memory}
\begin{tabular}{@{}cccccc@{}}
\toprule
$N$ & Initial cost & Recent-record cost & Recent cutoff & $B_{N,2}$ & Sustained\\
\midrule
$3^{*}$ & 0.1084 & 0.3075 & 2 & 3.0748 & No\\
$4$ & 0.1733 & 0.4279 & 2 & 2.3357 & Yes\\
$5$ & 0.2361 & 0.4928 & 2 & 1.8733 & Yes\\
$6$ & 0.2822 & 0.5256 & 3 & 1.5345 & Yes\\
$7$ & 0.3172 & 0.5486 & 3 & 1.2777 & Yes\\
$8$ & 0.3436 & 0.5711 & 3 & 1.0799 & Yes\\
$9^{\dagger}$ & 0.3644 & 0.5908 & 3 & 0.9221 & No\\
\bottomrule
\end{tabular}
\end{table}

\begin{table}[!htbp]
\centering
\caption{Actual recommendation cells at memory seven. All displayed records are slack.}
\label{tab:multistate-records}
\begin{tabular}{@{}lcccc@{}}
\toprule
&\multicolumn{2}{c}{Quota $Q=2$}&\multicolumn{2}{c}{Quota $Q=3$}\\
\cmidrule(lr){2-3}\cmidrule(l){4-5}
Record & Cost & Favorable cell & Cost & Favorable cell\\
\midrule
000000 & 0.3172 & $\{\theta_2,\theta_3\}$ & 0.1756 & $\{\theta_2,\theta_3\}$\\
100000 & 0.3172 & $\{\theta_2,\theta_3\}$ & 0.1756 & $\{\theta_2,\theta_3\}$\\
010000 & 0.4188 & $\{\theta_2,\theta_3\}$ & 0.1756 & $\{\theta_2,\theta_3\}$\\
001000 & 0.4704 & $\{\theta_2,\theta_3\}$ & 0.2202 & $\{\theta_2,\theta_3\}$\\
000100 & 0.4979 & $\{\theta_2,\theta_3\}$ & 0.2658 & $\{\theta_2,\theta_3\}$\\
000010 & 0.5193 & $\{\theta_3\}$ & 0.3023 & $\{\theta_2,\theta_3\}$\\
000001 & 0.5486 & $\{\theta_3\}$ & 0.3461 & $\{\theta_2,\theta_3\}$\\
\bottomrule
\end{tabular}
\end{table}

\subsection{Notation and Continuation Values}\label{app:notation}
The symbols $V$ and $W$ denote unnormalized discounted sums; multiplication by
$1-\beta_S$ gives normalized payoffs. Distinguish the following policy values.
\begin{center}
\begin{tabularx}{\textwidth}{@{}p{0.22\textwidth}X@{}}
\toprule
Symbol & Continuation problem\\
\midrule
$V^{C,\mathrm{IR}}$ & Prescribed high-state recommendations under perpetual compliance.\\
$V^C$ & Prescribed high-state recommendations at slack histories, with optimal exits at binding histories.\\
$\widehat V^{\mathrm{IR}}$ & Optimal admissible policy under perpetual compliance.\\
$\widehat W$ & Optimal policy allowing both admissible choices and terminal violations.\\
\bottomrule
\end{tabularx}
\end{center}
Prescribed and optimal values coincide when the prescribed policy satisfies all
sender incentive constraints. They can differ for candidates that fail those
constraints. Immediate reversion (IR) assigns action $0$ to a violation; belief
updating (BU) evaluates a violation using its induced posterior. All violations
are followed by permanent babbling from the next period.

For a stationary admissible pure policy $\phi$, define
\[
q_\phi(\nu)=\sum_i\mu_i\mathbf 1\{\phi(\theta_i,\nu)=1\},\qquad
R_\phi(\nu)=\sum_i\mu_i g_i\mathbf 1\{\phi(\theta_i,\nu)=1\}.
\]
Its transition matrix is
\[
P_\phi(\nu,\nu')=(1-q_\phi(\nu))\mathbf 1\{\nu'=T_0\nu\}
+q_\phi(\nu)\mathbf 1\{\nu'=T_1\nu\},
\]
with $q_\phi(\nu)=0$ where $1$ is prohibited. The vector of unnormalized continuation values is
\begin{equation}
\boldsymbol V^\phi=(I-\beta_SP_\phi)^{-1}\boldsymbol R_\phi.
\label{eq:finite-matrix-value}
\end{equation}
The inverse exists because $P_\phi$ is stochastic and $\beta_S<1$. For a fixed completed policy, use the substochastic transition matrix between active histories and include violation-period payoffs in $R_\phi$. This formula also gives the continuation values when breakdown is followed by zero payoffs.

\paragraph{The prescribed two-state rule.}
Use the two-state setup of Section~\ref{sec:binary-environment}: $\Pr(H)=p$, $g_H=1$, and $g_L=\ell$. The prescribed rule under perpetual compliance is
\begin{equation}
\phi^{C,\mathrm{IR}}(\theta,\nu)
=\mathbf 1\{\theta=H,\ \nu\in\mathcal B(\mathcal V)\}.
\label{eq:extension-committed}
\end{equation}
For a full quota, $\mathcal B(\mathcal V)=\{\nu:|\nu|<Q\}$, so $q_\phi(\nu)=R_\phi(\nu)=p\mathbf 1\{|\nu|<Q\}$. Equation~\eqref{eq:finite-matrix-value} gives its value $V^{C,\mathrm{IR}}$, independently of $\ell$. This is a prescribed-policy value, including for noncredible candidates; it equals the optimal admissible value $\widehat V^{\mathrm{IR}}$ when the rule is credible. Write
\begin{align}
c^{C,\mathrm{IR}}(\nu)&=\beta_S[V^{C,\mathrm{IR}}(T_0\nu)-V^{C,\mathrm{IR}}(T_1\nu)],\label{eq:binary-cost}\\
d^{C,\mathrm{IR}}(\nu)&=\ell-c^{C,\mathrm{IR}}(\nu).\label{eq:binary-one-period}
\end{align}

\subsection{Full-quota comparisons at fixed patience}
\label{app:quota-numerics}\label{app:immediate-reversion}\label{sec:binary-illustration}

\begin{figure}[t]
\centering
\includegraphics[width=\textwidth]{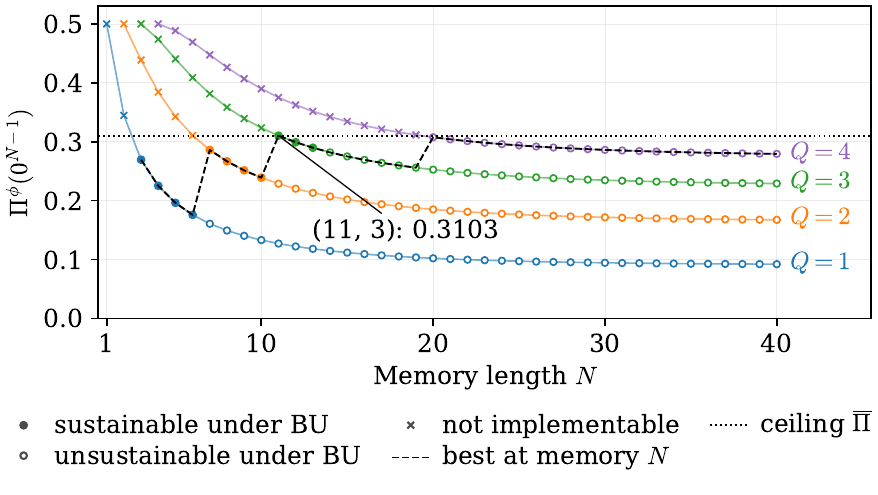}
\caption{Normalized initial values of the separating rule under full quotas and perpetual compliance at $p=1/2$, $\beta_S=0.9$, and $\ell=0.379$. Circles are implementable rules, filled when also sustainable under belief updating; crosses are not implementable and may lie above the ceiling.}
\label{fig:binary-memory-curves}
\end{figure}

Set $p=1/2$, $\beta_S=0.9$, and $\ell=0.379$, as in Figure~\ref{fig:binary-memory-curves}.
Table~\ref{tab:binary-peaks} evaluates each quota at its shortest
IR-implementing memory. The first payoff column evaluates perpetual compliance;
the remaining columns evaluate the candidate completed policy at the same pair.
All payoffs in this table are unnormalized.

\begin{table}[htbp]
\centering
\begin{tabular}{@{}ccrrrr@{}}
\toprule
$Q$ & $N_Q^{\mathrm{IR}}$ & $P_Q^{\mathrm{IR}}$ & $V^C(z)$ & $d^C(z)$ & $s^C$\\
\midrule
1 & 3  & 2.695418 & 2.695418 & $-0.081916$ & 1.183288\\
2 & 7  & 2.861076 & 2.861076 & $-0.048785$ & 0.689437\\
3 & 11 & 3.102816 & 3.102816 & $-0.000437$ & 0.335660\\
4 & 20 & 3.076342 & 3.175507 & $ 0.014101$ & ---\\
\bottomrule
\end{tabular}
\caption{Perpetual-compliance and completed values at each quota's shortest
IR-implementing memory. Here $z=0^{N-1}$ and $d^C(z)=\ell-c^C(z)$.
The first three completed policies are implementable and sustainable.
The fourth fails BU implementation.}
\label{tab:binary-peaks}
\end{table}

The lifetime waiting factor is $\rho=9/11$, and $\rho^5<\ell<\rho^4$, so the
largest IR-eligible quota is four. At the memories in Table~\ref{tab:binary-peaks},
the selected optimizer and the prescribed separating rule coincide. At the
preceding memories they differ, because separation fails credibility.
Table~\ref{tab:preceding-costs} distinguishes their costs. Each optimized cost is
below $\ell$, verifying the shortest implementing memories using the optimal-policy
test. The prescribed costs also fail the separating-rule test.

\begin{table}[htbp]
\centering
\begin{tabular}{@{}ccrr@{}}
\toprule
$Q$ & $N_Q^{\mathrm{IR}}-1$ & $\widehat c^{\mathrm{IR}}(z)$ & $c^{C,\mathrm{IR}}(z)$\\
\midrule
1 & 2 & 0.326605 & 0.310345\\
2 & 6 & 0.378750 & 0.378574\\
3 & 10 & 0.364436 & 0.352991\\
4 & 19 & 0.377764 & 0.376761\\
\bottomrule
\end{tabular}
\caption{Optimal-policy and prescribed-policy costs at the preceding memories.}
\label{tab:preceding-costs}
\end{table}

For BU candidates over $Q<N\le40$, no quota-four candidate is implementable.
Quotas one, two, and three are implementable from memories $3$, $7$, and $11$,
respectively. Their sustainable ranges are $3$--$6$, $7$--$10$, and $11$--$13$;
their longer-memory implementable completions allow breakdown. The best
implementable completed pair in this range is $(11,3)$.

For a two-state calibration with $p=1/2$, $\beta_S=0.9$, and $\ell=0.379$, Figure~\ref{fig:binary-memory-curves} plots the separating rule's normalized initial values under full quotas and perpetual compliance. Where the rule is implementable, this is the optimal full-quota value. The pair $(N,Q)=(11,3)$ gives the highest implementable payoff, $0.310282$, above the $0.307634$ of $(20,4)$, the largest eligible quota at its shortest implementing memory. Thus increasing the quota can require enough extra memory to reduce the best implementable payoff. The filled markers also record sustainability under belief updating, analyzed in Section~\ref{sec:sustainability-analysis}. The horizontal payoff benchmark is derived in Section~\ref{subsec:obedience-ceiling}.

\subsection{Patience thresholds and unsustainable communication}
\label{app:patience-threshold-numerics}\label{subsubsec:quota-illustration}

\begin{figure}[p]
\centering
\includegraphics[width=\textwidth]{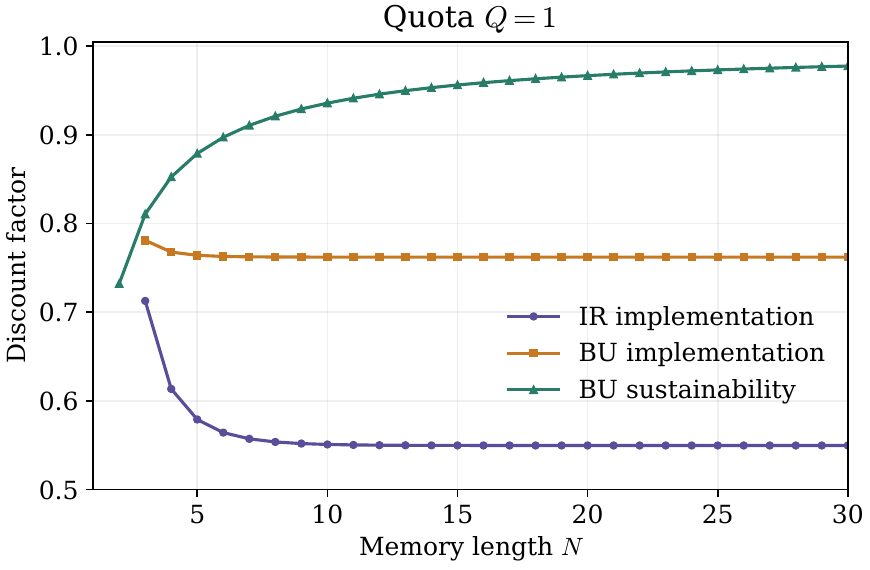}
\caption{Implementation and sustainability thresholds for quota one. An
implementation curve is absent when no discount factor below one qualifies.
Curves connect the evaluated integer memories.}
\label{fig:threshold-q1}
\end{figure}

\begin{figure}[p]
\centering
\includegraphics[width=\textwidth]{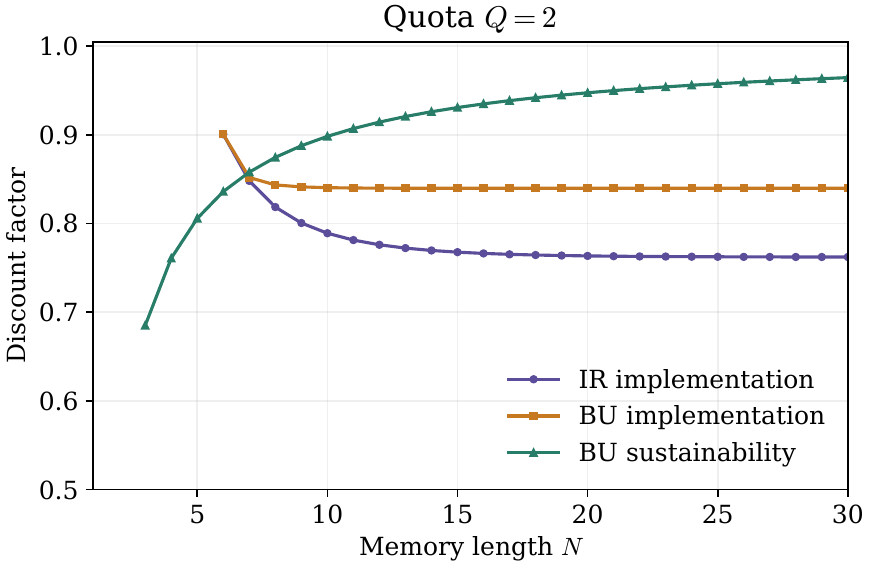}
\caption{Implementation and sustainability thresholds at quota two, $p=1/2$, and $\ell=0.379$. IR evaluates perpetual compliance; BU implementation includes optimal exit decisions. Curves connect the evaluated integer memories.}
\label{fig:threshold-q2}
\end{figure}

\begin{figure}[p]
\centering
\includegraphics[width=\textwidth]{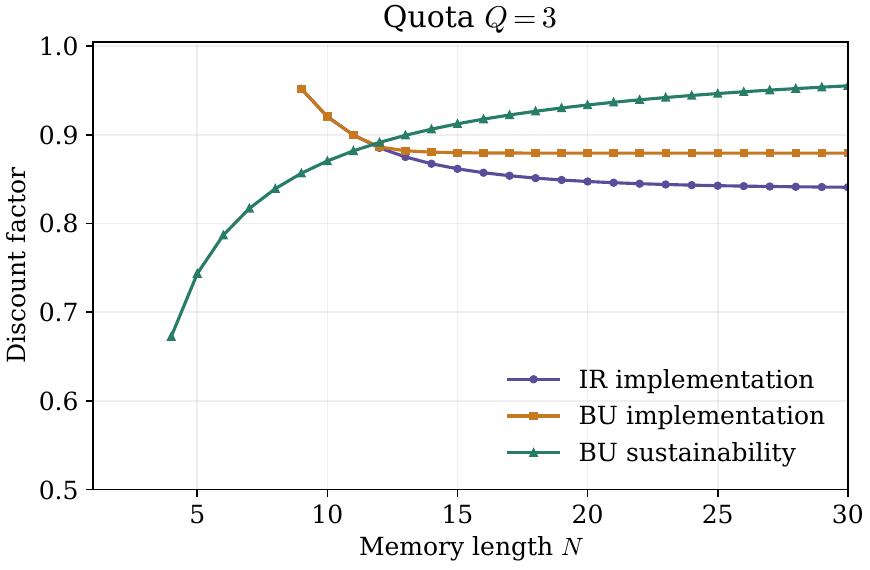}
\caption{Implementation and sustainability thresholds for quota three, with
the same primitives and conventions as Figure~\ref{fig:threshold-q1}.}
\label{fig:threshold-q3}
\end{figure}

Figures~\ref{fig:threshold-q1}, \ref{fig:threshold-q2}, and~\ref{fig:threshold-q3} compare three thresholds
at $p=1/2$ and $\ell=0.379$, for $Q<N\le30$ and $Q\in\{1,2,3\}$.
The IR threshold solves $c^{C,\mathrm{IR}}_{N,Q}(z)=\ell$.
The BU implementation threshold solves $c^C_{N,Q}(z)=\ell$, incorporating
optimal exit decisions. The BU sustainability threshold solves
\eqref{eq:binary-sustainability-root}. At an implementation threshold, the
tie-breaking convention selects recommendation $0$ in $L$; at a sustainability
threshold it selects a violation in $H$, so selected sustainability requires
strictly greater patience.

Longer memory delays replacement of used quota capacity. In this calibration,
both implementation thresholds fall with memory, while the sustainability
threshold rises. The exit option weakly lowers the initial continuation cost,
so BU implementation requires weakly more patience than IR implementation.
Consequently implementation and sustainability can favor an intermediate
memory. For $Q=2$ and $\beta_S=0.9$, both requirements hold exactly at
$N\in\{7,8,9,10\}$ over the displayed range.

\clearpage
\clearpage
\subsection{An Unsustainable Optimum}\label{app:unsustainable-optimum}
For a $Q=1$ illustration of the sustainability comparison in Section~\ref{subsec:memory-survival}, fix $p=1/2$, $\ell=0.379$, and $\beta_S=4/5$. Write $V_N=V^C_{N,1}(0^{N-1})$ and let $W_k$ be the value when the single inherited favorable message expires after $k$ further periods. For $N\geq3$, the binding-history choices give
\[
W_1=\tfrac45 V_N,\qquad
W_k=\tfrac12+\tfrac25 W_{k-1}\quad(2\leq k\leq N-1),
\qquad
V_N=\tfrac12+\tfrac25 W_{N-1}+\tfrac25 V_N.
\]
The sender continues in both states at $k=1$; at $k\geq2$, the sender ends communication in $H$ and continues in $L$. Solving these equations yields
\[
V_N=\frac{125-50x_N}{90-48x_N},\qquad
x_N=\left(\tfrac25\right)^{N-2},\qquad N\geq3.
\]
This solution lies in $25/18<V_N\leq175/118$, which verifies the stated choices: $\tfrac45 V_N>1$, whereas $\tfrac45 W_{k-1}<1$ and $\tfrac45 W_{k-1}>\ell$ for $k\geq2$. Since
\[
\frac{\mathrm d}{\mathrm dx}\frac{125-50x}{90-48x}
=\frac{1500}{(90-48x)^2}>0,
\]
the value strictly decreases with $N$. At $N=3$, $V_3=175/118\simeq1.483051$, $c^C_{3,1}(00)=24/59\simeq0.406780>\ell$, and $s^C_{3,1}=-3/59\simeq-0.050847$. The sole slack history has cost $c^C_{N,1}(0^{N-1})=1-\tfrac25 V_N$, which increases with $N$ and remains below $1$. Thus every $N\geq3$ is implementable under BU. Every such rule is unsustainable because the binding history with $k=2$ has continuation value $\tfrac{16}{25}V_N<1$. At $N=2$, the cost is $2/7<\ell$; $N=1$ is unrestricted and has zero continuation cost. Therefore $N=3$ maximizes the implementable BU payoff for $Q=1$ over all integer memories in this calibration. The numerical values for $3\leq N\leq30$ agree with this formula within $2\times10^{-11}$.

\subsection{A Continuous-State Illustration}
\label{app:continuous-example}

A lobbyist advises a policymaker whether to adopt a policy. Each period, the lobbyist privately observes its probability of success, $\theta_t$, drawn independently and uniformly from $[0,1]$. With $a=1$ denoting adoption, $u_S(\theta,a)=a(\theta+0.20)$ and $u_R(\theta,a)=a(\theta-0.55)$, so the lobbyist prefers adoption in every state, and probabilities and payoffs are evaluated by integration. Let $\beta_S=0.90$ and impose the quota $(N,Q)=(4,2)$, which allows at most two endorsements in four consecutive periods. For a three-message entering history $\nu$, recorded oldest first, the optimal rule recommends adoption exactly when $\theta$ exceeds a cutoff of $0.1001$ at $\nu\in\{000,100\}$, $0.2859$ at $\nu=010$, and $0.3575$ at $\nu=001$, rounded to four decimal places, and recommends $0$ at the binding histories. With one past endorsement, the lobbyist becomes more selective as that endorsement becomes more recent, as Proposition~\ref{prop:quota-cost-order} predicts: an older endorsement expires sooner, reducing the opportunity cost of recommending adoption today. The same quota and number of past endorsements thus induce different signaling partitions, which no fixed state cutoff can reproduce.

The cutoff at a slack history is $t_\nu=c(\nu)-0.20$, with $c(\nu)=\beta_S[V(T_0\nu)-V(T_1\nu)]$. Solving the optimal Bellman system gives $V(000)=V(100)=4.04919$, $V(001)=3.71575$, $V(010)=3.89926$, $V(011)=3.27984$, $V(101)=3.50933$, and $V(110)=3.64427$, and continuation costs $0.30009$, $0.55748$, $0.48587$, and $0.30009$ at $000$, $001$, $010$, and $100$. Receiver obedience to recommendation $1$ requires $(1+t_\nu)/2\ge0.55$, which holds at every slack history, and recommendation $0$ is also obedient. At the binding histories $011$, $101$, and $110$, the values of preserving communication exceed the maximal current sender gain $1.2$, so the arrangement is sustainable. The smallest cutoff exceeds the obedience threshold $0.10$ only by about $9\times10^{-5}$, so the calibration illustrates history-dependent cutoffs but is not robust to small increases in the receiver's threshold. The replication package solves the Bellman system by value iteration with a value-error bound below $2\times10^{-12}$.

\subsection{Computational Materials}\label{app:computational-materials}
The directory \path{simulation/} contains \path{quota_payoff.py}, \path{temporal_construction.py}, \path{patient_beta.py}, \path{plot_patient_beta.py}, \path{simulate_zoom.py}, and \path{plot_zoomed_comparison.py}, together with their JSON/CSV outputs. The script \path{simulation/plot_klr_simple.py} generates Figure~\ref{fig:klr-sawtooth} from the archived full-quota data and \path{simulation/long_memory_results/results.json}. The latter is generated by \path{simulation/long_memory.py}, which evaluates binary-weight prefixes with a forbidden-word automaton at $N=16,32,64,128$, checks Bellman residuals and incentives at every reachable slack state, and combines these constructions with the archived searches at $N\le12$ and full quotas optimized over all finite memory lengths. This is an attained lower bound, not an optimization over all temporally lower-closed restrictions. The script \path{simulation/plot_v191_figures.py} generates Figures~\ref{fig:selective-matched-memory}--\ref{fig:selective-gain} and preserves the earlier two-panel plot under a separate filename. Full-quota optima are exhaustive through $\beta_S=0.88$; selective searches remain lower bounds for their class.

The source package contains the three-state policy-iteration script, \path{replicate_multistate.py}, and its output \path{multistate_results.json}; the script also generates Figure~\ref{fig:multistate-linking}. It contains the submitted quota--memory figure generator, \path{figures/fig1_quota_memory.py}, and the computation script for the appendix, \path{replicate_supplement.py}, with its reported results in \path{supplement_results.json}. The latter computes continuation values, reproduces the numerical tables in the appendix, checks implementation and sustainability over the stated range, verifies the selective construction, and generates the patience-threshold figures. Policy values use contraction iteration with explicit value-error bounds, and threshold roots are bracketed. The script \path{replicate_order_check.py} reproduces Table~\ref{tab:order-check} and writes \path{order_check_results.json}. The script \path{replicate_main.py} reproduces the accelerator calculations in Section~\ref{app:accelerator-calculation}, the boundary examples in Appendix~\ref{app:exclusion-examples} and Section~\ref{app:no-lower-closure} (the latter in exact rational arithmetic), the continuous-state example in Section~\ref{app:continuous-example}, and the exact-attainment example in Section~\ref{app:benchmark-details}, and writes \path{main_results.json}.

\bibliographystyle{chicago}
\bibliography{main}
\end{document}